%
%
%
\def\unredoffs{} \def\redoffs{\voffset=-.31truein\hoffset=-.48truein}
\def\speclscape{}
%
%
%
%
%
\newbox\leftpage \newdimen\fullhsize \newdimen\hstitle \newdimen\hsbody
\tolerance=1000\hfuzz=2pt
\catcode`\@=11 
\ifx\hyperdef\UNd@FiNeD\def\hyperdef#1#2#3#4{#4}\def\hyperref#1#2#3#4{#4}\fi
\def\bigans{b }
\def\answ{b }
%
\ifx\answ\bigans\message{(This will come out unreduced.}
\magnification=1200\unredoffs\baselineskip=16pt plus 2pt minus 1pt
\hsbody=\hsize \hstitle=\hsize 
\else\message{(This will be reduced.} \let\l@r=L
\magnification=1000\baselineskip=16pt plus 2pt minus 1pt \vsize=7truein
\redoffs \hstitle=8truein\hsbody=4.75truein\fullhsize=10truein\hsize=\hsbody
\output={\ifnum\pageno=0 
  \shipout\vbox{\speclscape{\hsize\fullhsize\makeheadline}
    \hbox to \fullhsize{\hfill\pagebody\hfill}}\advancepageno
  \else
  \almostshipout{\leftline{\vbox{\pagebody\makefootline}}}\advancepageno
  \fi}
\def\almostshipout#1{\if L\l@r \count1=1 \message{[\the\count0.\the\count1]}
      \global\setbox\leftpage=#1 \global\let\l@r=R
 \else \count1=2
  \shipout\vbox{\speclscape{\hsize\fullhsize\makeheadline}
      \hbox to\fullhsize{\box\leftpage\hfil#1}}  \global\let\l@r=L\fi}
\fi
%
\newcount\yearltd\yearltd=\year\advance\yearltd by -2000

\def\Title#1#2{\nopagenumbers\abstractfont\hsize=\hstitle\rightline{#1}%
\vskip 1in\centerline{\titlefont #2}\abstractfont\vskip .5in\pageno=0}
\def\Date#1{\vfill\leftline{#1}\tenpoint\supereject\global\hsize=\hsbody%
\footline={\hss\tenrm\hyperdef\hypernoname{page}\folio\folio\hss}}%
%

\def\draftmode{\message{ DRAFTMODE }\def\draftdate{{\rm preliminary draft:
\number\month/\number\day/\number\yearltd\ \ \hourmin}}%
\headline={\hfil\draftdate}\writelabels\baselineskip=20pt plus 2pt minus 2pt
 {\count255=\time\divide\count255 by 60 \xdef\hourmin{\number\count255}
  \multiply\count255 by-60\advance\count255 by\time
  \xdef\hourmin{\hourmin:\ifnum\count255<10 0\fi\the\count255}}}
\def\nolabels{\def\wrlabeL##1{}\def\eqlabeL##1{}\def\reflabeL##1{}}
\def\writelabels{\def\wrlabeL##1{\leavevmode\vadjust{\rlap{\smash%
{\line{{\escapechar=` \hfill\rlap{\sevenrm\hskip.03in\string##1}}}}}}}%
\def\eqlabeL##1{{\escapechar-1\rlap{\sevenrm\hskip.05in\string##1}}}%
\def\reflabeL##1{\noexpand\llap{\noexpand\sevenrm\string\string\string##1}}}
\nolabels
%
\global\newcount\secno \global\secno=0
\global\newcount\meqno \global\meqno=1
\def\s@csym{}
\def\newsec#1{\global\advance\secno by1%
{\toks0{#1}\message{(\the\secno. \the\toks0)}}%
\global\subsecno=0\eqnres@t\let\s@csym\secsym\xdef\secn@m{\the\secno}\noindent
{\bf\hyperdef\hypernoname{section}{\the\secno}{\the\secno.} #1}%
\writetoca{{\string\hyperref{}{section}{\the\secno}{\the\secno.}} {#1}}%
\par\nobreak\medskip\nobreak}
\def\eqnres@t{\xdef\secsym{\the\secno.}\global\meqno=1\bigbreak\bigskip}
\def\sequentialequations{\def\eqnres@t{\bigbreak}}\xdef\secsym{}
\global\newcount\subsecno \global\subsecno=0
\def\subsec#1{\global\advance\subsecno by1%
{\toks0{#1}\message{(\s@csym\the\subsecno. \the\toks0)}}%
\ifnum\lastpenalty>9000\else\bigbreak\fi
\noindent{\it\hyperdef\hypernoname{subsection}{\secn@m.\the\subsecno}%
{\secn@m.\the\subsecno.} #1}\writetoca{\string\quad
{\string\hyperref{}{subsection}{\secn@m.\the\subsecno}{\secn@m.\the\subsecno.}}
{#1}}\par\nobreak\medskip\nobreak}
\def\appendix#1#2{\global\meqno=1\global\subsecno=0\xdef\secsym{\hbox{#1.}}%
\bigbreak\bigskip\noindent{\bf Appendix \hyperdef\hypernoname{appendix}{#1}%
{#1.} #2}{\toks0{(#1. #2)}\message{\the\toks0}}%
\xdef\s@csym{#1.}\xdef\secn@m{#1}%
\writetoca{\string\hyperref{}{appendix}{#1}{Appendix {#1.}} {#2}}%
\par\nobreak\medskip\nobreak}
%
%
\def\checkm@de#1#2{\ifmmode{\def\f@rst##1{##1}\hyperdef\hypernoname{equation}%
{#1}{#2}}\else\hyperref{}{equation}{#1}{#2}\fi}
\def\eqnn#1{\DefWarn#1\xdef #1{(\noexpand\relax\noexpand\checkm@de%
{\s@csym\the\meqno}{\secsym\the\meqno})}%
\wrlabeL#1\writedef{#1\leftbracket#1}\global\advance\meqno by1}
\def\f@rst#1{\c@t#1a\em@ark}\def\c@t#1#2\em@ark{#1}
\def\eqna#1{\DefWarn#1\wrlabeL{#1$\{\}$}%
\xdef #1##1{(\noexpand\relax\noexpand\checkm@de%
{\s@csym\the\meqno\noexpand\f@rst{##1}}{\hbox{$\secsym\the\meqno##1$}})}
\writedef{#1\numbersign1\leftbracket#1{\numbersign1}}\global\advance\meqno by1}
\def\eqn#1#2{\DefWarn#1%
\xdef #1{(\noexpand\hyperref{}{equation}{\s@csym\the\meqno}%
{\secsym\the\meqno})}$$#2\eqno(\hyperdef\hypernoname{equation}%
{\s@csym\the\meqno}{\secsym\the\meqno})\eqlabeL#1$$%
\writedef{#1\leftbracket#1}\global\advance\meqno by1}
\def\xeqn{\expandafter\xe@n}\def\xe@n(#1){#1}
\def\xeqna#1{\expandafter\xe@n#1}
\def\eqns#1{(\e@ns #1{\hbox{}})}
\def\e@ns#1{\ifx\UNd@FiNeD#1\message{eqnlabel \string#1 is undefined.}%
\xdef#1{(?.?)}\fi{\let\hyperref=\relax\xdef\next{#1}}%
\ifx\next\em@rk\def\next{}\else%
\ifx\next#1\xeqn#1\else\def\n@xt{#1}\ifx\n@xt\next#1\else\xeqna#1\fi
\fi\let\next=\e@ns\fi\next}

\def\DefWarn#1{\ifx\UNd@FiNeD#1\else
\immediate\write16{*** WARNING: the label \string#1 is already defined ***}\fi}
%
\newskip\footskip\footskip14pt plus 1pt minus 1pt 
\def\footnotefont{\ninepoint}\def\f@t#1{\footnotefont #1\@foot}
\def\f@@t{\baselineskip\footskip\bgroup\footnotefont\aftergroup\@foot\let\next}
\setbox\strutbox=\hbox{\vrule height9.5pt depth4.5pt width0pt}
\global\newcount\ftno \global\ftno=0
\def\foot{\global\advance\ftno by1\def\foot@rg{\hyperref{}{footnote}%
{\the\ftno}{\the\ftno}\xdef\foot@rg{\noexpand\hyperdef\noexpand\hypernoname%
{footnote}{\the\ftno}{\the\ftno}}}\footnote{$^{\foot@rg}$}}
%
\newwrite\ftfile
\def\footend{\def\foot{\global\advance\ftno by1\chardef\wfile=\ftfile
\hyperref{}{footnote}{\the\ftno}{$^{\the\ftno}$}%
\ifnum\ftno=1\immediate\openout\ftfile=\jobname.fts\fi%
\immediate\write\ftfile{\noexpand\smallskip%
\noexpand\item{\noexpand\hyperdef\noexpand\hypernoname{footnote}
{\the\ftno}{f\the\ftno}:\ }\pctsign}\findarg}%
\def\footatend{\vfill\eject\immediate\closeout\ftfile{\parindent=20pt
\centerline{\bf Footnotes}\nobreak\bigskip\input \jobname.fts }}}
\def\footatend{}
%
%
\global\newcount\refno \global\refno=1
\newwrite\rfile
\def\ref{[\hyperref{}{reference}{\the\refno}{\the\refno}]\nref}
\def\nref#1{\DefWarn#1%
\xdef#1{[\noexpand\hyperref{}{reference}{\the\refno}{\the\refno}]}%
\writedef{#1\leftbracket#1}%
\ifnum\refno=1\immediate\openout\rfile=\jobname.refs\fi
\chardef\wfile=\rfile\immediate\write\rfile{\noexpand\item{[\noexpand\hyperdef%
\noexpand\hypernoname{reference}{\the\refno}{\the\refno}]\ }%
\reflabeL{#1\hskip.31in}\pctsign}\global\advance\refno by1\findarg}
\def\findarg#1#{\begingroup\obeylines\newlinechar=`\^^M\pass@rg}
{\obeylines\gdef\pass@rg#1{\writ@line\relax #1^^M\hbox{}^^M}%
\gdef\writ@line#1^^M{\expandafter\toks0\expandafter{\striprel@x #1}%
\edef\next{\the\toks0}\ifx\next\em@rk\let\next=\endgroup\else\ifx\next\empty%
\else\immediate\write\wfile{\the\toks0}\fi\let\next=\writ@line\fi\next\relax}}
\def\striprel@x#1{} \def\em@rk{\hbox{}}
\def\lref{\begingroup\obeylines\lr@f}
\def\lr@f#1#2{\DefWarn#1\gdef#1{\let#1=\UNd@FiNeD\ref#1{#2}}\endgroup\unskip}

\def\addref#1{\immediate\write\rfile{\noexpand\item{}#1}} 
\def\listrefs{\footatend\vfill\supereject\immediate\closeout\rfile\writestoppt
\baselineskip=\footskip\centerline{{\bf References}}\bigskip{\parindent=20pt%
\frenchspacing\escapechar=` \input \jobname.refs\vfill\eject}\nonfrenchspacing}
\def\startrefs#1{\immediate\openout\rfile=\jobname.refs\refno=#1}
\def\xref{\expandafter\xr@f}\def\xr@f[#1]{#1}
\def\refs#1{\count255=1[\r@fs #1{\hbox{}}]}
\def\r@fs#1{\ifx\UNd@FiNeD#1\message{reflabel \string#1 is undefined.}%
\nref#1{need to supply reference \string#1.}\fi%
\vphantom{\hphantom{#1}}{\let\hyperref=\relax\xdef\next{#1}}%
\ifx\next\em@rk\def\next{}%
\else\ifx\next#1\ifodd\count255\relax\xref#1\count255=0\fi%
\else#1\count255=1\fi\let\next=\r@fs\fi\next}
%

%
\newwrite\ffile\global\newcount\figno \global\figno=1
\def\fig{fig.~\hyperref{}{figure}{\the\figno}{\the\figno}\nfig}
\def\nfig#1{\DefWarn#1%
\xdef#1{fig.~\noexpand\hyperref{}{figure}{\the\figno}{\the\figno}}%
\writedef{#1\leftbracket fig.\noexpand~\xfig#1}%
\ifnum\figno=1\immediate\openout\ffile=\jobname.figs\fi\chardef\wfile=\ffile%
{\let\hyperref=\relax
\immediate\write\ffile{\noexpand\medskip\noexpand\item{Fig.\ %
\noexpand\hyperdef\noexpand\hypernoname{figure}{\the\figno}{\the\figno}. }
\reflabeL{#1\hskip.55in}\pctsign}}\global\advance\figno by1\findarg}
\def\listfigs{\vfill\eject\immediate\closeout\ffile{\parindent40pt
\baselineskip14pt\centerline{{\bf Figure Captions}}\nobreak\medskip
\escapechar=` \input \jobname.figs\vfill\eject}}
\def\xfig{\expandafter\xf@g}\def\xf@g fig.\penalty\@M\ {}
\def\figs#1{figs.~\f@gs #1{\hbox{}}}
\def\f@gs#1{{\let\hyperref=\relax\xdef\next{#1}}\ifx\next\em@rk\def\next{}\else
\ifx\next#1\xfig #1\else#1\fi\let\next=\f@gs\fi\next}
\def\figin{\epsfcheck\figin}\def\figins{\epsfcheck\figins}
\def\epsfcheck{\ifx\epsfbox\UNd@FiNeD
\message{(NO epsf.tex, FIGURES WILL BE IGNORED)}
\gdef\figin##1{\vskip2in}\gdef\figins##1{\hskip.5in}
\else\message{(FIGURES WILL BE INCLUDED)}%
\gdef\figin##1{##1}\gdef\figins##1{##1}\fi}
\def\DefWarn#1{}
\def\figinsert{\goodbreak\midinsert}
\def\ifig#1#2#3{\DefWarn#1\xdef#1{fig.~\noexpand\hyperref{}{figure}%
{\the\figno}{\the\figno}}\writedef{#1\leftbracket fig.\noexpand~\xfig#1}%
\figinsert\figin{\centerline{#3}}\medskip\centerline{\vbox{\baselineskip12pt
\advance\hsize by -1truein\noindent\wrlabeL{#1=#1}\footnotefont%
{\bf Fig.~\hyperdef\hypernoname{figure}{\the\figno}{\the\figno}:} #2}}
\bigskip\endinsert\global\advance\figno by1}
\newwrite\lfile
{\escapechar-1\xdef\pctsign{\string\%}\xdef\leftbracket{\string\{}
\xdef\rightbracket{\string\}}\xdef\numbersign{\string\#}}
\def\writedefs{\immediate\openout\lfile=\jobname.defs \def\writedef##1{%
{\let\hyperref=\relax\let\hyperdef=\relax\let\hypernoname=\relax
 \immediate\write\lfile{\string\def\string##1\rightbracket}}}}%
\def\writestop{\def\writestoppt{\immediate\write\lfile{\string\pageno
 \the\pageno\string\startrefs\leftbracket\the\refno\rightbracket
 \string\def\string\secsym\leftbracket\secsym\rightbracket
 \string\secno\the\secno\string\meqno\the\meqno}\immediate\closeout\lfile}}
\def\writestoppt{}\def\writedef#1{}
\def\seclab#1{\DefWarn#1%
\xdef #1{\noexpand\hyperref{}{section}{\the\secno}{\the\secno}}%
\writedef{#1\leftbracket#1}\wrlabeL{#1=#1}}
\def\subseclab#1{\DefWarn#1%
\xdef #1{\noexpand\hyperref{}{subsection}{\secn@m.\the\subsecno}%
{\secn@m.\the\subsecno}}\writedef{#1\leftbracket#1}\wrlabeL{#1=#1}}
\def\applab#1{\DefWarn#1%
\xdef #1{\noexpand\hyperref{}{appendix}{\secn@m}{\secn@m}}%
\writedef{#1\leftbracket#1}\wrlabeL{#1=#1}}
\newwrite\tfile \def\writetoca#1{}
\def\leaderfill{\leaders\hbox to 1em{\hss.\hss}\hfill}
\def\writetoc{\immediate\openout\tfile=\jobname.toc
   \def\writetoca##1{{\edef\next{\write\tfile{\noindent ##1
   \string\leaderfill {\string\hyperref{}{page}{\noexpand\number\pageno}%
                       {\noexpand\number\pageno}} \par}}\next}}}
\newread\ch@ckfile
\def\listtoc{\immediate\closeout\tfile\immediate\openin\ch@ckfile=\jobname.toc
\ifeof\ch@ckfile\message{no file \jobname.toc, no table of contents this pass}%
\else\closein\ch@ckfile\centerline{\bf Contents}\nobreak\medskip%
{\baselineskip=12pt\footnotefont\parskip=0pt\catcode`\@=11\input\jobname.toc
\catcode`\@=12\bigbreak\bigskip}\fi}
\catcode`\@=12 
%
\edef\tfontsize{\ifx\answ\bigans scaled\magstep3\else scaled\magstep4\fi}
\font\titlerm=cmr10 \tfontsize \font\titlerms=cmr7 \tfontsize
\font\titlermss=cmr5 \tfontsize \font\titlei=cmmi10 \tfontsize
\font\titleis=cmmi7 \tfontsize \font\titleiss=cmmi5 \tfontsize
\font\titlesy=cmsy10 \tfontsize \font\titlesys=cmsy7 \tfontsize
\font\titlesyss=cmsy5 \tfontsize \font\titleit=cmti10 \tfontsize
\skewchar\titlei='177 \skewchar\titleis='177 \skewchar\titleiss='177
\skewchar\titlesy='60 \skewchar\titlesys='60 \skewchar\titlesyss='60
\def\titlefont{\def\rm{\fam0\titlerm}
\textfont0=\titlerm \scriptfont0=\titlerms \scriptscriptfont0=\titlermss
\textfont1=\titlei \scriptfont1=\titleis \scriptscriptfont1=\titleiss
\textfont2=\titlesy \scriptfont2=\titlesys \scriptscriptfont2=\titlesyss
\textfont\itfam=\titleit \def\it{\fam\itfam\titleit}\rm}
 \ifx\answ\bigans\else scaled\magstep1\fi
\ifx\answ\bigans\def\abstractfont{\tenpoint}\else
\font\absit=cmti10 scaled \magstep1
\font\abssl=cmsl10 scaled \magstep1
\font\absrm=cmr10 scaled\magstep1 \font\absrms=cmr7 scaled\magstep1
\font\absrmss=cmr5 scaled\magstep1 \font\absi=cmmi10 scaled\magstep1
\font\absis=cmmi7 scaled\magstep1 \font\absiss=cmmi5 scaled\magstep1
\font\abssy=cmsy10 scaled\magstep1 \font\abssys=cmsy7 scaled\magstep1
\font\abssyss=cmsy5 scaled\magstep1 \font\absbf=cmbx10 scaled\magstep1
\skewchar\absi='177 \skewchar\absis='177 \skewchar\absiss='177
\skewchar\abssy='60 \skewchar\abssys='60 \skewchar\abssyss='60
\def\abstractfont{\def\rm{\fam0\absrm}
\textfont0=\absrm \scriptfont0=\absrms \scriptscriptfont0=\absrmss
\textfont1=\absi \scriptfont1=\absis \scriptscriptfont1=\absiss
\textfont2=\abssy \scriptfont2=\abssys \scriptscriptfont2=\abssyss
\textfont\itfam=\absit \def\it{\fam\itfam\absit}\def\footnotefont{\tenpoint}%
\textfont\slfam=\abssl \def\sl{\fam\slfam\abssl}%
\textfont\bffam=\absbf \def\bf
{\fam\bffam\absbf}\rm}\fi
\def\tenpoint{\def\rm{\fam0\tenrm}
\textfont0=\tenrm \scriptfont0=\sevenrm \scriptscriptfont0=\fiverm
\textfont1=\teni  \scriptfont1=\seveni  \scriptscriptfont1=\fivei
\textfont2=\tensy \scriptfont2=\sevensy \scriptscriptfont2=\fivesy
\textfont\itfam=\tenit \def\it{\fam\itfam\tenit}\def\footnotefont{\ninepoint}%
\textfont\bffam=\tenbf \def\bf{\fam\bffam\tenbf}\def\sl{\fam\slfam\tensl}\rm}
\font\ninerm=cmr9 \font\sixrm=cmr6 \font\ninei=cmmi9 \font\sixi=cmmi6
\font\ninesy=cmsy9 \font\sixsy=cmsy6 \font\ninebf=cmbx9
\font\nineit=cmti9 \font\ninesl=cmsl9 \skewchar\ninei='177
\skewchar\sixi='177 \skewchar\ninesy='60 \skewchar\sixsy='60
\def\ninepoint{\def\rm{\fam0\ninerm}
\textfont0=\ninerm \scriptfont0=\sixrm \scriptscriptfont0=\fiverm
\textfont1=\ninei \scriptfont1=\sixi \scriptscriptfont1=\fivei
\textfont2=\ninesy \scriptfont2=\sixsy \scriptscriptfont2=\fivesy
\textfont\itfam=\ninei \def\it{\fam\itfam\nineit}\def\sl{\fam\slfam\ninesl}%
\textfont\bffam=\ninebf \def\bf{\fam\bffam\ninebf}\rm}
%
%
\def\noblackbox{\overfullrule=0pt}
\hyphenation{anom-aly anom-alies coun-ter-term coun-ter-terms}
\def\inv{^{\raise.15ex\hbox{${\scriptscriptstyle -}$}\kern-.05em 1}}

\def\Dsl{\,\raise.15ex\hbox{/}\mkern-13.5mu D} 
\def\dsl{\raise.15ex\hbox{/}\kern-.57em\partial}

\def\lspace{\ifx\answ\bigans{}\else\qquad\fi}
\def\lbspace{\ifx\answ\bigans{}\else\hskip-.2in\fi} 

\def\boxeqn#1{\vcenter{\vbox{\hrule\hbox{\vrule\kern3pt\vbox{\kern3pt
	\hbox{${\displaystyle #1}$}\kern3pt}\kern3pt\vrule}\hrule}}}
\def\mbox#1#2{\vcenter{\hrule \hbox{\vrule height#2in
		\kern#1in \vrule} \hrule}}  

\def\darr#1{\raise1.5ex\hbox{$\leftrightarrow$}\mkern-16.5mu #1}

\def\roughly#1{\raise.3ex\hbox{$#1$\kern-.75em\lower1ex\hbox{$\sim$}}}

\input amssym
\input epsf

\def\IZ{\relax\ifmmode\mathchoice
{\hbox{\cmss Z\kern-.4em Z}}{\hbox{\cmss Z\kern-.4em Z}} {\lower.9pt\hbox{\cmsss Z\kern-.4em Z}}
{\lower1.2pt\hbox{\cmsss Z\kern-.4em Z}}\else{\cmss Z\kern-.4em Z}\fi}

\newif\ifdraft\draftfalse
\newif\ifinter\interfalse
\ifdraft\draftmode\else\interfalse\fi
\def\journal#1&#2(#3){\unskip, \sl #1\ \bf #2 \rm(19#3) }
\def\andjournal#1&#2(#3){\sl #1~\bf #2 \rm (19#3) }

\def\ie{{\it i.e.}}
\def\eg{{\it e.g.}}

\def\frac#1#2{{#1\over#2}}

\def\ds{\displaystyle}

\def\inbar{\,\vrule height1.5ex width.4pt depth0pt}
\def\IC{\relax\hbox{$\inbar\kern-.3em{\rm C}$}}
\def\IR{\relax{\rm I\kern-.18em R}}
\def\IP{\relax{\rm I\kern-.18em P}}

%
%


%
\catcode`\@=11
\def\slash#1{\mathord{\mathpalette\c@ncel{#1}}}
\overfullrule=0pt

\def\CC{{\cal C}}

\def\S{\hbox{$\bb S$}}

\def\Z{\hbox{$\bb Z$}}

\def\underrel#1\over#2{\mathrel{\mathop{\kern\z@#1}\limits_{#2}}}

\catcode`\@=12


%

\def\mod{{\rm mod}}


\def\[{[}
\def\]{]}

\def\comment#1{ }

%
\def\draftnote#1{\ifdraft{\baselineskip2ex
                 \vbox{\kern1em\hrule\hbox{\vrule\kern1em\vbox{\kern1ex
                 \noindent \underbar{NOTE}: #1
             \vskip1ex}\kern1em\vrule}\hrule}}\fi}
\def\internote#1{\ifinter{\baselineskip2ex
                 \vbox{\kern1em\hrule\hbox{\vrule\kern1em\vbox{\kern1ex
                 \noindent \underbar{Internal Note}: #1
             \vskip1ex}\kern1em\vrule}\hrule}}\fi}

%

%
%

%

\def\inv{^{-1}}



\def\b{\beta}

\def\CSg{{\rm CS_{grav}}}

\def\bC{{\bf C}}

\def\bT{{\bf T}}


\def\bb{
\font\tenmsb=msbm10
\font\sevenmsb=msbm7
\font\fivemsb=msbm5
\textfont1=\tenmsb
\scriptfont1=\sevenmsb
\scriptscriptfont1=\fivemsb
}




\def\wt{\widetilde}
\def\wh{\widehat}
\def\bar{\overline}
\def\b{\bar}
\def\bsq#1{{{\b{#1}}^{\lower 2.5pt\hbox{$\scriptstyle 2$}}}}
\def\bexp#1#2{{{\b{#1}}^{\lower 2.5pt\hbox{$\scriptstyle #2$}}}}
\def\dotexp#1#2{{{#1}^{\lower 2.5pt\hbox{$\scriptstyle #2$}}}}


\def\Tr{\mathop{\rm Tr}}

\def\rt2{\sqrt{2}}

\def\mod{{\rm mod}}



\def\CA{{\cal A}}
\def\CB{{\cal B}}
\def\CC{{\cal C}}

\def\CH{{\cal H}}

\def\CL{{\cal L}}
\def\CM{{\cal M}}
\def\CN{{\cal N}}
\def\CO{{\cal O}}
\def\CP{{\cal P}}

\def\CT{{\cal T}}

\def\CW{{\cal W}}


\def\1{{\ds 1}}

\def\C{\hbox{$\bb C$}}

\def\Z{\hbox{$\bb Z$}}

\def\P{\hbox{$\bb P$}}
\def\S{\hbox{$\bb S$}}


\noblackbox

\def\unit{\relax{\rm 1\kern-.26em I}}
\def\nada{\relax{\rm 0\kern-.30em l}}

\def\mod{{\rm \ mod \ }}

\noblackbox
\def\IL{\relax{\rm I\kern-.18em L}}
\def\IH{\relax{\rm I\kern-.18em H}}
\def\IR{\relax{\rm I\kern-.18em R}}
\def\IC{\relax\hbox{$\inbar\kern-.3em{\rm C}$}}
\def\IZ{\relax\ifmmode\mathchoice
{\hbox{\cmss Z\kern-.4em Z}}{\hbox{\cmss Z\kern-.4em Z}} {\lower.9pt\hbox{\cmsss Z\kern-.4em Z}}
{\lower1.2pt\hbox{\cmsss Z\kern-.4em Z}}\else{\cmss Z\kern-.4em Z}\fi}
\def\CM {{\cal M}}

\def\partialslash{\not{\hbox{\kern-2pt $\partial$}}}
\def\cO{{\cal O}}

\def\cP{{\cal P}}

\font\manual=manfnt \def\dbend{\lower3.5pt\hbox{\manual\char127}}

\def\IZ{\relax\ifmmode\mathchoice
{\hbox{\cmss Z\kern-.4em Z}}{\hbox{\cmss Z\kern-.4em Z}} {\lower.9pt\hbox{\cmsss Z\kern-.4em Z}}
{\lower1.2pt\hbox{\cmsss Z\kern-.4em Z}}\else{\cmss Z\kern-.4em Z}\fi}
\def\half{{1\over 2}}

\def\bar{\overline}

\def\rt2{\sqrt{2}}
\def\irt2{{1\over\sqrt{2}}}

\def\slashchar#1{\setbox0=\hbox{$#1$}           
   \dimen0=\wd0                                 
   \setbox1=\hbox{/} \dimen1=\wd1               
   \ifdim\dimen0>\dimen1                        
      \rlap{\hbox to \dimen0{\hfil/\hfil}}      
      #1                                        
   \else                                        
      \rlap{\hbox to \dimen1{\hfil$#1$\hfil}}   
      /                                         
   \fi}

\def\gcd{{\rm gcd}}
\def\lcm{{\rm lcm}}

\def\figcaption#1#2{\DefWarn#1\xdef#1{Figure~\noexpand\hyperref{}{figure}%
{\the\figno}{\the\figno}}\writedef{#1\leftbracket Figure\noexpand~\xfig#1}%
\medskip\centerline{{\footnotefont\bf Figure~\hyperdef\hypernoname{figure}{\the\figno}{\the\figno}:}  #2 \wrlabeL{#1=#1}}%
\global\advance\figno by1}

%
%

\lref\AbanovQZ{
  A.~G.~Abanov and P.~B.~Wiegmann,
  ``Theta terms in nonlinear sigma models,''
Nucl.\ Phys.\ B {\bf 570}, 685 (2000).
[hep-th/9911025].
}

\lref\AffleckAS{
  I.~Affleck, J.~A.~Harvey and E.~Witten,
  ``Instantons and (Super)Symmetry Breaking in (2+1)-Dimensions,''
Nucl.\ Phys.\ B {\bf 206}, 413 (1982).
}

\lref\AharonyBX{
  O.~Aharony, A.~Hanany, K.~A.~Intriligator, N.~Seiberg and M.~J.~Strassler,
  ``Aspects of $N=2$ supersymmetric gauge theories in three-dimensions,''
Nucl.\ Phys.\ B {\bf 499}, 67 (1997).
[hep-th/9703110].
}

\lref\AharonyGP{
  O.~Aharony,
  ``IR duality in $d = 3$ $N=2$ supersymmetric $USp(2N_c)$ and $U(N_c)$ gauge theories,''
Phys.\ Lett.\ B {\bf 404}, 71 (1997).
[hep-th/9703215].
}

\lref\AharonyCI{
  O.~Aharony and I.~Shamir,
  ``On $O(N_c)$ $d=3$ ${\cal N}{=}2$ supersymmetric QCD Theories,''
JHEP {\bf 1112}, 043 (2011).
[arXiv:1109.5081 [hep-th]].
}

\lref\AharonyJZ{
  O.~Aharony, G.~Gur-Ari and R.~Yacoby,
  ``$d=3$ Bosonic Vector Models Coupled to Chern-Simons Gauge Theories,''
JHEP {\bf 1203}, 037 (2012).
[arXiv:1110.4382 [hep-th]].
}

\lref\AharonyNH{
  O.~Aharony, G.~Gur-Ari and R.~Yacoby,
  ``Correlation Functions of Large $N$ Chern-Simons-Matter Theories and Bosonization in Three Dimensions,''
JHEP {\bf 1212}, 028 (2012).
[arXiv:1207.4593 [hep-th]].
}

\lref\AharonyNS{
  O.~Aharony, S.~Giombi, G.~Gur-Ari, J.~Maldacena and R.~Yacoby,
  ``The Thermal Free Energy in Large $N$ Chern-Simons-Matter Theories,''
JHEP {\bf 1303}, 121 (2013).
[arXiv:1211.4843 [hep-th]].
}

\lref\AharonyHDA{
  O.~Aharony, N.~Seiberg and Y.~Tachikawa,
  ``Reading between the lines of four-dimensional gauge theories,''
JHEP {\bf 1308}, 115 (2013).
[arXiv:1305.0318 [hep-th]].
}

\lref\AharonyDHA{
  O.~Aharony, S.~S.~Razamat, N.~Seiberg and B.~Willett,
  ``3d dualities from 4d dualities,''
JHEP {\bf 1307}, 149 (2013).
[arXiv:1305.3924 [hep-th]].
}

\lref\AharonyKMA{
  O.~Aharony, S.~S.~Razamat, N.~Seiberg and B.~Willett,
  ``3d dualities from 4d dualities for orthogonal groups,''
JHEP {\bf 1308}, 099 (2013)
[arXiv:1307.0511 [hep-th]].
}

\lref\AharonyMJS{
  O.~Aharony,
  ``Baryons, monopoles and dualities in Chern-Simons-matter theories,''
JHEP {\bf 1602}, 093 (2016).
[arXiv:1512.00161 [hep-th]].
}

\lref\AharonyJVV{
  O.~Aharony, F.~Benini, P.~S.~Hsin and N.~Seiberg,
  ``Chern-Simons-matter dualities with $SO$ and $USp$ gauge groups,''
JHEP {\bf 1702}, 072 (2017).
[arXiv:1611.07874 [cond-mat.str-el]].
}

\lref\WangTXT{
  C.~Wang, A.~Nahum, M.~A.~Metlitski, C.~Xu and T.~Senthil,
  ``Deconfined quantum critical points: symmetries and dualities,''
[arXiv:1703.02426 [cond-mat.str-el]].
}

\lref\AlvarezGaumeNF{
  L.~Alvarez-Gaume, S.~Della Pietra and G.~W.~Moore,
  ``Anomalies and Odd Dimensions,''
Annals Phys.\  {\bf 163}, 288 (1985).
}

\lref\AnninosUI{
  D.~Anninos, T.~Hartman and A.~Strominger,
  ``Higher Spin Realization of the dS/CFT Correspondence,''
[arXiv:1108.5735 [hep-th]].
}

\lref\AnninosHIA{
  D.~Anninos, R.~Mahajan, D�.~Radicevic and E.~Shaghoulian,
  ``Chern-Simons-Ghost Theories and de Sitter Space,''
JHEP {\bf 1501}, 074 (2015).
[arXiv:1405.1424 [hep-th]].
}

\lref\AtiyahJF{
  M.~F.~Atiyah, V.~K.~Patodi and I.~M.~Singer,
  ``Spectral Asymmetry in Riemannian Geometry, I,''
  Math.\ Proc.\ Camb.\ Phil.\ Soc.\ {\bf 77} (1975) 43--69.
}

\lref\BanksZN{
  T.~Banks and N.~Seiberg,
  ``Symmetries and Strings in Field Theory and Gravity,''
Phys.\ Rev.\ D {\bf 83}, 084019 (2011).
[arXiv:1011.5120 [hep-th]].
}

\lref\BarkeshliIDA{
  M.~Barkeshli and J.~McGreevy,
  ``Continuous transition between fractional quantum Hall and superfluid states,''
Phys.\ Rev.\ B {\bf 89}, 235116 (2014).
}

\lref\BeemMB{
  C.~Beem, T.~Dimofte and S.~Pasquetti,
  ``Holomorphic Blocks in Three Dimensions,''
[arXiv:1211.1986 [hep-th]].
}

\lref\BeniniMF{
  F.~Benini, C.~Closset and S.~Cremonesi,
  ``Comments on 3d Seiberg-like dualities,''
JHEP {\bf 1110}, 075 (2011).
[arXiv:1108.5373 [hep-th]].
}

\lref\BernardXY{
  D.~Bernard,
  ``String Characters From {Kac-Moody} Automorphisms,''
  Nucl.\ Phys.\ B {\bf 288}, 628 (1987).
}

\lref\BhattacharyaZY{
  J.~Bhattacharya, S.~Bhattacharyya, S.~Minwalla and S.~Raju,
  ``Indices for Superconformal Field Theories in 3,5 and 6 Dimensions,''
JHEP {\bf 0802}, 064 (2008).
[arXiv:0801.1435 [hep-th]].
}

\lref\deBoerMP{
  J.~de Boer, K.~Hori, H.~Ooguri and Y.~Oz,
  ``Mirror symmetry in three-dimensional gauge theories, quivers and D-branes,''
Nucl.\ Phys.\ B {\bf 493}, 101 (1997).
[hep-th/9611063].
}

\lref\deBoerKA{
  J.~de Boer, K.~Hori, Y.~Oz and Z.~Yin,
  ``Branes and mirror symmetry in N=2 supersymmetric gauge theories in three-dimensions,''
Nucl.\ Phys.\ B {\bf 502}, 107 (1997).
[hep-th/9702154].
}

\lref\BondersonPLA{
  P.~Bonderson, C.~Nayak and X.~L.~Qi,
  ``A time-reversal invariant topological phase at the surface of a 3D topological insulator,''
J.\ Stat.\ Mech.\  {\bf 2013}, P09016 (2013).
}

\lref\BorokhovIB{
  V.~Borokhov, A.~Kapustin and X.~k.~Wu,
  ``Topological disorder operators in three-dimensional conformal field theory,''
JHEP {\bf 0211}, 049 (2002).
[hep-th/0206054].
}

\lref\GaiottoYUP{
  D.~Gaiotto, A.~Kapustin, Z.~Komargodski and N.~Seiberg,
  ``Theta, Time Reversal, and Temperature,''
[arXiv:1703.00501 [hep-th]].
}

\lref\BorokhovCG{
  V.~Borokhov, A.~Kapustin and X.~k.~Wu,
  ``Monopole operators and mirror symmetry in three-dimensions,''
JHEP {\bf 0212}, 044 (2002).
[hep-th/0207074].
}

\lref\Browder{
  W.~Browder and E.~Thomas,
  ``Axioms for the generalized Pontryagin cohomology operations,''
  Quart.\ J.\ Math.\ Oxford {\bf 13}, 55--60 (1962).
}

\lref\debult{
  F.~van~de~Bult,
  ``Hyperbolic Hypergeometric Functions,''
University of Amsterdam Ph.D. thesis
}

\lref\Camperi{
  M.~Camperi, F.~Levstein and G.~Zemba,
  ``The Large N Limit Of Chern-simons Gauge Theory,''
  Phys.\ Lett.\ B {\bf 247} (1990) 549.
}

\lref\ChenCD{
  W.~Chen, M.~P.~A.~Fisher and Y.~S.~Wu,
  ``Mott transition in an anyon gas,''
Phys.\ Rev.\ B {\bf 48}, 13749 (1993).
[cond-mat/9301037].
}

\lref\ChenJHA{
  X.~Chen, L.~Fidkowski and A.~Vishwanath,
  ``Symmetry Enforced Non-Abelian Topological Order at the Surface of a Topological Insulator,''
Phys.\ Rev.\ B {\bf 89}, no. 16, 165132 (2014).
[arXiv:1306.3250 [cond-mat.str-el]].
}

\lref\ChengPDN{
  M.~Cheng and C.~Xu,
  ``Series of (2+1)-dimensional stable self-dual interacting conformal field theories,''
Phys.\ Rev.\ B {\bf 94}, 214415 (2016). 
[arXiv:1609.02560 [cond-mat.str-el]].
}

\lref\ClossetVG{
  C.~Closset, T.~T.~Dumitrescu, G.~Festuccia, Z.~Komargodski and N.~Seiberg,
  ``Contact Terms, Unitarity, and F-Maximization in Three-Dimensional Superconformal Theories,''
JHEP {\bf 1210}, 053 (2012).
[arXiv:1205.4142 [hep-th]].
}

\lref\ClossetVP{
  C.~Closset, T.~T.~Dumitrescu, G.~Festuccia, Z.~Komargodski and N.~Seiberg,
  ``Comments on Chern-Simons Contact Terms in Three Dimensions,''
JHEP {\bf 1209}, 091 (2012).
[arXiv:1206.5218 [hep-th]].
}

\lref\ClossetRU{
  C.~Closset, T.~T.~Dumitrescu, G.~Festuccia and Z.~Komargodski,
  ``Supersymmetric Field Theories on Three-Manifolds,''
JHEP {\bf 1305}, 017 (2013).
[arXiv:1212.3388 [hep-th]].
}

\lref\CveticXN{
  M.~Cvetic, T.~W.~Grimm and D.~Klevers,
  ``Anomaly Cancellation And Abelian Gauge Symmetries In F-theory,''
JHEP {\bf 1302}, 101 (2013).
[arXiv:1210.6034 [hep-th]].
}

\lref\DaiKQ{
  X.~z.~Dai and D.~S.~Freed,
  ``eta invariants and determinant lines,''
J.\ Math.\ Phys.\  {\bf 35}, 5155 (1994), Erratum: [J.\ Math.\ Phys.\  {\bf 42}, 2343 (2001)].
[hep-th/9405012].
}

\lref\DasguptaZZ{
  C.~Dasgupta and B.~I.~Halperin,
  ``Phase Transition in a Lattice Model of Superconductivity,''
Phys.\ Rev.\ Lett.\  {\bf 47}, 1556 (1981).
}

\lref\DaviesUW{
  N.~M.~Davies, T.~J.~Hollowood, V.~V.~Khoze and M.~P.~Mattis,
  ``Gluino condensate and magnetic monopoles in supersymmetric gluodynamics,''
Nucl.\ Phys.\ B {\bf 559}, 123 (1999).
[hep-th/9905015].
}

\lref\DaviesNW{
  N.~M.~Davies, T.~J.~Hollowood and V.~V.~Khoze,
  ``Monopoles, affine algebras and the gluino condensate,''
J.\ Math.\ Phys.\  {\bf 44}, 3640 (2003).
[hep-th/0006011].
}

\lref\DimoftePY{
  T.~Dimofte, D.~Gaiotto and S.~Gukov,
  ``3-Manifolds and 3d Indices,''
[arXiv:1112.5179 [hep-th]].
}

\lref\DolanQI{
  F.~A.~Dolan and H.~Osborn,
  ``Applications of the Superconformal Index for Protected Operators and q-Hypergeometric Identities to N=1 Dual Theories,''
Nucl.\ Phys.\ B {\bf 818}, 137 (2009).
[arXiv:0801.4947 [hep-th]].
}

\lref\DolanRP{
  F.~A.~H.~Dolan, V.~P.~Spiridonov and G.~S.~Vartanov,
  ``From 4d superconformal indices to 3d partition functions,''
Phys.\ Lett.\ B {\bf 704}, 234 (2011).
[arXiv:1104.1787 [hep-th]].
}

\lref\DouglasEX{
  M.~R.~Douglas,
  ``Chern-Simons-Witten theory as a topological Fermi liquid,''
[hep-th/9403119].
}

\lref\EagerHX{
  R.~Eager, J.~Schmude and Y.~Tachikawa,
  ``Superconformal Indices, Sasaki-Einstein Manifolds, and Cyclic Homologies,''
[arXiv:1207.0573 [hep-th]].
}

\lref\ElitzurFH{
  S.~Elitzur, A.~Giveon and D.~Kutasov,
  ``Branes and N=1 duality in string theory,''
Phys.\ Lett.\ B {\bf 400}, 269 (1997).
[hep-th/9702014].
}

\lref\KapustinLWA{
  A.~Kapustin and R.~Thorngren,
  ``Anomalies of discrete symmetries in three dimensions and group cohomology,''
Phys.\ Rev.\ Lett.\  {\bf 112}, no. 23, 231602 (2014).
[arXiv:1403.0617 [hep-th]].
}
\lref\KapustinZVA{
  A.~Kapustin and R.~Thorngren,
  ``Anomalies of discrete symmetries in various dimensions and group cohomology,''
[arXiv:1404.3230 [hep-th]].
}

\lref\ElitzurHC{
  S.~Elitzur, A.~Giveon, D.~Kutasov, E.~Rabinovici and A.~Schwimmer,
  ``Brane dynamics and N=1 supersymmetric gauge theory,''
Nucl.\ Phys.\ B {\bf 505}, 202 (1997).
[hep-th/9704104].
}

\lref\EssinRQ{
  A.~M.~Essin, J.~E.~Moore and D.~Vanderbilt,
  ``Magnetoelectric polarizability and axion electrodynamics in crystalline insulators,''
Phys.\ Rev.\ Lett.\  {\bf 102}, 146805 (2009).
[arXiv:0810.2998 [cond-mat.mes-hall]].
}

\lref\slthreeZ{
  J.~Felder, A.~Varchenko,
  ``The elliptic gamma function and $SL(3,Z) \times Z^3$,'' $\;\;$
[arXiv:math/0001184].
}

\lref\FestucciaWS{
  G.~Festuccia and N.~Seiberg,
  ``Rigid Supersymmetric Theories in Curved Superspace,''
JHEP {\bf 1106}, 114 (2011).
[arXiv:1105.0689 [hep-th]].
}

\lref\FidkowskiJUA{
  L.~Fidkowski, X.~Chen and A.~Vishwanath,
  ``Non-Abelian Topological Order on the Surface of a 3D Topological Superconductor from an Exactly Solved Model,''
Phys.\ Rev.\ X {\bf 3}, no. 4, 041016 (2013).
[arXiv:1305.5851 [cond-mat.str-el]].
}

\lref\FradkinTT{
  E.~H.~Fradkin and F.~A.~Schaposnik,
  ``The Fermion - boson mapping in three-dimensional quantum field theory,''
Phys.\ Lett.\ B {\bf 338}, 253 (1994).
[hep-th/9407182].
}

\lref\GaddeEN{
  A.~Gadde, L.~Rastelli, S.~S.~Razamat and W.~Yan,
  ``On the Superconformal Index of N=1 IR Fixed Points: A Holographic Check,''
JHEP {\bf 1103}, 041 (2011).
[arXiv:1011.5278 [hep-th]].
}

\lref\GaddeIA{
  A.~Gadde and W.~Yan,
  ``Reducing the 4d Index to the $S^3$ Partition Function,''
JHEP {\bf 1212}, 003 (2012).
[arXiv:1104.2592 [hep-th]].
}

\lref\GaddeDDA{
  A.~Gadde and S.~Gukov,
  ``2d Index and Surface operators,''
[arXiv:1305.0266 [hep-th]].
}

\lref\GaiottoAK{
  D.~Gaiotto and E.~Witten,
  ``S-Duality of Boundary Conditions In N=4 Super Yang-Mills Theory,''
Adv.\ Theor.\ Math.\ Phys.\  {\bf 13}, no. 3, 721 (2009).
[arXiv:0807.3720 [hep-th]].
}

\lref\GaiottoBE{
  D.~Gaiotto, G.~W.~Moore and A.~Neitzke,
  ``Framed BPS States,''
[arXiv:1006.0146 [hep-th]].
}

\lref\GaiottoKFA{
  D.~Gaiotto, A.~Kapustin, N.~Seiberg and B.~Willett,
  ``Generalized Global Symmetries,''
JHEP {\bf 1502}, 172 (2015).
[arXiv:1412.5148 [hep-th]].
}

\lref\GeraedtsPVA{
  S.~D.~Geraedts, M.~P.~Zaletel, R.~S.~K.~Mong, M.~A.~Metlitski, A.~Vishwanath and O.~I.~Motrunich,
  ``The half-filled Landau level: the case for Dirac composite fermions,''
Science {\bf 352}, 197 (2016).
[arXiv:1508.04140 [cond-mat.str-el]].
}

\lref\GiombiYA{
  S.~Giombi and X.~Yin,
  ``On Higher Spin Gauge Theory and the Critical $O(N)$ Model,''
Phys.\ Rev.\ D {\bf 85}, 086005 (2012).
[arXiv:1105.4011 [hep-th]].
}

\lref\GiombiKC{
  S.~Giombi, S.~Minwalla, S.~Prakash, S.~P.~Trivedi, S.~R.~Wadia and X.~Yin,
  ``Chern-Simons Theory with Vector Fermion Matter,''
Eur.\ Phys.\ J.\ C {\bf 72}, 2112 (2012).
[arXiv:1110.4386 [hep-th]].
}

\lref\GiombiMS{
  S.~Giombi and X.~Yin,
  ``The Higher Spin/Vector Model Duality,''
J.\ Phys.\ A {\bf 46}, 214003 (2013).
[arXiv:1208.4036 [hep-th]].
}

\lref\GiombiZWA{
  S.~Giombi, V.~Gurucharan, V.~Kirilin, S.~Prakash and E.~Skvortsov,
  ``On the Higher-Spin Spectrum in Large $N$ Chern-Simons Vector Models,''
JHEP {\bf 1701}, 058 (2017).
[arXiv:1610.08472 [hep-th]].
}

\lref\GiveonSR{
  A.~Giveon and D.~Kutasov,
  ``Brane dynamics and gauge theory,''
Rev.\ Mod.\ Phys.\  {\bf 71}, 983 (1999).
[hep-th/9802067].
}

\lref\GiveonZN{
  A.~Giveon and D.~Kutasov,
  ``Seiberg Duality in Chern-Simons Theory,''
Nucl.\ Phys.\ B {\bf 812}, 1 (2009).
[arXiv:0808.0360 [hep-th]].
}

\lref\GoddardQE{
  P.~Goddard, J.~Nuyts and D.~I.~Olive,
  ``Gauge Theories and Magnetic Charge,''
Nucl.\ Phys.\ B {\bf 125}, 1 (1977).
}

\lref\GoddardVK{
  P.~Goddard, A.~Kent and D.~I.~Olive,
  ``Virasoro Algebras and Coset Space Models,''
Phys.\ Lett.\ B {\bf 152}, 88 (1985).
}

\lref\GreenDA{
  D.~Green, Z.~Komargodski, N.~Seiberg, Y.~Tachikawa and B.~Wecht,
  ``Exactly Marginal Deformations and Global Symmetries,''
JHEP {\bf 1006}, 106 (2010).
[arXiv:1005.3546 [hep-th]].
}

\lref\GurPCA{
  G.~Gur-Ari and R.~Yacoby,
  ``Three Dimensional Bosonization From Supersymmetry,''
JHEP {\bf 1511}, 013 (2015).
[arXiv:1507.04378 [hep-th]].
}

\lref\GurAriXFF{
  G.~Gur-Ari, S.~A.~Hartnoll and R.~Mahajan,
  ``Transport in Chern-Simons-Matter Theories,''
JHEP {\bf 1607}, 090 (2016).
[arXiv:1605.01122 [hep-th]].
}

\lref\HalperinMH{
  B.~I.~Halperin, P.~A.~Lee and N.~Read,
  ``Theory of the half filled Landau level,''
Phys.\ Rev.\ B {\bf 47}, 7312 (1993).
}

\lref\HamaEA{
  N.~Hama, K.~Hosomichi and S.~Lee,
  ``SUSY Gauge Theories on Squashed Three-Spheres,''
JHEP {\bf 1105}, 014 (2011).
[arXiv:1102.4716 [hep-th]].
}

\lref\Hasegawa{
K.~Hasegawa,
  ``Spin Module Versions of Weyl's Reciprocity Theorem for Classical Kac-Moody Lie Algebras - An Application to Branching Rule Duality,''
Publ.\ Res.\ Inst.\ Math.\ Sci.\ {\bf 25}, 741-828 (1989).
}

\lref\HoriDK{
  K.~Hori and D.~Tong,
  ``Aspects of Non-Abelian Gauge Dynamics in Two-Dimensional N=(2,2) Theories,''
JHEP {\bf 0705}, 079 (2007).
[hep-th/0609032].
}

\lref\HoriPD{
  K.~Hori,
  ``Duality In Two-Dimensional (2,2) Supersymmetric Non-Abelian Gauge Theories,''
[arXiv:1104.2853 [hep-th]].
}

\lref\HsinBLU{
  P.~S.~Hsin and N.~Seiberg,
  ``Level/rank Duality and Chern-Simons-Matter Theories,''
JHEP {\bf 1609}, 095 (2016).
[arXiv:1607.07457 [hep-th]].
}

\lref\HullMS{
  C.~M.~Hull and B.~J.~Spence,
  ``The Geometry of the gauged sigma model with Wess-Zumino term,''
Nucl.\ Phys.\ B {\bf 353}, 379 (1991).
}

\lref\HwangQT{
  C.~Hwang, H.~Kim, K.~-J.~Park and J.~Park,
  ``Index computation for 3d Chern-Simons matter theory: test of Seiberg-like duality,''
JHEP {\bf 1109}, 037 (2011).
[arXiv:1107.4942 [hep-th]].
}

\lref\HwangHT{
  C.~Hwang, K.~-J.~Park and J.~Park,
  ``Evidence for Aharony duality for orthogonal gauge groups,''
JHEP {\bf 1111}, 011 (2011).
[arXiv:1109.2828 [hep-th]].
}

\lref\HwangJH{
  C.~Hwang, H.~-C.~Kim and J.~Park,
  ``Factorization of the 3d superconformal index,''
[arXiv:1211.6023 [hep-th]].
}

\lref\ImamuraSU{
  Y.~Imamura and S.~Yokoyama,
  ``Index for three dimensional superconformal field theories with general R-charge assignments,''
JHEP {\bf 1104}, 007 (2011).
[arXiv:1101.0557 [hep-th]].
}

\lref\ImamuraUW{
  Y.~Imamura,
 ``Relation between the 4d superconformal index and the $S^3$ partition function,''
JHEP {\bf 1109}, 133 (2011).
[arXiv:1104.4482 [hep-th]].
}

\lref\ImamuraWG{
  Y.~Imamura and D.~Yokoyama,
 ``N=2 supersymmetric theories on squashed three-sphere,''
Phys.\ Rev.\ D {\bf 85}, 025015 (2012).
[arXiv:1109.4734 [hep-th]].
}

\lref\ImamuraRQ{
  Y.~Imamura and D.~Yokoyama,
 ``$S^3/Z_n$ partition function and dualities,''
JHEP {\bf 1211}, 122 (2012).
[arXiv:1208.1404 [hep-th]].
}

\lref\InbasekarTSA{
  K.~Inbasekar, S.~Jain, S.~Mazumdar, S.~Minwalla, V.~Umesh and S.~Yokoyama,
  ``Unitarity, crossing symmetry and duality in the scattering of ${\cal N}{=}1 $ SUSY matter Chern-Simons theories,''
JHEP {\bf 1510}, 176 (2015).
[arXiv:1505.06571 [hep-th]].
}

\lref\IntriligatorID{
  K.~A.~Intriligator and N.~Seiberg,
  ``Duality, monopoles, dyons, confinement and oblique confinement in supersymmetric SO(N(c)) gauge theories,''
Nucl.\ Phys.\ B {\bf 444}, 125 (1995).
[hep-th/9503179].
}

\lref\IntriligatorNE{
  K.~A.~Intriligator and P.~Pouliot,
  ``Exact superpotentials, quantum vacua and duality in supersymmetric SP(N(c)) gauge theories,''
Phys.\ Lett.\ B {\bf 353}, 471 (1995).
[hep-th/9505006].
}

\lref\IntriligatorER{
  K.~A.~Intriligator and N.~Seiberg,
  ``Phases of N=1 supersymmetric gauge theories and electric - magnetic triality,''
In *Los Angeles 1995, Future perspectives in string theory* 270-282.
[hep-th/9506084].
}

\lref\IntriligatorAU{
  K.~A.~Intriligator and N.~Seiberg,
  ``Lectures on supersymmetric gauge theories and electric - magnetic duality,''
Nucl.\ Phys.\ Proc.\ Suppl.\  {\bf 45BC}, 1 (1996).
[hep-th/9509066].
}

\lref\IntriligatorEX{
  K.~A.~Intriligator and N.~Seiberg,
  ``Mirror symmetry in three-dimensional gauge theories,''
Phys.\ Lett.\ B {\bf 387}, 513 (1996).
[hep-th/9607207].
}

\lref\IntriligatorLCA{
  K.~Intriligator and N.~Seiberg,
  ``Aspects of 3d ${\cal N}{=}2$ Chern-Simons-Matter Theories,''
JHEP {\bf 1307}, 079 (2013).
[arXiv:1305.1633 [hep-th]].
}

\lref\IvanovFN{
   E.~A.~Ivanov,
   ``Chern-Simons matter systems with manifest N=2 supersymmetry,''
Phys.\ Lett.\ B {\bf 268}, 203 (1991).
}

\lref\JafferisUN{
  D.~L.~Jafferis,
  ``The Exact Superconformal R-Symmetry Extremizes Z,''
JHEP {\bf 1205}, 159 (2012).
[arXiv:1012.3210 [hep-th]].
}

\lref\JafferisZI{
  D.~L.~Jafferis, I.~R.~Klebanov, S.~S.~Pufu and B.~R.~Safdi,
  ``Towards the F-Theorem: N=2 Field Theories on the Three-Sphere,''
JHEP {\bf 1106}, 102 (2011).
[arXiv:1103.1181 [hep-th]].
}

\lref\JainTX{
  J.~K.~Jain,
  ``Composite fermion approach for the fractional quantum Hall effect,''
Phys.\ Rev.\ Lett.\  {\bf 63}, 199 (1989).
}

\lref\JainPY{
  S.~Jain, S.~Minwalla, T.~Sharma, T.~Takimi, S.~R.~Wadia and S.~Yokoyama,
  ``Phases of large $N$ vector Chern-Simons theories on $S^2 {\times} S^1$,''
JHEP {\bf 1309}, 009 (2013).
[arXiv:1301.6169 [hep-th]].
}

\lref\JainGZA{
  S.~Jain, S.~Minwalla and S.~Yokoyama,
  ``Chern Simons duality with a fundamental boson and fermion,''
JHEP {\bf 1311}, 037 (2013).
[arXiv:1305.7235 [hep-th]].
}

\lref\JainNZA{
  S.~Jain, M.~Mandlik, S.~Minwalla, T.~Takimi, S.~R.~Wadia and S.~Yokoyama,
  ``Unitarity, Crossing Symmetry and Duality of the S-matrix in large $N$ Chern-Simons theories with fundamental matter,''
JHEP {\bf 1504}, 129 (2015).
[arXiv:1404.6373 [hep-th]].
}

\lref\KachruRMA{
  S.~Kachru, M.~Mulligan, G.~Torroba and H.~Wang,
  ``Mirror symmetry and the half-filled Landau level,''
Phys.\ Rev.\ B {\bf 92}, 235105 (2015).
[arXiv:1506.01376 [cond-mat.str-el]].
}

\lref\KachruRUI{
  S.~Kachru, M.~Mulligan, G.~Torroba and H.~Wang,
  ``Bosonization and Mirror Symmetry,''
Phys.\ Rev.\ D {\bf 94}, no. 8, 085009 (2016).
[arXiv:1608.05077 [hep-th]].
}

\lref\KachruAON{
  S.~Kachru, M.~Mulligan, G.~Torroba and H.~Wang,
  ``Nonsupersymmetric dualities from mirror symmetry,''
Phys.\ Rev.\ Lett.\  {\bf 118}, 011602 (2017).
[arXiv:1609.02149 [hep-th]].
}

\lref\KajantieVY{
  K.~Kajantie, M.~Laine, T.~Neuhaus, A.~Rajantie and K.~Rummukainen,
  ``Duality and scaling in three-dimensional scalar electrodynamics,''
Nucl.\ Phys.\ B {\bf 699}, 632 (2004).
[hep-lat/0402021].
}

\lref\KapustinHA{
  A.~Kapustin and M.~J.~Strassler,
  ``On mirror symmetry in three-dimensional Abelian gauge theories,''
JHEP {\bf 9904}, 021 (1999).
[hep-th/9902033].
}

\lref\KapustinPY{
  A.~Kapustin,
  ``Wilson-'t Hooft operators in four-dimensional gauge theories and S-duality,''
Phys.\ Rev.\ D {\bf 74}, 025005 (2006).
[hep-th/0501015].
}

\lref\KapustinKZ{
  A.~Kapustin, B.~Willett and I.~Yaakov,
  ``Exact Results for Wilson Loops in Superconformal Chern-Simons Theories with Matter,''
JHEP {\bf 1003}, 089 (2010).
[arXiv:0909.4559 [hep-th]].
}

\lref\KapustinSim{
A.~Kapustin,  2010 Simons Workshop talk, a video of this talk can be found at
{\tt
http://media.scgp.stonybrook.edu/video/video.php?f=20110810\_1\_qtp.mp4}
}

\lref\KapustinXQ{
  A.~Kapustin, B.~Willett and I.~Yaakov,
  ``Nonperturbative Tests of Three-Dimensional Dualities,''
JHEP {\bf 1010}, 013 (2010).
[arXiv:1003.5694 [hep-th]].
}

\lref\KapustinGH{
  A.~Kapustin,
  ``Seiberg-like duality in three dimensions for orthogonal gauge groups,''
arXiv:1104.0466 [hep-th].
}

\lref\KapustinJM{
  A.~Kapustin and B.~Willett,
  ``Generalized Superconformal Index for Three Dimensional Field Theories,''
[arXiv:1106.2484 [hep-th]].
}

\lref\KapustinVZ{
  A.~Kapustin, H.~Kim and J.~Park,
  ``Dualities for 3d Theories with Tensor Matter,''
JHEP {\bf 1112}, 087 (2011).
[arXiv:1110.2547 [hep-th]].
}

\lref\KapustinGUA{
  A.~Kapustin and N.~Seiberg,
  ``Coupling a QFT to a TQFT and Duality,''
JHEP {\bf 1404}, 001 (2014).
[arXiv:1401.0740 [hep-th]].
}

\lref\KarchUX{
  A.~Karch,
  ``Seiberg duality in three-dimensions,''
Phys.\ Lett.\ B {\bf 405}, 79 (1997).
[hep-th/9703172].
}

\lref\KarchSXI{
  A.~Karch and D.~Tong,
  ``Particle-Vortex Duality from 3d Bosonization,''
Phys.\ Rev.\ X {\bf 6}, 031043 (2016). 
[arXiv:1606.01893 [hep-th]].
}

\lref\KarchAUX{
  A.~Karch, B.~Robinson and D.~Tong,
  ``More Abelian Dualities in 2+1 Dimensions,''
JHEP {\bf 1701}, 017 (2017).
[arXiv:1609.04012 [hep-th]].
}

\lref\KimWB{
  S.~Kim,
  ``The Complete superconformal index for N=6 Chern-Simons theory,''
Nucl.\ Phys.\ B {\bf 821}, 241 (2009), [Erratum-ibid.\ B {\bf 864}, 884 (2012)].
[arXiv:0903.4172 [hep-th]].
}

\lref\KimCMA{
  H.~Kim and J.~Park,
  ``Aharony Dualities for 3d Theories with Adjoint Matter,''
[arXiv:1302.3645 [hep-th]].
}

\lref\KinneyEJ{
  J.~Kinney, J.~M.~Maldacena, S.~Minwalla and S.~Raju,
  ``An Index for 4 dimensional super conformal theories,''
Commun.\ Math.\ Phys.\  {\bf 275}, 209 (2007).
[hep-th/0510251].
}

\lref\KlebanovJA{
  I.~R.~Klebanov and A.~M.~Polyakov,
  ``AdS dual of the critical $O(N)$ vector model,''
Phys.\ Lett.\ B {\bf 550}, 213 (2002).
[hep-th/0210114].
}

\lref\KrattenthalerDA{
  C.~Krattenthaler, V.~P.~Spiridonov, G.~S.~Vartanov,
  ``Superconformal indices of three-dimensional theories related by mirror symmetry,''
JHEP {\bf 1106}, 008 (2011).
[arXiv:1103.4075 [hep-th]].
}

\lref\McG{
  S. M. Kravec, J. McGreevy, and B. Swingle,
  ``All-Fermion Electrodynamics And Fermion Number Anomaly Inflow,''
arXiv:1409.8339.
}

\lref\KutasovVE{
  D.~Kutasov,
  ``A Comment on duality in N=1 supersymmetric nonAbelian gauge theories,''
Phys.\ Lett.\ B {\bf 351}, 230 (1995).
[hep-th/9503086].
}

\lref\KutasovNP{
  D.~Kutasov and A.~Schwimmer,
  ``On duality in supersymmetric Yang-Mills theory,''
Phys.\ Lett.\ B {\bf 354}, 315 (1995).
[hep-th/9505004].
}

\lref\KutasovSS{
  D.~Kutasov, A.~Schwimmer and N.~Seiberg,
  ``Chiral rings, singularity theory and electric - magnetic duality,''
Nucl.\ Phys.\ B {\bf 459}, 455 (1996).
[hep-th/9510222].
}

\lref\LeeVP{
  K.~-M.~Lee and P.~Yi,
  ``Monopoles and instantons on partially compactified D-branes,''
Phys.\ Rev.\ D {\bf 56}, 3711 (1997).
[hep-th/9702107].
}

\lref\LeeVU{
  K.~-M.~Lee,
  ``Instantons and magnetic monopoles on R**3 x S**1 with arbitrary simple gauge groups,''
Phys.\ Lett.\ B {\bf 426}, 323 (1998).
[hep-th/9802012].
}

\lref\MaldacenaSS{
  J.~M.~Maldacena, G.~W.~Moore and N.~Seiberg,
  ``D-brane charges in five-brane backgrounds,''
JHEP {\bf 0110}, 005 (2001).
[hep-th/0108152].
}

\lref\MaldacenaJN{
  J.~Maldacena and A.~Zhiboedov,
  ``Constraining Conformal Field Theories with A Higher Spin Symmetry,''
J.\ Phys.\ A {\bf 46}, 214011 (2013).
[arXiv:1112.1016 [hep-th]].
}

\lref\MetlitskiEKA{
  M.~A.~Metlitski and A.~Vishwanath,
  ``Particle-vortex duality of 2d Dirac fermion from electric-magnetic duality of 3d topological insulators,''
[arXiv:1505.05142 [cond-mat.str-el]].
}

\lref\MetlitskiBPA{
  M.~A.~Metlitski, C.~L.~Kane and M.~P.~A.~Fisher,
  ``Symmetry-respecting topologically ordered surface phase of three-dimensional electron topological insulators,''
Phys.\ Rev.\ B {\bf 92}, no. 12, 125111 (2015).
}

\lref\MetlitskiYQA{
  M.~A.~Metlitski,
  ``$S$-duality of $u(1)$ gauge theory with $\theta =\pi$ on non-orientable manifolds: Applications to topological insulators and superconductors,''
[arXiv:1510.05663 [hep-th]].
}

\lref\MetlitskiDHT{
  M.~A.~Metlitski, A.~Vishwanath and C.~Xu,
  ``Duality and bosonization of (2+1)d Majorana fermions,''
  arXiv:1611.05049 [cond-mat.str-el].
}

\lref\MinwallaSCA{
  S.~Minwalla and S.~Yokoyama,
  ``Chern Simons Bosonization along RG Flows,''
JHEP {\bf 1602}, 103 (2016).
[arXiv:1507.04546 [hep-th]].
}

\lref\MlawerUV{
  E.~J.~Mlawer, S.~G.~Naculich, H.~A.~Riggs and H.~J.~Schnitzer,
  ``Group level duality of WZW fusion coefficients and Chern-Simons link observables,''
Nucl.\ Phys.\ B {\bf 352}, 863 (1991).
}

\lref\MSN{
G.~W.~Moore and N.~Seiberg,
  ``Naturality in Conformal Field Theory,''
  Nucl.\ Phys.\ B {\bf 313}, 16 (1989).}

\lref\MooreYH{
  G.~W.~Moore and N.~Seiberg,
  ``Taming the Conformal Zoo,''
Phys.\ Lett.\ B {\bf 220}, 422 (1989).
}

\lref\MoritaCS{
  T.~Morita and V.~Niarchos,
  ``F-theorem, duality and SUSY breaking in one-adjoint Chern-Simons-Matter theories,''
Nucl.\ Phys.\ B {\bf 858}, 84 (2012).
[arXiv:1108.4963 [hep-th]].
}

\lref\MrossIDY{
  D.~F.~Mross, J.~Alicea and O.~I.~Motrunich,
  ``Explicit derivation of duality between a free Dirac cone and quantum electrodynamics in (2+1) dimensions,''
[arXiv:1510.08455 [cond-mat.str-el]].
}

\lref\MulliganGLM{
  M.~Mulligan, S.~Raghu and M.~P.~A.~Fisher,
  ``Emergent particle-hole symmetry in the half-filled Landau level,''
[arXiv:1603.05656 [cond-mat.str-el]].
}

\lref\MuruganZAL{
  J.~Murugan and H.~Nastase,
  ``Particle-vortex duality in topological insulators and superconductors,''
[arXiv:1606.01912 [hep-th]].
}

\lref\NaculichPA{
  S.~G.~Naculich, H.~A.~Riggs and H.~J.~Schnitzer,
  ``Group Level Duality in {WZW} Models and {Chern-Simons} Theory,''
Phys.\ Lett.\ B {\bf 246}, 417 (1990).
}

\lref\NaculichNC{
  S.~G.~Naculich and H.~J.~Schnitzer,
  ``Level-rank duality of the U(N) WZW model, Chern-Simons theory, and 2-D qYM theory,''
JHEP {\bf 0706}, 023 (2007).
[hep-th/0703089 [HEP-TH]].
}

\lref\NakaharaNW{
  M.~Nakahara,
  ``Geometry, topology and physics,''
Boca Raton, USA: Taylor and Francis (2003) 573 p.
}

\lref\NakanishiHJ{
  T.~Nakanishi and A.~Tsuchiya,
  ``Level rank duality of WZW models in conformal field theory,''
Commun.\ Math.\ Phys.\  {\bf 144}, 351 (1992).
}

\lref\NguyenZN{
  A.~K.~Nguyen and A.~Sudbo,
  ``Topological phase fluctuations, amplitude fluctuations, and criticality in extreme type II superconductors,''
Phys.\ Rev.\ B {\bf 60}, 15307 (1999).
[cond-mat/9907385].
}

\lref\NiarchosJB{
  V.~Niarchos,
  ``Seiberg Duality in Chern-Simons Theories with Fundamental and Adjoint Matter,''
JHEP {\bf 0811}, 001 (2008).
[arXiv:0808.2771 [hep-th]].
}

\lref\NiarchosAA{
  V.~Niarchos,
  ``R-charges, Chiral Rings and RG Flows in Supersymmetric Chern-Simons-Matter Theories,''
JHEP {\bf 0905}, 054 (2009).
[arXiv:0903.0435 [hep-th]].
}

\lref\NiarchosAH{
  V.~Niarchos,
  ``Seiberg dualities and the 3d/4d connection,''
JHEP {\bf 1207}, 075 (2012).
[arXiv:1205.2086 [hep-th]].
}

\lref\NiemiRQ{
  A.~J.~Niemi and G.~W.~Semenoff,
  ``Axial Anomaly Induced Fermion Fractionization and Effective Gauge Theory Actions in Odd Dimensional Space-Times,''
Phys.\ Rev.\ Lett.\  {\bf 51}, 2077 (1983).
}

\lref\VOstrik{
  V.~Ostrik and M.~Sun,
  ``Level-Rank Duality Via Tensor Categories,''
Comm. Math. Phys. 326 (2014) 49-61.
[arXiv:1208.5131 [math-ph]].
}

\lref\ParkWTA{
  J.~Park and K.~J.~Park,
  ``Seiberg-like Dualities for 3d ${\cal N}{=}2$ Theories with $SU(N)$ gauge group,''
JHEP {\bf 1310}, 198 (2013).
[arXiv:1305.6280 [hep-th]].
}

\lref\PaulyAMA{
  C.~Pauly,
  ``Strange duality revisited,''
Math.\ Res.\ Lett.\  {\bf 21}, 1353 (2014).
}

\lref\PeskinKP{
  M.~E.~Peskin,
  ``Mandelstam 't Hooft Duality in Abelian Lattice Models,''
Annals Phys.\  {\bf 113}, 122 (1978).
}

\lref\PolyakovFU{
  A.~M.~Polyakov,
  ``Quark Confinement and Topology of Gauge Groups,''
Nucl.\ Phys.\ B {\bf 120}, 429 (1977).
}

\lref\PolyakovMD{
  A.~M.~Polyakov,
  ``Fermi-Bose Transmutations Induced by Gauge Fields,''
Mod.\ Phys.\ Lett.\ A {\bf 3}, 325 (1988).
}

\lref\PotterCDN{
  A.~C.~Potter, M.~Serbyn and A.~Vishwanath,
  ``Thermoelectric transport signatures of Dirac composite fermions in the half-filled Landau level,''
Phys.\ Rev.\ X {\bf 6}, 031026 (2016).
[arXiv:1512.06852 [cond-mat.str-el]].
}

\lref\QiEW{
  X.~L.~Qi, T.~Hughes and S.~C.~Zhang,
  ``Topological Field Theory of Time-Reversal Invariant Insulators,''
Phys.\ Rev.\ B {\bf 78}, 195424 (2008).
[arXiv:0802.3537 [cond-mat.mes-hall]].
}

\lref\RabinoviciMJ{
  E.~Rabinovici, A.~Schwimmer and S.~Yankielowicz,
  ``Quantization in the Presence of {Wess-Zumino} Terms,''
Nucl.\ Phys.\ B {\bf 248}, 523 (1984).
}

\lref\RadicevicYLA{
  D.~Radicevic,
  ``Disorder Operators in Chern-Simons-Fermion Theories,''
JHEP {\bf 1603}, 131 (2016).
[arXiv:1511.01902 [hep-th]].
}

\lref\RadicevicWQN{
  D.~Radicevic, D.~Tong and C.~Turner,
  ``Non-Abelian 3d Bosonization and Quantum Hall States,''
JHEP {\bf 1612}, 067 (2016).
[arXiv:1608.04732 [hep-th]].
}

\lref\RazamatUV{
  S.~S.~Razamat,
  ``On a modular property of N=2 superconformal theories in four dimensions,''
JHEP {\bf 1210}, 191 (2012).
[arXiv:1208.5056 [hep-th]].
}

\lref\RedlichDV{
  A.~N.~Redlich,
  ``Parity Violation and Gauge Noninvariance of the Effective Gauge Field Action in Three-Dimensions,''
Phys.\ Rev.\ D {\bf 29}, 2366 (1984).
}

\lref\Rehren{
  K.-H.~Rehren,
  ``Algebraic Conformal QFT'',
3rd Meeting of the French-Italian Research Team on Noncommutative Geometry and Quantum Physics Vietri sul Mare, 2009.
}

\lref\RomelsbergerEG{
  C.~Romelsberger,
  ``Counting chiral primaries in N = 1, d=4 superconformal field theories,''
Nucl.\ Phys.\ B {\bf 747}, 329 (2006).
[hep-th/0510060].
}

\lref\RoscherWOX{
  D.~Roscher, E.~Torres and P.~Strack,
  ``Dual QED$_3$ at "$N_F = 1/2$" is an interacting CFT in the infrared,''
[arXiv:1605.05347 [cond-mat.str-el]].
}

\lref\SafdiRE{
  B.~R.~Safdi, I.~R.~Klebanov and J.~Lee,
  ``A Crack in the Conformal Window,''
[arXiv:1212.4502 [hep-th]].
}

\lref\SeibergBZ{
  N.~Seiberg,
  ``Exact results on the space of vacua of four-dimensional SUSY gauge theories,''
Phys.\ Rev.\ D {\bf 49}, 6857 (1994).
[hep-th/9402044].
}

\lref\SeibergPQ{
  N.~Seiberg,
  ``Electric - magnetic duality in supersymmetric nonAbelian gauge theories,''
Nucl.\ Phys.\ B {\bf 435}, 129 (1995).
[hep-th/9411149].
}

\lref\SeibergNZ{
  N.~Seiberg and E.~Witten,
  ``Gauge dynamics and compactification to three-dimensions,''
In *Saclay 1996, The mathematical beauty of physics* 333-366.
[hep-th/9607163].
}

\lref\SeibergQD{
  N.~Seiberg,
  ``Modifying the Sum Over Topological Sectors and Constraints on Supergravity,''
JHEP {\bf 1007}, 070 (2010).
[arXiv:1005.0002 [hep-th]].
}

\lref\SeibergRSG{
  N.~Seiberg and E.~Witten,
  ``Gapped Boundary Phases of Topological Insulators via Weak Coupling,''
PTEP {\bf 2016}, 12C101 (2016). 
[arXiv:1602.04251 [cond-mat.str-el]].
}

\lref\SeibergGMD{
  N.~Seiberg, T.~Senthil, C.~Wang and E.~Witten,
  ``A Duality Web in 2+1 Dimensions and Condensed Matter Physics,''
Annals Phys.\  {\bf 374}, 395 (2016).
[arXiv:1606.01989 [hep-th]].
}

\lref\SenthilJK{
  T.~Senthil and M.~P.~A.~Fisher,
  ``Competing orders, non-linear sigma models, and topological terms in quantum magnets,''
Phys.\ Rev.\ B {\bf 74}, 064405 (2006).
[cond-mat/0510459].
}

\lref\SezginRT{
  E.~Sezgin and P.~Sundell,
  ``Massless higher spins and holography,''
Nucl.\ Phys.\ B {\bf 644}, 303 (2002), Erratum: [Nucl.\ Phys.\ B {\bf 660}, 403 (2003)].
[hep-th/0205131].
}

\lref\SezginPT{
  E.~Sezgin and P.~Sundell,
  ``Holography in 4D (super) higher spin theories and a test via cubic scalar couplings,''
JHEP {\bf 0507}, 044 (2005).
[hep-th/0305040].
}

\lref\ShajiIS{
  N.~Shaji, R.~Shankar and M.~Sivakumar,
  ``On Bose-fermi Equivalence in a U(1) Gauge Theory With {Chern-Simons} Action,''
Mod.\ Phys.\ Lett.\ A {\bf 5}, 593 (1990).
}

\lref\Shamirthesis{
  I.~Shamir,
  ``Aspects of three dimensional Seiberg duality,''
  M.~Sc. thesis submitted to the Weizmann Institute of Science, April 2010.
}

\lref\ShenkerZF{
  S.~H.~Shenker and X.~Yin,
  ``Vector Models in the Singlet Sector at Finite Temperature,''
[arXiv:1109.3519 [hep-th]].
}

\lref\SonXQA{
  D.~T.~Son,
  ``Is the Composite Fermion a Dirac Particle?,''
Phys.\ Rev.\ X {\bf 5}, 031027 (2015).
[arXiv:1502.03446 [cond-mat.mes-hall]].
}

\lref\SpiridonovZR{
  V.~P.~Spiridonov and G.~S.~Vartanov,
  ``Superconformal indices for N = 1 theories with multiple duals,''
Nucl.\ Phys.\ B {\bf 824}, 192 (2010).
[arXiv:0811.1909 [hep-th]].
}

\lref\SpiridonovZA{
  V.~P.~Spiridonov and G.~S.~Vartanov,
  ``Elliptic Hypergeometry of Supersymmetric Dualities,''
Commun.\ Math.\ Phys.\  {\bf 304}, 797 (2011).
[arXiv:0910.5944 [hep-th]].
}

\lref\SpiridonovHF{
  V.~P.~Spiridonov and G.~S.~Vartanov,
  ``Elliptic hypergeometry of supersymmetric dualities II. Orthogonal groups, knots, and vortices,''
[arXiv:1107.5788 [hep-th]].
}

\lref\SpiridonovWW{
  V.~P.~Spiridonov and G.~S.~Vartanov,
  ``Elliptic hypergeometric integrals and 't Hooft anomaly matching conditions,''
JHEP {\bf 1206}, 016 (2012).
[arXiv:1203.5677 [hep-th]].
}

\lref\StrasslerFE{
  M.~J.~Strassler,
  ``Duality, phases, spinors and monopoles in $SO(N)$ and $spin(N)$ gauge theories,''
JHEP {\bf 9809}, 017 (1998).
[hep-th/9709081].
}

\lref\VasilievVF{
  M.~A.~Vasiliev,
  ``Holography, Unfolding and Higher-Spin Theory,''
J.\ Phys.\ A {\bf 46}, 214013 (2013).
[arXiv:1203.5554 [hep-th]].
}

\lref\VerstegenAT{
  D.~Verstegen,
  ``Conformal embeddings, rank-level duality and exceptional modular invariants,''
Commun.\ Math.\ Phys.\  {\bf 137}, 567 (1991).
}

\lref\WangUKY{
  C.~Wang, A.~C.~Potter and T.~Senthil,
  ``Gapped symmetry preserving surface state for the electron topological insulator,''
Phys.\ Rev.\ B {\bf 88}, no. 11, 115137 (2013).
[arXiv:1306.3223 [cond-mat.str-el]].
}

\lref\MPS{
  C. Wang, A. C. Potter, and T. Senthil,
  ``Classification Of Interacting Electronic Topological Insulators In Three Dimensions,''
Science {\bf 343} (2014) 629,
[arXiv:1306.3238].
}

\lref\WangLCA{
  C.~Wang and T.~Senthil,
  ``Interacting fermionic topological insulators/superconductors in three dimensions,''
Phys.\ Rev.\ B {\bf 89}, no. 19, 195124 (2014), Erratum: [Phys.\ Rev.\ B {\bf 91}, no. 23, 239902 (2015)].
[arXiv:1401.1142 [cond-mat.str-el]].
}

\lref\WangQMT{
  C.~Wang and T.~Senthil,
  ``Dual Dirac Liquid on the Surface of the Electron Topological Insulator,''
Phys.\ Rev.\ X {\bf 5}, no. 4, 041031 (2015). [arXiv:1505.05141 [cond-mat.str-el]].
}

\lref\WangFQL{
  C.~Wang and T.~Senthil,
  ``Half-filled Landau level, topological insulator surfaces, and three-dimensional quantum spin liquids,''
Phys.\ Rev.\ B {\bf 93}, no. 8, 085110 (2016). [arXiv:1507.08290 [cond-mat.st-el]].
}

\lref\WangGQJ{
  C.~Wang and T.~Senthil,
  ``Composite fermi liquids in the lowest Landau level,''
Phys.\ Rev.\ B {\bf 94}, 245107 (2016). 
[arXiv:1604.06807 [cond-mat.str-el]].
}

\lref\WangCTO{
  C.~Wang and T.~Senthil,
  ``Time-Reversal Symmetric $U(1)$ Quantum Spin Liquids,''
Phys.\ Rev.\ X {\bf 6}, no. 1, 011034 (2016).
}

\lref\Whitehead{
  J.~H.~C.~Whitehead,
  ``On simply connected, 4-dimensional polyhedra,''
Comm.\ Math.\ Helv.\ {\bf 22} (1949) 48.
}

\lref\WilczekDU{
  F.~Wilczek,
  ``Magnetic Flux, Angular Momentum, and Statistics,''
Phys.\ Rev.\ Lett.\  {\bf 48}, 1144 (1982).
}

\lref\WilczekCY{
  F.~Wilczek and A.~Zee,
  ``Linking Numbers, Spin, and Statistics of Solitons,''
Phys.\ Rev.\ Lett.\  {\bf 51}, 2250 (1983).
}

\lref\WillettGP{
  B.~Willett and I.~Yaakov,
  ``${\cal N}{=}2$ Dualities and $Z$-extremization in Three Dimensions,''
arXiv:1104.0487 [hep-th].
}

\lref\WittenHF{
  E.~Witten,
  ``Quantum Field Theory and the Jones Polynomial,''
Commun.\ Math.\ Phys.\  {\bf 121}, 351 (1989).
}

\lref\WittenXI{
  E.~Witten,
  ``The Verlinde algebra and the cohomology of the Grassmannian,''
In *Cambridge 1993, Geometry, topology, and physics* 357-422.
[hep-th/9312104].
}

\lref\WittenGF{
  E.~Witten,
  ``On S duality in Abelian gauge theory,''
Selecta Math.\  {\bf 1}, 383 (1995).
[hep-th/9505186].
}

\lref\WittenDS{
  E.~Witten,
  ``Supersymmetric index of three-dimensional gauge theory,''
In *Shifman, M.A. (ed.): The many faces of the superworld* 156-184.
[hep-th/9903005].
}

\lref\WittenYA{
  E.~Witten,
  ``SL(2,Z) action on three-dimensional conformal field theories with Abelian symmetry,''
In *Shifman, M. (ed.) et al.: From fields to strings, vol. 2* 1173-1200.
[hep-th/0307041].
}

\lref\WittenABA{
  E.~Witten,
  ``Fermion Path Integrals And Topological Phases,''
[arXiv:1508.04715 [cond-mat.mes-hall]].
}

\lref\WuGE{
  T.~T.~Wu and C.~N.~Yang,
  ``Dirac Monopole Without Strings: Monopole Harmonics,''
Nucl.\ Phys.\ B {\bf 107}, 365 (1976).
}

\lref\XuNXA{
  F.~Xu,
  ``Algebraic coset conformal field theories,''
Commun.\ Math.\ Phys.\  {\bf 211}, 1 (2000).
[math/9810035].
}

\lref\XuLXA{
  C.~Xu and Y.~Z.~You,
  ``Self-dual Quantum Electrodynamics as Boundary State of the three dimensional Bosonic Topological Insulator,''
Phys.\ Rev.\ B {\bf 92}, 220416 (2015). 
[arXiv:1510.06032 [cond-mat.str-el]].
}

\lref\ZupnikRY{
   B.~M.~Zupnik and D.~G.~Pak,
   ``Topologically Massive Gauge Theories In Superspace,''
Sov.\ Phys.\ J.\  {\bf 31}, 962 (1988).
}

\lref\ZwiebelWA{
  B.~I.~Zwiebel,
  ``Charging the Superconformal Index,''
JHEP {\bf 1201}, 116 (2012).
[arXiv:1111.1773 [hep-th]].
}

%
%

\vskip-60pt
{\hfil SISSA 06/2017/FISI}
\Title{} {\vbox{\centerline{Comments on Global Symmetries, Anomalies,}\centerline{}\centerline{and Duality in $(2+1)d$}
}}

\vskip-15pt

\centerline{Francesco Benini${}^{1,2}$, Po-Shen Hsin${}^3$, and Nathan Seiberg${}^1$}
\vskip15pt
\centerline{\it ${}^1$ School of Natural Sciences, Institute for Advanced Study, Princeton, NJ 08540, USA}
\centerline{\it ${}^2$ {SISSA \&\ INFN, via Bonomea 265, 34136 Trieste, Italy}}
\centerline{\it ${}^3$ Department of Physics, Princeton University, Princeton, NJ 08544, USA}
\vskip25pt

\noindent
We analyze in detail the global symmetries of various $(2+1)d$ quantum field theories and couple them to classical background gauge fields.  A proper identification of the global symmetries allows us to consider all non-trivial bundles of those background fields, thus finding more subtle observables.  The global symmetries exhibit interesting 't~Hooft anomalies.  These allow us to constrain the IR behavior of the theories and provide powerful constraints on conjectured dualities.

\bigskip
\Date{February 2017}


\newif\ifbf\bffalse
\let\BF=\bf
\def\bf{\bftrue\BF}
\font\bfit=cmbxti10
\def\mbf#1{\ifbf{\hbox{\bfit#1}}\else#1\fi}


%
%

\newsec{Introduction}

Recently, a convergence of ideas from condensed matter physics \refs{\PeskinKP\DasguptaZZ\BarkeshliIDA\SonXQA\PotterCDN-\WangGQJ},
supersymmetric quantum field theory \refs{\IntriligatorEX\deBoerMP\AharonyBX\AharonyGP\GiveonZN\KapustinGH
\WillettGP\BeniniMF\AharonyCI\IntriligatorLCA\AharonyDHA\ParkWTA-\AharonyKMA},
and string theory \refs{\SezginRT\KlebanovJA\GiombiYA\AharonyJZ\GiombiKC\MaldacenaJN\AharonyNH
\GiombiMS\AharonyNS\JainPY\JainGZA
\JainNZA\InbasekarTSA\MinwallaSCA-\GurAriXFF}
has led to a large set of new boson/fermion dualities \refs{\AharonyMJS\KarchSXI\MuruganZAL\SeibergGMD\HsinBLU\RadicevicWQN\KachruRUI
\KachruAON\KarchAUX\MetlitskiDHT-\AharonyJVV}
and new fermion/fermion dualities \refs{\XuLXA,\SeibergGMD,\HsinBLU,\ChengPDN,\AharonyJVV}.%
\foot{See \GiombiZWA\ for some recent tests.}
Our goal in this note is to further explore these theories.  In particular, we will focus on various aspects of their global symmetries.

The main boson/fermion dualities that we will study are \refs{\AharonyMJS,\HsinBLU}%
\foot{We will follow the notation and conventions of \refs{\SeibergRSG,\SeibergGMD,\HsinBLU,\AharonyJVV} and will not repeat them here.}
\eqn\SUUduality{\eqalign{
SU(N)_k \ {\rm with}\ N_f\ {\rm scalars} \quad &\longleftrightarrow \quad U(k)_{-N+{N_f\over 2}} \ {\rm with}\ N_f\ {\rm fermions} \cr
U(N)_k \ {\rm with}\ N_f\ {\rm scalars} \quad &\longleftrightarrow \quad SU(k)_{-N+{N_f\over 2}} \ {\rm with}\ N_f\ {\rm fermions} \cr
U(N)_{k,k\pm N} \ {\rm with}\ N_f\ {\rm scalars} \quad &\longleftrightarrow \quad U(k)_{-N+{N_f\over 2}, -N\mp k+{N_f\over 2}} \ {\rm with}\ N_f\ {\rm fermions}
}}
conjectured to hold for $N_f \leq N$ (our notation is $U(N)_k \equiv U(N)_{k,k}$), and \AharonyJVV
\eqn\SOSPduality{\eqalign{
SO(N)_k \ {\rm with}\ N_f\ {\rm real\ scalars} \quad &\longleftrightarrow \quad SO(k)_{-N+{N_f\over 2}} \ {\rm with}\ N_f\ {\rm real\ fermions} \cr
USp(2N)_k \ {\rm with}\ N_f\ {\rm scalars} \quad &\longleftrightarrow \quad USp(2k)_{-N+{N_f\over 2}} \ {\rm with}\ N_f\ {\rm fermions}
}}
for $N_f \leq N$ in the $USp$ case, and $N_f \leq N-2$ if $k=1$, $N_f \leq N -1$ if $k=2$, and $N_f \leq N$ if $k>2$ in the $SO$ case.

In the fermionic theories the only interactions are gauge interactions. On the contrary, the scalar theories also have generic quartic potential terms compatible with a given global symmetry. Hence, it is important to specify what symmetry we impose, as different choices in general lead to different fixed points.  We will discuss it in more details below.

\ifig\figDualities{Four UV theories, related by dualities, flow to the same IR fixed point. The two UV theories with $SU(2)$ gauge symmetry have a global $SO(3)$ symmetry, and therefore the IR theory also has that global symmetry. The two UV theories with $U(1)$ gauge symmetry have only $O(2)$ global symmetry. The duality implies that they have an enhanced quantum $SO(3)$ global symmetry in the IR.}%
{\epsfxsize4.5in\epsfbox{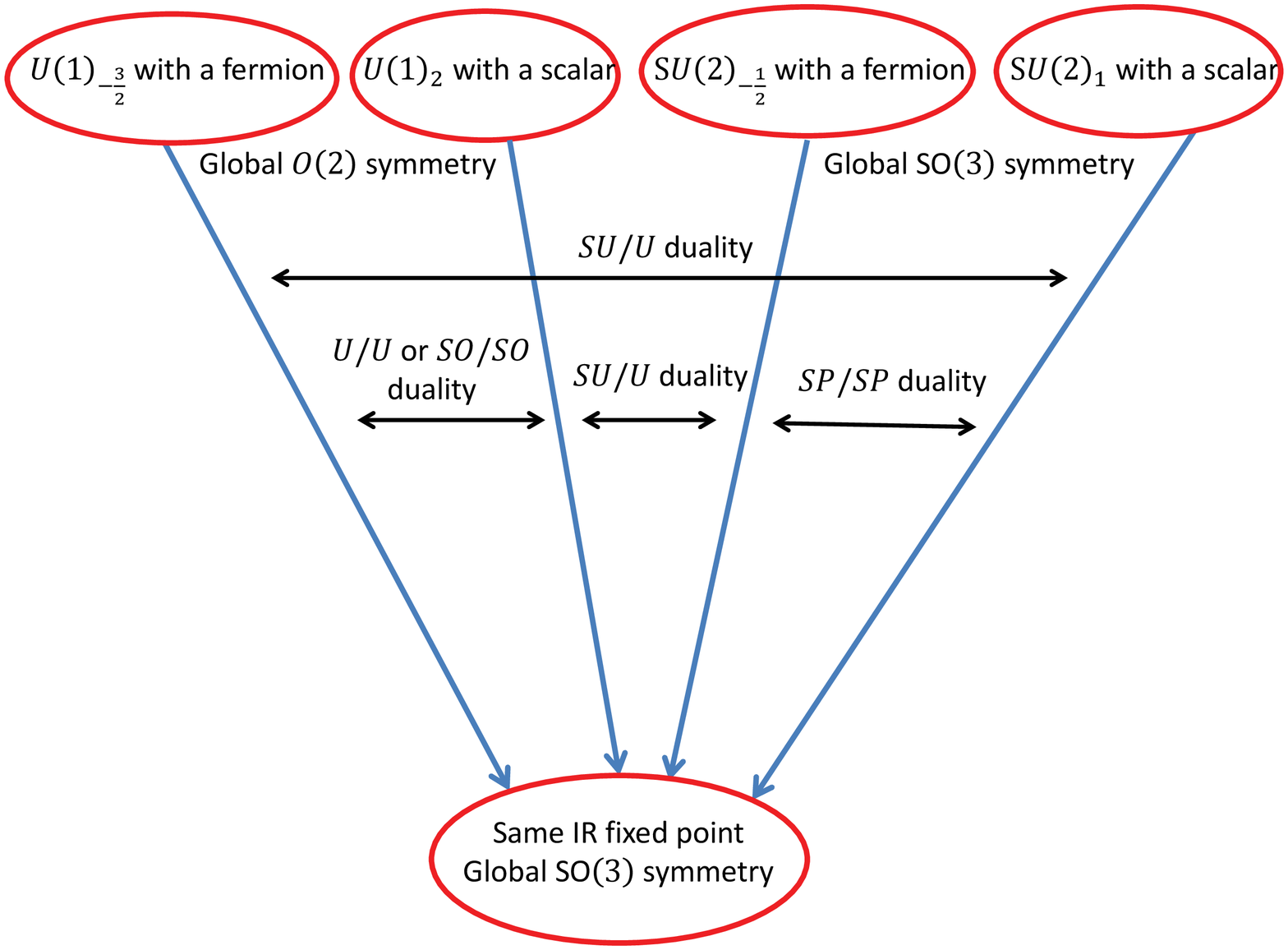}}

Some special examples of the dualities in \SUUduality\ and \SOSPduality\ lead to
\eqn\specialcases{\eqalign{
U(1)_0 \ {\rm with\ a\ scalar} \qquad &\longleftrightarrow \qquad {\rm a\ scalar}  \cr
U(1)_1 \ {\rm with\ a\ scalar} \qquad &\longleftrightarrow \qquad {\rm a\ free\ fermion}  \cr
U(1)_2 \ {\rm with\ a\ scalar}  \qquad &\longleftrightarrow \qquad \!\!\!\!\!\!\!\!\! \underbrace{SU(2)_1 \ {\rm with\ a\ scalar}}_{{\rm enhanced}\ SO(3)\ {\rm global\ symmetry}} \!\!\!\!\!\!\!\! \;.
}}
The first duality is the celebrated particle/vortex duality of \refs{\PeskinKP,\DasguptaZZ}. The second duality maps an interacting bosonic theory to a free fermion \SeibergGMD.  The theory in the third duality has a quantum $SO(3)$ global symmetry \AharonyJVV\ (see \figDualities). In all these cases the monopole operator of $U(1)_k$ in the theory on the left side of the duality, whose spin is $k\over 2$, is an important operator in the theory on the right side.  It is the scalar in the first case, it is the free fermion in the second case, and it is the new current of the enhanced $SO(3)$ symmetry in the third case.%
\foot{One might wonder whether the theory of $U(1)_3$ with a scalar, which has a monopole operator of spin $3\over 2$ and a global $U(1)$ symmetry, could have $\CN=2$ supersymmetry in the IR. It has a dual description as $SO(3)_{-{3\over 2}} \cong SU(2)_{-3}/\Z_2$ with a fermion in the adjoint \AharonyJVV, which seems to have $\CN=1$ supersymmetry.  However, this supersymmetric theory is expected to be gapped \WittenDS\ with a low energy $SO(3)_{-1} \cong SU(2)_{-2}/\Z_2$ trivial TQFT.  As we vary the fermion mass, we can find a transition to another gapped phase with a TQFT \hbox{$SO(3)_{-2} \cong SU(2)_{-4}/\Z_2 \leftrightarrow SU(3)_{-1} \leftrightarrow U(1)_3$}.  The duality statement could mean that the theory at this transition point is dual to the $U(1)_3$ theory with a scalar.  However, since we needed to change the fermion mass from the supersymmetric point, we broke supersymmetry explicitly and there is no reason to believe that the IR theory at the transition point is supersymmetric.  Alternatively, if the supersymmetric $\CN=1$ $SO(3)_{-{3\over 2}}$ theory is actually gapless, it could be dual to the $U(1)_3$ theory with a scalar, in which case it will also have enhanced $\CN=2$ supersymmetry.}

All these dualities are IR dualities.  We start at short distances with a renormalizable Lagrangian and impose some global symmetry on its terms.  Then, we scan the relevant deformations that are consistent with the global symmetry.  These are typically mass terms, but there are also others.  For generic values of these parameters the low-energy theory is gapped.  As these parameters are varied there could be phase transitions between different phases and the phase transition points occur at fine-tuned values of the scanned parameters.  We will assume that, as we vary these parameters, the phase transitions can be second order.  Then the long-distance physics is described by a fixed point of the renormalization group, which is a continuum conformal field theory.  The statement of the IR duality is about this fixed point and its neighborhood.  If, on the other hand, the IR theory is always gapped with possible first-order transitions between phases, the statement of the duality is significantly weaker and it applies only to the gapped phases.

\subsec{Global symmetries}

Our starting point is to identify the correct global symmetry of a quantum field theory. For the moment we ignore discrete symmetries like
time reversal ${\bf T}$ and higher-form global symmetries \refs{\KapustinGUA,\GaiottoKFA}.  We will discuss them later.

We should distinguish between the global symmetry of the UV theory $G^{\rm UV}$ and the global symmetry of the IR theory $G^{\rm IR}$.  Although there might be elements in $G^{\rm UV}$ that do not act on the IR degrees of freedom, we should still pay attention to them in the IR.  The IR effective action might contain topological local counterterms for background gauge fields coupled to those elements.

Conversely, there could be new elements in $G^{\rm IR}$ that are not present in the UV.  These lead to an accidental or quantum symmetry in the IR.  These symmetries are approximate and are violated by higher-dimension operators in the IR theory.  Examples of such quantum symmetries are common in $(1+1)d$ field theories and have played an important role in supersymmetric dualities, in particular in $(2+1)d$ mirror symmetry \refs{\IntriligatorEX\deBoerMP}.

A noteworthy simple example \AharonyJVV\ is summarized in \figDualities, where four different UV theories, some with $G^{\rm UV}=O(2)$ and some with $G^{\rm UV}=SO(3)$, flow to the very same IR fixed point with $SO(3)$ global symmetry (we will discuss this example in Section 3).

These considerations are extremely important in the context of duality.  Two dual theories $\CT_A$ and $\CT_B$ that flow to the same IR fixed point must have the same global symmetries. In some cases the UV symmetries are the same $G^{\rm UV}_A = G^{\rm UV}_B$.  But it is also common that the UV symmetries are different $G^{\rm UV}_A \not= G^{\rm UV}_B$, and yet they are enhanced to the same IR symmetry $G^{\rm IR}$. Again, the example in \figDualities\ demonstrates it and gives interesting consistency checks on the various dualities. We will see several examples of that in Section 3.

When we discuss the global symmetry $G$ (either $G^{\rm UV}$ or $G^{\rm IR}$) we should make sure that it acts faithfully on the operators.  Specifically, we will see many examples where all the local gauge-invariant operators in the theory transform in certain representations of the naive global symmetry group $G_{\rm naive}$, but the true global symmetry $G$---which acts faithfully---is a quotient $G=G_{\rm naive}/\CC$ by an appropriate $\CC$.

A key tool in the analysis of a quantum field theory is its coupling to background gauge fields for the global symmetry. If we misidentify the global symmetry and couple the system to background $G_{\rm naive}$ gauge fields, we miss important observables.  In particular, if all the local operators transform trivially under $\CC\subset G_{\rm naive}$ we can couple the system to $G=G_{\rm naive}/\CC$ bundles, which are not $G$ bundles.

For example, consider the $SU(2)_1$ theory with a scalar in \figDualities.  The naive global symmetry is $G_{\rm naive}=SU(2)$.  However, in this case all gauge-invariant operators in the theory are in integer isospin representations of this group and therefore the true global symmetry is $G=G_{\rm naive}/\Z_2= SO(3)$.  This means that the system can be coupled to additional background fields---$SO(3)$ gauge fields, which are not $SU(2)$ gauge fields. The response to such more subtle backgrounds leads to interesting observables, which give us more diagnostics of the theory.

More explicitly, we can couple the matter fields to
\eqn\prodquo{
\big( SU(2)_{\rm dyn} \times SU(2)_{\rm naive}\big)/\Z_2
}
gauge fields.  This is consistent because the matter fields do not sense the $\Z_2$ quotient.  The expression \prodquo\ means that when the classical fields are ordinary $SU(2)_{\rm naive}$ gauge fields, the dynamical gauge fields are ordinary $SU(2)_{\rm dyn}$ gauge fields.  However, when the classical fields are nontrivial $SO(3)$ gauge fields ({\it i.e.}\ $SO(3)$ fields with nontrivial second Stiefel-Whitney class $w_2$), also the dynamical fields are in $SO(3)$ bundles.  This demonstrates that by using $SO(3)$ background fields we can probe more twisted sectors of the dynamical fields.

Below we will see many generalizations of this example.  We will encounter dynamical fields $b$ for a gauge group $G_{\rm dyn}$ and background fields $B$ for the true global symmetry of the model $G=G_{\rm naive}/\CC$ for some $\CC$. As in the example \prodquo, the dynamical and classical fields can be combined to a gauge field $\CB$ with group
\eqn\prodquog{
\big( G_{\rm dyn} \times G_{\rm naive} \big)/\CC \;.
}
If the classical fields $B$ are in $G_{\rm naive}$ bundles, the dynamical fields $b$ are in $G_{\rm dyn}$ bundles.  But when $B$ are in nontrivial $G=G_{\rm naive}/\CC$ bundles, also the dynamical fields $b$ are in $G_{\rm dyn}/\CC$ bundles rather than in $G_{\rm dyn}$ bundles. The consistency of the theory under gauge transformations in \prodquog\ and possible anomalies in these transformations will be extremely important below.

We should point out that the authors of \refs{\KapustinLWA,\KapustinZVA} have examined such anomalies for discrete groups from a different perspective.

We will be particularly interested in the theories in \SUUduality\ and \SOSPduality, so let us discuss their UV symmetry $G^{\rm UV}$. In the fermionic case that is the actual UV symmetry of the theory, while in the bosonic case that is the symmetry that we impose on the quartic potential. The naive UV global symmetry $G_{\rm naive}$ is%
\foot{In this discussion we mostly neglect time-reversal symmetry ${\bf T}$.}
$U(N_f) \rtimes \Z_2^{\bf C}$ (where the second factor is charge conjugation) in the theories with $SU$ gauge group, $SU(N_f) \times U(1)_M \rtimes \Z_2^{\bf C}$ (where the second factor is the magnetic symmetry) for $U$ gauge group, $O(N_f) \times \Z_2^M \times \Z_2^{\bf C}$ for $SO$ gauge group,%
\foot{In this case $\Z_2^{\bf C}$ is an element of $O(N)$ not connected to the identity. When $N, N_f$ are both odd, $\Z_2^{\bf C}$ is already contained in $O(N_f)$ (up to a gauge transformation), and should not be listed as an independent symmetry.}
and $USp(2N_f)$ for $USp$ gauge group. However, we will find that the faithfully-acting symmetry $G^{\rm UV}$ is (we will not discuss the $SO$ case here):
\eqn\faithfulG{\eqalign{
{\rm Theory} \qquad\qquad& {\rm Global\ Symmetry}\ G^{\rm UV} \cr
SU(N)_{r} \ {\rm with}\ N_f\ {\rm scalars\ or\ fermions} \qquad& \big( U(N_f)/\Z_N \big) \rtimes \Z_2^{\bf C} \cr
U(N)_k \ {\rm with}\ N_f\ {\rm scalars} \qquad& \big( U(N_f)/\Z_k \big) \rtimes \Z_2^{\bf C} \cr
U(k)_{-N + \frac{N_f}2} \ {\rm with}\ N_f\ {\rm fermions} \qquad & \big( U(N_f)/\Z_N \big) \rtimes \Z_2^{\bf C} \cr
USp(2N)_{r} \ {\rm with}\ N_f\ {\rm scalars\ or\ fermions} \qquad& USp(2N_f)/\Z_2
}}
where $r$ is an integer in the theory with scalars and an integer plus $\frac{N_f}2$ in the theory with fermions.  Here by $U(N_f)/\Z_N$ we mean the quotient by $e^{2\pi i/N} \unit$. In the special cases of $U(N)_0$ with $N_f$ scalars and $U(k)_{\frac{N_f}2}$ with $N_f$ fermions, the global symmetry is $SU(N_f)/\Z_{N_f} \times U(1)_M \rtimes \Z_2^{\bf C}$ which is isomorphic%
\foot{More generally $U(N_f)/\Z_N$ is isomorphic to $U(N_f)/\Z_{N + N_f}$. Representing $U(N_f)/\Z_N$ as $\big(g\in SU(N_f), u \in U(1) \big)$ with $(g,u) \sim \big( e^{2\pi i /N_f} g, e^{-2\pi i /N_f} u \big) \sim \big( g, e^{2\pi i /N} u\big)$, the isomorphism is $(g,u) \to \big( g , v = u^\frac{N}{N+N_f} \big)$. The identifications map to $(g,v) \sim \big( e^\frac{2\pi i}{N_f} g, e^{-\frac{2\pi i}{N_f}} v \big) \sim \big( g, e^\frac{2\pi i}{N+N_f} v \big)$ that represent $U(N_f)/\Z_{N+N_f}$. See also footnote 10.}
to $U(N_f)/ \Z_{N_f} \rtimes \Z_2^{\bf C}$. One should be careful at small values of the ranks. For instance, $SU(2)_r$ with $N_f$ fermions has $USp(2N_f)/\Z_2$ symmetry as manifest in the $USp(2)_r$ description, while the symmetry of $SU(2)_r$ with $N_f$ scalars depends on what we impose on the quartic potential. This will be analyzed in Section 3.

\subsec{Anomalies}

It is often the case that the global symmetry $G$ has 't Hooft anomalies.  This means that the correlation functions at separated points are $G$ invariant, but the contact terms in correlation functions cannot be taken to be $G$ invariant.  Related to that is the fact that the system with nonzero background gauge fields for $G$ is not invariant under $G$ gauge transformations.  Often, this lack of $G$ gauge invariance of background fields can be avoided by coupling the system to a higher-dimensional bulk theory with appropriate bulk terms.

Let us discuss it more explicitly.  Since we denote the classical gauge fields by uppercase letters, $A$, $B$, etc., we will denote the coefficients of their Chern-Simons counterterms \refs{\ClossetVG,\ClossetVP} by $K$.\foot{We will use uppercase $N$ in the gauge group of dynamical fields and Chern-Simons levels of dynamical fields depending on $N$ and $N_f$.  We hope that this will not cause confusion.} They should be distinguished from the Chern-Simons coefficients of dynamical fields $a$, $b$, etc., which we denote by lower case $k$.  It is important that $k$ and $K$ should be properly normalized as $(2+1)d$ terms.  As we will see below, it is often the case that the proper normalization of these coefficients involves a nontrivial relation between $K$ and $k$.

It might happen that imposing the entire symmetry $G$ there is no consistent value of $K$.  In that case we say that $G$ has an 't~Hooft anomaly and we have two options.  First, we consider only a subgroup or a multiple cover of $ G$ and turn on background fields only for that group.  Alternatively, we allow gauge fields for the entire global symmetry group $G$, but extend them to a $(3+1)d$ bulk. In this case the partition function has a dependence on how the background fields are extended to the bulk. It is important, however, that the dynamical gauge fields are not extended to the bulk.

We will not present a general analysis of such anomalies.  Instead, we will first mention two well known examples.  Then make some general comments, and later in the body of the paper we will discuss more sophisticated examples.

A well known typical example in which we can preserve only a subgroup $\wh G \subset G$ is the time-reversal anomaly of $(2+1)d$ free fermions.  Here $G$ includes a global $U(1)$ symmetry and time reversal, but they have a mixed anomaly.  One common option is to preserve $\wh G=U(1)$, but not time reversal.  Alternatively, in the topological insulator we extend the background $U(1)$ gauge field to the bulk and we turn on a $(3+1)d$ $\theta$-parameter equal to $\pi$ \refs{\QiEW,\EssinRQ}, such that the entire global symmetry $G$ is preserved.

In this case the bulk term with $\theta=\pi$ is time-reversal invariant on a closed four-manifold, but not when the manifold has a boundary: a time-reversal transformation shifts the Lagrangian by a $U(1)_1$ Chern-Simons term.  This is an anomaly in time reversal.  The fermion theory on the boundary has exactly the opposite anomaly, such that they cancel each other and the combined $(3+1)d$ theory is anomaly free.

Another well known example, where we can preserve a multiple cover $ G_{\rm naive}$ of the global symmetry $G$, is the following.  Consider a quantum mechanical particle moving on $\S^2$ with a Wess-Zumino term with coefficient $k$.  (This is the problem of a charged particle on $\S^2$ with magnetic flux $k$.)  The global symmetry of the problem is $G=SO(3)$, but as we will soon review, for odd $k$ this symmetry is anomalous.

One way to represent the theory uses two complex degrees of freedom $z^i$ with a potential forcing $\left| z^1 \right|^2+ \left| z^2\right|^2 =1$.  This system has an $O(4)$ global symmetry.  Next we introduce a dynamical $U(1)$ gauge field $b$ coupled to the phase rotation of $z^i$.  The resulting theory is the $\C\P^1$ model whose target space is a sphere.  We can add to the theory the analog of a Chern-Simons term, which is simply a coupling $kb$.  In terms of the effective $\C\P^1$ model this is a Wess-Zumino term with coefficient $k$ \RabinoviciMJ.  The spectrum of the theory is well known: it is $\oplus_{j} \CH_j$, where $\CH_j$ is the isospin $j$ representation of $SU(2)$ and the sum over $j$ runs over $j={k\over 2}, {k\over 2} +1,...$

Naively, the global symmetry is $G_{\rm naive} = SU(2)$ which rotates $z^i$.  However, the global symmetry that acts faithfully is $G=SU(2)/\Z_2\cong SO(3)$.  To see that, note that the coordinates $z^i$ are coupled to a $U(1)$ gauge field $b$, can be further coupled to an $SU(2)$ classical field $B$, but then  $b$ and $B$ combine into a
\eqn\CBgauge{
U(2)=\big( U(1)_{\rm dyn} \times SU(2)_{\rm naive} \big)/\Z_2
}
gauge field $\CB$.  The expression \CBgauge\ shows that the element in the center of $SU(2)_{\rm naive}$ is not a global symmetry transformation---it acts as a gauge transformation.  Hence, the global symmetry that $B$ couples to is really $SO(3)$.  Indeed, all gauge-invariant local operators are in $SO(3)$ representations.

For even $k$ the Hilbert space includes integer $j$ representations and represents $SO(3)$ faithfully.  In this case there is no anomaly.  But for odd $k$ all the states in the Hilbert space have half-integer $j$ and the global symmetry acts projectively---it represents the double cover $G_{\rm naive}$.

What should we do about this anomaly?  One option is to say that the global $SO(3)$ symmetry acts projectively, or equivalently, the global symmetry is $SU(2)$.  A more interesting option is to introduce a $(1+1)d$ bulk $\wt\CM_2$ (with boundary the original timeline), and add a bulk term that depends on the $SO(3)$ gauge field $B$.

Explicitly, the original degrees of freedom $z^i$ couple to a $U(2)$ gauge field $\CB$. Therefore, the CS term $kb$ should be written as ${k\over 2} \Tr \CB$.  Although this is properly normalized as a CS term for a $U(1)$ gauge field $b$, for odd $k$ it is not properly normalized for a $U(2)$ gauge field $\CB$.  However, we can extend $\CB$ to the $(1+1)d$ bulk and replace the ill-defined contribution to the functional integral $e^{i {k\over 2} \int \Tr \CB}$ by the gauge-invariant expression
\eqn\bulkt{
e^{i{k\over 2} \int_{\rlap{\raise -1.5pt  \hbox{$\widetilde{\phantom{m}}$}} \CM_2} \Tr F_\CB} \;,
}
where $F_\CB$ is the field strength of $\CB$.

We should check whether \bulkt\ depends on the bulk values of the fields.  A standard way to do that is to replace the bulk $\wt\CM_2$ by another bulk $\wt\CM_2'$ with the same boundary, and to consider the integral in \bulkt\ over the closed manifold $\CM_2$ constructed gluing $\wt\CM_2$ with the orientation-reversal of $\wt\CM_2'$:
\eqn\bulktd{
e^{i{k\over 2} \int_{\CM_2} \Tr F_\CB} = e^{ik \int_{\CM_2} db} \;.
}
If this is always $1$ then \bulkt\ is independent of the bulk fields.  Had $b$ been an ordinary $U(1)$ gauge field, this would have been $1$ for every $b$ in the bulk.  But since the bulk involves nontrivial $SO(3)$ bundles, the gauge field $b$ can have half-integral periods (it is a spin$_c$ connection) and \bulktd\ can be $\pm 1$, thus showing that it depends on the bulk values of $b$.  However, this does not mean that the term \bulkt\ is not a valid term.  In fact its dependence on $b$ is completely fixed in terms of the $SO(3)$ gauge field $B$:
\eqn\bulktdw{
e^{i{k\over 2} \int_{\CM_2} \Tr F_\CB} = e^{ik \pi \int_{\CM_2} w_2(B)} \;,
}
where $w_2 \in H^2(\CM_2,\Z_2)$ is the second Stiefel-Whitney class of the $SO(3)$ bundle.

In other words, the bulk term \bulkt\ can be viewed as a $(1+1)d$ discrete $\theta$-parameter for the background $SO(3)$ gauge field, which is independent of the bulk values of the dynamical field $b$ as long as we keep the background fixed.  (It does depend on the boundary values of $b$.) The addition of such a bulk term, which depends on the background $SO(3)$ fields, is familiar in the famous Haldane chain.%
\foot{We thank E.~Witten for a useful discussion about the Haldane chain.}

The perspective on this phenomenon that we will use below is the following. The boundary theory---in this example a particle in the background of an odd-charge magnetic monopole---is anomalous and its action is not well-defined in $(0+1)d$ in the presence of $G$ background fields. To make it well-defined, we extend the background fields to a $(1+1)d$ bulk, making sure that there is no dependence on the extension of the dynamical gauge fields at fixed background. Then the bulk term $\int w_2(B)$ is well-defined mod~$2$ on a closed two-manifold, and it captures the dependence of the partition function on the extension of the background field $B$. On a manifold with boundary the definition of $\int w_2(B) \mod 2$ depends on additional data.  It is anomalous.  This anomaly is exactly canceled by the anomaly in the boundary theory, such that the combined system is well defined.

Below we will see higher-dimensional generalizations of these examples. Using the notation discussed around \prodquog, the dynamical fields $b$ will typically have Chern-Simons couplings $k$ while the background fields $B$ will have Chern-Simons couplings $K$.  In addition, for $U(1)$ factors in the two groups there can be mixed Chern-Simons couplings.  The way to properly define these couplings is by writing them as $(3+1)d$ bulk terms of the form $\theta \Tr F_\CB\wedge F_\CB$ or $\theta \Tr F_\CB\wedge\Tr F_\CB$ with various $\theta$'s, where $F_\CB$ are the field strengths of the gauge fields $\CB$.
In addition, we will also encounter discrete $\theta$-parameters, like those in \AharonyHDA.
In this form we have a well defined expression for gauge fields of $(G_{\rm dyn} \times G_{\rm naive})/\CC$.

As in the quantum mechanical example of a particle on $\S^2$, it is crucial that these bulk terms must be independent of the bulk values of $b$ at fixed $B$.  This guarantees that $b$ is a dynamical field living on the boundary.  If the bulk terms are also independent of the bulk values of $B$, we say that the global symmetry $G$ is anomaly free.  Instead, if there is a dependence on the bulk values of $B$, the global symmetry suffers from 't Hooft anomalies.

As in the same quantum mechanical example, we can check the independence of the bulk values of $b$ and characterize the dependence on the bulk values of $B$ by considering the bulk terms on a closed four-manifold $\CM_4$.  Then the integrals of the various $F\wedge F$ terms of $\CB$ should be expressed in terms of characteristic classes of $B$. These characteristic classes characterize the 't~Hooft anomalies.

Some of these characteristic classes are related to various discrete $\theta$-parameters.  We have already seen such a discrete $\theta$-parameter in \bulktdw. Below we will encounter the discrete $\theta$-parameter of \AharonyHDA, which is associated with the Pontryagin square operation $\CP(w_2)$ \refs{\Whitehead,\Browder}.  As in \refs{\KapustinGUA,\GaiottoKFA}, these can be represented by a two-form field $B$ with a $(3+1)d$ coupling $B\wedge B$.  This coupling is gauge invariant on a manifold without boundary.  But when a boundary is present, this term has an anomaly.  The anomaly is canceled by having an appropriate boundary theory, which has the opposite anomaly.  For a $(0+1)d$ boundary we have already seen that around \bulktdw, while below we will see examples with a $(2+1)d$ boundary.

It is well known that in $(3+1)$ dimensions, 't Hooft anomaly matching conditions lead to powerful consistency constraints on the IR behavior of a theory and on its possible dual descriptions.  Consider first the simpler case of $G^{\rm UV}_A = G^{\rm UV}_B$.  Then the 't Hooft anomaly, which is the obstruction on the theory to be purely $(2+1)$-dimensional, must be the same on the two sides of the duality.  In other words, if we need to couple the theory to a $(3+1)d$ bulk and add some bulk terms with coefficients $\theta$, these bulk terms should be the same in the two dual theories. Such $\theta $-parameters can be ordinary or discrete ones.  More precisely, $\theta$ should be the same, but the boundary counterterms $K^A$ and $K^B$ in the two theories can be different, provided they are properly quantized. This condition is the same as the celebrated 't Hooft anomaly matching.

In the more interesting case that $G^{\rm UV}_A \not= G^{\rm UV}_B$, we can use the constraint in the UV by coupling background fields to the common subgroup $G^{\rm UV}_A \bigcap G^{\rm UV}_B$.  Their $\theta$ must be the same on the two sides of the duality.  The IR theory can then be coupled to $G^{\rm IR}_A = G^{\rm IR}_B$ gauge fields and this analysis also allows us to determine the value of $\theta $ for these fields.  Again, we will see examples of that below.

\subsec{Outline}

In Section 2 we check 't~Hooft anomaly matching in the dualities \SUUduality-\SOSPduality.  This is both an example of our methods and a nontrivial new test of those dualities.

In Section 3 we focus on some interesting special cases of the dualities with gauge group $U(1) \cong SO(2)$ and $SU(2) \cong USp(2)$, either in the fermionic or the bosonic side. Such theories participate in more than one duality in \SUUduality-\SOSPduality. This leads to new tests of the dualities and to deeper insights into their dynamics. We also use those special cases to analyze theories with a surprising quantum $SO(3)$ global symmetry in the IR, as in \figDualities.

In Section 4 we follow \refs{\XuLXA,\HsinBLU} and consider in detail a fermion/fermion duality that leads to an enhanced $O(4) $ global symmetry.  We extend previous discussions of this system by paying close attention to the global structure of the global symmetry and to the counterterms.  This allows us to find the precise anomaly in $O(4)$ and time-reversal, and to restore those symmetries by adding appropriate bulk terms.

In Section 5 we analyze the phase diagram of systems with global $SO(5)$ symmetry and clarify some possible confusions about various fixed points with that global symmetry.

Appendix A derives the induced Wess-Zumino term in the model of Section 5, while Appendix B describes carefully the duality of \ChengPDN\ paying attention to the proper quantization of CS couplings, to the spin/charge relation, to the global structure of the symmetry group, and to the bulk terms. In Appendix C we discuss more examples of 't~Hooft anomalies.


\newsec{'t~Hooft Anomalies and Matching
\seclab\SecAnomaly}

We start by determining the 't~Hooft anomalies in the following theories:
\eqn\SUU{\eqalign{
SU(N)_k {\rm\ with}\ N_f\ \Phi \qquad&\longleftrightarrow\qquad U(k)_{-N + \frac{N_f}2} {\rm\ with}\ N_f \ \Psi \cr
U(N)_k {\rm\ with}\ N_f\ \Phi \qquad&\longleftrightarrow\qquad SU(k)_{-N + \frac{N_f}2} {\rm\ with}\ N_f \ \Psi \; .
}}
The dualities are valid only for $N_f \leq N$, but we will determine the symmetries and anomalies for generic integer values of $N$, $k$, $N_f$.  Here and in the following, to be concise, we indicate complex scalars as $\Phi$, real scalars as $\phi$, complex fermions as $\Psi$ and real fermions as $\psi$.

All four theories have a naive global symmetry $SU(N_f) \times U(1) \rtimes \Z_2^{\bf C}$, where the last factor is charge conjugation. In the theories with $SU$ gauge group, the first two factors combine into a manifest $U(N_f)$ acting on the scalars or fermions. In the theories with $U$ gauge group, $SU(N_f)$ acts on the scalars or fermions, while the Abelian factor is the magnetic $U(1)_M$, whose charge is the monopole number. However the faithfully-acting symmetry $G$ is a quotient thereof, which as we will soon see is $\big( U(N_f)/\Z_N \big)\rtimes \Z_2^{\bf C}$ in the first line of \SUU\ and $\big( U(N_f)/\Z_k \big)\rtimes \Z_2^{\bf C}$ in the second line, as summarized in \faithfulG.%
\foot{In the special cases of $U(k)_{N_f/2}$ with $N_f$ $\Psi$ and $U(N)_0$ with $N_f$ $\Phi$ the global symmetry is $SU(N_f)/\Z_{N_f} \times U(1)_M \rtimes \Z_2^{\bf C}$, which is isomorphic to $U(N_f)/ \Z_{N_f} \rtimes \Z_2^{\bf C}$ (see footnote 6). The scalar theory is also time-reversal invariant. More care has to be used in the case of $SU(2)$ gauge group, as explained in Section 3.}
For $N_f\leq N$ this is a check of the dualities.

There might be an obstruction---an 't~Hooft anomaly---to turning on background gauge fields for $G$. We will show that the obstruction is the same on the two sides of the dualities, thus providing a nontrivial check of them.

\subsec{Global symmetry}

The first step is to identify the global symmetry that acts faithfully on the four theories in \SUU. To do that, we analyze the local gauge-invariant operators.

Let us start with $SU(N)_k$ with $N_f$ scalars. There is a $U(N_f)$ symmetry that acts on the scalars in the fundamental representation, but only $U(N_f)/\Z_N$ acts faithfully on gauge invariants. In the absence of a magnetic symmetry, monopole operators do not change this result (since GNO flux configurations \GoddardQE\ are continuously connected to the vacuum). There is also a charge-conjugation symmetry $\Z_2^{\bf C}$ that exchanges the fundamental with the antifundamental representation, therefore the symmetry is $\big( U(N_f) /\Z_N \big) \rtimes \Z_2^{\bf C}$. By the same argument, $SU(k)_{-N + \frac{N_f}2}$ with $N_f$ fermions has $\big( U(N_f)/\Z_k \big)\rtimes \Z_2^{\bf C}$ symmetry.

Next consider $U(k)_{-N + \frac{N_f}2}$ with $N_f$ fermions. The bare CS level is $-(N-N_f)$. There is an $SU(N_f)$ symmetry that acts on the fermions in the fundamental representation and a $U(1)_M$ magnetic symmetry, whose charge is the monopole number. A monopole configuration of monopole number $Q_M$ has gauge charge $(N_f-N)Q_M$ under $U(1) \subset U(k)$. To form gauge invariants we dress the monopole with fermionic fields and their conjugates, and the net number of fundamentals minus antifundamentals is $(N-N_f)Q_M$. Therefore the operators are in $SU(N_f)$ representations of $N_f$-ality $NQ_M \mod N_f$. We can then combine $SU(N_f)$ with an $N$-fold multiple cover of $U(1)_M$ to form $U(N_f)$, and gauge-invariant local operators give representations of $U(N_f)/\Z_N$. Going to the multiple cover is natural from the point of view of the duality, since a monopole of charge $1$ under $U(1)_M$ is mapped to a baryon of charge $N$ under $U(1) \subset U(N_f)$ \RadicevicYLA.%
\foot{Instead of going to the multiple cover and then take the $\Z_N$ quotient, we can represent the symmetry group as $\big( SU(N_f) \times U(1)_M \big)/\Z_{N_f}$ where $\Z_{N_f}$ is generated by \hbox{$g = (e^{-2\pi i /N_f}\unit, e^{2\pi i N/N_f})$}. This $\Z_{N_f}$ has a $\Z_d$ subgroup generated by $g^{N_f/d}$, where $d=\gcd(N,N_f)$, which acts only on $SU(N_f)$. Therefore, only the representations of $SU(N_f)/\Z_d$ appear.}
Charge conjugation acts both on $SU(N_f)$ and $U(1)_M$, therefore the full symmetry is $\big( U(N_f) / \Z_N \big) \rtimes \Z_2^{\bf C}$. In the case that $N=0$ the symmetry is $\big( U(N_f)/\Z_{N_f} \big)\rtimes \Z_2^{\bf C}$.

By the same argument, $U(N)_k$ with $N_f$ scalars has a faithfully-acting symmetry $\big( U(N_f)/\Z_k \big)\rtimes \Z_2^{\bf C}$. When $k=0$ the symmetry is $\big( U(N_f)/\Z_{N_f} \big) \rtimes \Z_2^{\bf C} \times \Z_2^{\bf T}$.

\subsec{Background fields}

Now we turn on a background for the $SU(N_f) \times U(1)$ symmetry of the four theories in \SUU, which can always be done, and analyze under what conditions the background gauge fields can be extended to $U(N_f)/\Z_N$ or $U(N_f)/\Z_k$ bundles.

Consider $SU(N)_k$ with $N_f$ scalars. Turning on background gauge fields with generic CS counterterms we obtain the theory
\eqn\sunkPhiwithbackground{
\frac{SU(N)_k \times SU(N_f)_L \times U(1)_J}{\Z_N \times \Z_{N_f}} \ \ {\rm with}\ \Phi {\rm\ in}\ ({\bf N}, {\bf N_f}, 1) \;.
}
The $\Z_N$ quotient acts anti-diagonally on $SU(N)$ and the Abelian factor by a phase rotation $e^{2\pi i/N}$, while $\Z_{N_f}$ acts anti-diagonally on $SU(N_f)$ and the Abelian factor by $e^{2\pi i/N_f}$.
The quantization conditions on CS counterterms are
\eqn\anomaly{
L \in \Z \;,\qquad J - Nk \in N^2 \Z \;,\qquad J  - N_f L \in N_f^2 \Z \;,\qquad J \in N N_f\Z \;.
}
The first condition comes from the $SU(N_f)$ factor. The second and third conditions come from the separate quotients by $\Z_N$ and $\Z_{N_f}$, respectively. The last condition ensures that the generators of $\Z_N$ and $\Z_{N_f}$ have trivial braiding and one can take the simultaneous quotient.

The equations in \anomaly\ have solutions in $L, J$, if and only if $k=0 \mod {\rm gcd}(N,N_f)$. If this is not the case, there is an 't~Hooft anomaly and the theory with background is not consistent in $(2+1)d$. One can make sense of the theory on the boundary of a $(3+1)d$ bulk, but then there is an unavoidable dependence on how the classical background fields are extended to the bulk. We will express the anomaly below.

Now consider $U(k)_{-N + \frac{N_f}2}$ with $N_f$ fermions. With background gauge fields we obtain
\eqn\uknPsiwithbackground{
U(k)_{-N + \frac{N_f}2} \times SU(N_f)_{L + \frac k2} \ \ {\rm with}\ \Psi {\rm\ in}\ ({\bf k}, {\bf N_f}),\ {\rm magnetic}\ U(1)_{K_f} {\rm\ and}\ / \Z_{N_f} \;.
}
We stress that the magnetic $U(1)$ is coupled to $U(k)$ by a mixed CS term.
The quotient by $\Z_{N_f}$ acts on $SU(N_f)$ and the two Abelian factors. We have chosen to parametrize the CS counterterms in a way that matches the dual description \sunkPhiwithbackground\ when the duality is valid. Then the topological symmetry $U(1)_{K_f}$ has CS counterterm%
\foot{It can be interpreted as the $\Z_N$ quotient of $U(1)_{J-Nk}$.}
$K_f = \frac{J - Nk}{N^2}$. To see that, we mass deform the scalar theory by $\pm|\Phi|^2$ and the fermionic theory by $\mp\bar\Psi\Psi$. The two resulting topological theories are identified, exploiting level-rank duality on the dynamical fields.%
\foot{We cannot use level-rank duality on the background fields, which are not integrated over in the path-integral.}
The map of CS counterterms for the $U(1)$ global symmetry was already discussed in \HsinBLU.

The quantization condition for $SU(N_f)_{L+\frac k2}$ with $k$ fermions is $L \in \Z$, which reproduces the first condition in \anomaly. The topological factor $U(1)_{K_f}$ determines the condition $K_f \in \Z$, which reproduces the second one in \anomaly.
To understand the $\Z_{N_f}$ quotient, consider the Abelian factors:
\eqn\kmatrix{
\CL_{\rm Abelian} = - \frac{k(N-N_f)}{4\pi} \hat a d\hat a + \frac k{2\pi} \hat a d B + \frac{K_f}{4\pi} B dB \;,
}
where we have indicated as $\hat a\unit_k$ the Abelian factor in $U(k)$. The equations of motion are as follows (neglecting the matter contribution):
\eqn\eom{\eqalign{
k(N_f-N)\, d\hat a + k\, dB &= 0 \cr
k\, d\hat a + K_f\, dB &=0 \;.
}}
We are after a $\Z_{N_f}$ one-form symmetry---then the matter contribution is canceled by a rotation in the center of $SU(N_f)$. An integer linear combination of the equations in \eom\ gives $N_f k\, d\hat a + \frac JN dB = 0$, which describes a $\Z_{N_f}$ one-form symmetry, if and only if $J \in N N_f \Z$. This reproduces the fourth condition in \anomaly. The generator of the one-form symmetry is the line
\eqn\ZNfgen{
\CW_{N_f}= k\oint \hat a + \frac{J}{NN_f} \oint B \qquad{\rm with\ spin}\qquad S\big(\CW_{N_f}\big) = \frac{J + N_f k}{2N_f^2} \quad \mod{1} \;.
}
To perform the $\Z_{N_f}$ quotient in the fermionic theory we combine $\CW_{N_f}$ with the $\Z_{N_f}$ generator of $SU(N_f)$. The spin of the latter is $-(L + k)/2N_f \mod{1}$ (since the bare CS counterterm of $SU(N_f)$ is $L + k$). The $\Z_{N_f}$ quotient is well-defined if its generator has integer or half-integer total spin,
\eqn\anomalyferm{
\frac{J + N_f k}{2N_f^2} - \frac{L + k}{2N_f} \,\in\, \frac{\Z}2 \;,
}
which reproduces the third condition in \anomaly. Thus the 't~Hooft anomaly is the same on the two sides of the duality \SUU.

The discussion in the other two cases is similar. Consider $SU(k)_{-N + \frac{N_f}2}$ with $N_f$ fermions first. Turning on background gauge fields we have
\eqn\suknPsiwithbackground{
\frac{SU(k)_{-N + \frac{N_f}2} \times SU(N_f)_{L + \frac k2} \times U(1)_{J + \frac{kN_f}2} }{\Z_k \times \Z_{N_f}} \ \ {\rm with}\ \Psi {\rm\ in}\ ({\bf k}, {\bf N_f}, 1) \;.
}
Taking into account the bare CS levels, the quantization conditions are
\eqn\anomalyII{
L \in \Z \;,\qquad J + kN \in k^2 \Z \;,\qquad J - N_f L \in N_f^2 \Z \;,\qquad J \in kN_f \Z \;.
}
They have solutions, if and only if $N=0\mod {\rm gcd}(k,N_f)$, otherwise there is an 't~Hooft anomaly.

Next consider $U(N)_k$ with $N_f$ scalars. With background gauge fields we have
\eqn\unkPhiwithbackground{
U(N)_k \times SU(N_f)_L \ \ {\rm with}\ \Phi {\rm\ in}\ ({\bf N}, {\bf N_f}),\ {\rm magnetic}\ U(1)_{K_s} {\rm\ and}\ / \Z_{N_f} \;.
}
The CS counterterms are chosen to match with those in \suknPsiwithbackground\ when the duality is valid, with $K_s = \frac{J+kN}{k^2}$. The $SU(N_f)$ and $U(1)$ factors give the quantization conditions $L\in\Z$ and $K_s\in\Z$, respectively. An integer linear combination of the equations of motion for the Abelian factors is $(J/k)\, dB=0$ (where $B$ is the $U(1)_{K_s}$ background gauge field) which describes a $\Z_{N_f}$ one-form symmetry, if and only if $J \in kN_f\Z$. Such a one-form symmetry is generated by the line $\CW_{N_f} = \frac{J}{kN_f} \oint B$ with spin $J/2N_f^2 \mod 1$. This has to be combined with the generator in $SU(N_f)_L$, and the condition that the total spin be in $\frac12\Z$ reproduces the third condition in \anomalyII. Thus, all conditions in \anomalyII\ are reproduced and the anomaly matches across the duality.

We should emphasize again that if we are only interested in the naive global symmetry group $G_{\rm naive} = SU(N_f)\times U(1)$, which does not act faithfully, there is no problem turning on background gauge fields.  The issue is only in considering gauge fields of the quotient group.  In that case we can attach the system to a bulk, extend the fields to the bulk and replace the Chern-Simons terms by $F\wedge F$ type terms there.  Then the point is that the resulting theory depends on the extension.  From this perspective, the 't~Hooft anomaly matching is the statement that we can use the same bulk with the same background fields there and attach to it either of the two dual theories on the boundary.

Consider the theory $SU(N)_k$ with $N_f$ scalars in \sunkPhiwithbackground. To express the dependence on the bulk fields, we proceed as follows. A $U(N_f)/\Z_N$ bundle can be represented by two correlated bundles, $PSU(N_f)$ and $U(1)/\Z_{D}$, where we set $D = \lcm(N,N_f) = \frac{NN_f}d$ and $d = \gcd(N,N_f)$. We define $w_2^{(N_f)} \in H^2(\CM_4, \Z_{N_f})$ as the second Stiefel-Whitney class of the $PSU(N_f)$ bundle, and $\wt F = DF$ (in terms of the $U(1)$ field strength $F$) as the well-defined field strength of the $U(1)/\Z_D$ bundle. Then the correlation between the two bundles is expressed by the fact that
\eqn\correlationbundles{
\frac{\wt F}{2\pi} = \frac{N_f}d \, w_2^{(N)} + \frac{N}d \, w_2^{(N_f)} \quad\mod D
}
for some class $w_2^{(N)} \in H^2(\CM_4,\Z_N)$. Such a class is the obstruction to lift a $U(N_f)/\Z_N$ bundle to a $U(N_f)$ bundle.

Now consider a general bundle for the group in \sunkPhiwithbackground. The $PSU(N)$ bundle associated to the dynamical fields is correlated with the $U(N_f)/\Z_N$ bundle such that their Stiefel-Whitney classes are equal: $w_2\big( PSU(N) \big) = w_2^{(N)}$. Therefore the dependence on the bulk fields is completely fixed by the classical $U(N_f)/\Z_N$ background. Such a dependence is described by
\eqn\SUUanomaly{
S_{\rm anom} = 2\pi  \int_{\CM_4} \bigg[ - \frac kN \, \frac{\CP(w_2^{(N)})}2 - \frac L{N_f} \, \frac{\CP(w_2^{(N_f)})}2 + \frac{J}{D^2} \frac{\wt F^2}{8\pi^2} \bigg] \;.
}
The integral is on a closed spin four-manifold $\CM_4$, and $\cP$ is the Pontryagin square operation \refs{\Whitehead,\Browder} such that $\cP(w_2^{(N)})/2 \in H^4(\CM_4, \Z_N)$, etc. (for more details see \AharonyHDA\ and references therein). We say that $e^{iS_{\rm anom}}$ captures the phase dependence of the partition function on the bulk extension of the $U(N_f)/\Z_N$ bundle, in the sense that given two different extensions one can glue them into a closed manifold $\CM_4$ and then $e^{iS_{\rm anom}}$ is the relative phase of the two partition functions.

If we choose $J \in D\Z$, then we can substitute the square of \correlationbundles\ into \SUUanomaly\ to obtain%
\foot{If $J\not\in D\Z$ then \SUUanomaly\ contains more information than $w_2^{(N)}$ and $w_2^{(N_f)}$.}
\eqn\SUUanomalyI{
S_{\rm anom} = 2\pi \int_{\CM_4} \bigg[ \frac{J - Nk}{N^2} \; \frac{\CP(w_2^{(N)})}2 + \frac{J - N_f L}{N_f^2} \; \frac{\CP(w_2^{(N_f)})}2 + \frac{J}{NN_f} w_2^{(N)} \cup w_2^{(N_f)} \bigg] \;,
}
which is well-defined modulo $2\pi$. From this expression it is clear that if we can solve the constraints in \anomaly, then $e^{iS_{\rm anom}}=1$ and there is no anomaly. On the other hand, it is always possible to make a suitable choice of $L,J$ such that $S_{\rm anom}$ reduces to
\eqn\SUUanomalyII{
S_{\rm anom} = - 2\pi \, \frac{(k {\rm\ mod}\ d)}N \int_{\CM_4} \frac{\CP(w_2^{(N)})}2 \;.
}
We can regard this as a minimal expression for the anomaly.

As we have shown, the anomaly in $U(k)_{-N+\frac{N_f}2}$ with $N_f$ fermions is the same as in \SUUanomaly. However one has to remember that the $U(1)$ in \SUUanomaly\ is an $N$-fold multiple cover of $U(1)_M$. The special case $N=0$ is discussed in Appendix C. The other two cases are similar, with an obvious substitution of parameters, and are presented in Appendix C.

Although we checked the anomaly matching separately for the two dualities, in fact they are related by performing $S, T$ operations on the $U(1)$ symmetry \refs{\WittenYA,\HsinBLU}. Since the operations add equal terms on both sides, the change in the bulk dependence on both sides must be equal, and thus the anomaly must still match. The anomaly also matches for other dualities obtained from them by $S,T$ operations, such as the last two dualities in \SUUduality.

In general, the anomaly is characterized by bulk terms that are meaningful on closed manifolds, but anomalous when there is a boundary.%
\foot{We thank Dan Freed for a useful discussion about this point.}
This is true for the anomaly \SUUanomaly\ where ${\cal P}(w_2)$ is meaningful only on a closed manifold, and it is also true for the two examples discussed in Section 1.2.

Although we do not need it for the dualities, it is nice to demonstrate our general analysis of the anomaly by specializing it to a $U(1)$ gauge theory of scalars with $k=0$.  Ignoring charge conjugation, the global symmetry is $\big(SU(N_f)/\Z_{N_f}\big)\times U(1)_M$.  The scalars are coupled to a $U(N_f)$ gauge field $B$, where the $U(1)\subset U(N_f)$ gauge field $b$ is dynamical.  More precisely, $b$ satisfies $N_f b =\Tr B$.  Therefore, the coupling to the magnetic $U(1)_M$ background field $B_M$ is the ill-defined expression ${1\over 2\pi N_f} (\Tr B) d B_M$ that needs to be moved to the bulk. This highlights that the global symmetry suffers from 't Hooft anomalies, which are characterized by the bulk term
\eqn\anomalyUzero{
S_{\rm anom} = 2\pi \int \frac1{N_f} \, w_2\big( PSU(N_f) \big) \cup \frac{dB_M}{2\pi} \;.
}
This discussion is analogous to a similar example in \GaiottoYUP. See Appendix C for more details.

\subsec{Symplectic gauge group}

We conclude this section by briefly analyzing the 't~Hooft anomalies in the two theories
\eqn\USp{
USp(2N)_k \ {\rm with}\ N_f \ \Phi \qquad\longleftrightarrow\qquad USp(2k)_{-N + \frac{N_f}2} \ {\rm with}\ N_f \ \Psi \;.
}
Again, the dualities are valid only for $N_f \leq N$, but we will study these theories for generic integer values of $N, k, N_f$.  Since there is no magnetic symmetry, the faithfully-acting symmetry $G$ is the one acting on gauge invariants constructed out of the scalars or fermions, which is $USp(2N_f)/\Z_2$ in both cases.

Coupling the two theories to a generic background, we obtain
\eqn\USpwithbackground{\eqalign{
& \frac{USp(2N)_k \times USp(2N_f)_L}{\Z_2} {\rm\ with}\ \Phi {\rm\ in}\ ({\bf 2N}, {\bf 2N_f}) \qquad\longleftrightarrow\cr
&\qquad\qquad\qquad\qquad \frac{ USp(2k)_{-N + \frac{N_f}2} \times USp(2N_f)_{L + \frac k2} }{\Z_2} {\rm\ with}\ \Psi {\rm\ in}\ ({\bf 2k}, {\bf 2N_f}) \;.
}}
Recall that the scalars and fermions are in a pseudo-real representation, therefore they are subject to a symplectic reality condition. The CS counterterms are chosen in such a way that they match when the theories are dual. The quantization conditions are
\eqn\anomalyUSp{
Nk + N_fL \in 2\Z
}
together with $L\in\Z$ in both theories. This provides 't~Hooft anomaly matching for the duality \AharonyJVV.

When $Nk$ is odd and $N_f$ is even, \anomalyUSp\ cannot be solved and we have an `t~Hooft anomaly. The anomaly is captured by the bulk term
\eqn\sofiveanomaly{
S_{\rm anom} = \pi \int_{\CM_4} \frac{\CP(w_2)}2 \;,
}
where $w_2$ is the second Stiefel-Whitney class of the $USp(2N_f)/\Z_2$ bundle.
Given two different extensions of the bundle, $e^{iS_{\rm anom}}=\pm1$ (evaluated on their gluing $\CM_4$) is the relative sign of the two partition functions.


\newsec{Quantum Global Symmetries from Special Dualities
\seclab\SecDualities}

Infrared dualities provide alternative descriptions of the same IR physics. It might happen that one description, say $\CT_A$, makes a symmetry transformation manifest all along its RG flow, while the same symmetry is not present in the other description, say $\CT_B$. Then, duality predicts that $\CT_B$ develops the symmetry quantum mechanically in the IR, because of strong coupling. In this section we survey various dualities at our disposal \refs{\AharonyMJS,\SeibergGMD,\HsinBLU,\AharonyJVV} and examine in what cases they predict a quantum enhancement of the global symmetry in the IR.

The theories we consider have $N_f$ scalars or fermions in the fundamental representation. For gauge group $U$ they have a naive global symmetry $SU(N_f) \times U(1)_M \rtimes \Z_2^{\bf C}$, for gauge group $SU$ have symmetry $U(N_f) \rtimes \Z_2^{\bf C}$, for gauge group $SO$ have $O(N_f) \times \Z_2^M \times \Z_2^{\bf C}$, and for gauge group $USp$ have $USp(2N_f)$. In addition, they might have time-reversal symmetry depending on the CS level. We have analyzed in Section \SecAnomaly\ how Chern-Simons interactions determine the faithfully-acting subgroup, and the result for large enough values of $N$ is summarized in \faithfulG.

The special cases $U(1) \cong SO(2)$ and $SU(2)\cong USp(2)$ need special attention. Fermionic theories with $SO(2)$ gauge group have $SU(N_f)\times U(1)_M \rtimes \Z_2^{\bf C}$ naive symmetry (as seen in the $U(1)$ language) and fermionic theories with $SU(2)$ gauge group have $USp(2N_f)$ (as seen in the $USp(2)$ language).

For scalar theories there are two subtleties to take into account. First, when using these theories in dualities $N_f$ is restricted ($N_f\le N$ in $SU/U$ and $USp$ dualities, while $N_f\le N-2$ for $k=1$, $N_f\le N-1$ for $k=2$, and $N_f\le N$ for $k>2$ in $SO$ dualities).  Second, in the scalar theories we turn on quartic couplings in the UV and we must analyze their global symmetries.

The $U(1) \cong SO(2)$ theory with one scalar ($N_f=1$) has only a single gauge-invariant quartic coupling $(\Phi\Phi^\dagger)^2$. The theory preserves a $U(1)_M$ global symmetry, not present in generic $SO(N)$ theories.
The theory with $N_f=2$ scalars has a single gauge-invariant $SU(2)$-invariant quartic coupling $(\Phi^i \Phi^\dagger_i)^2$. However, this theory does not participate in the $SU/U$ dualities. The $SO$ dualities use this theory for $k>2$, but they require only $SO(2)\subset SU(2)$ invariance (in addition to the $U(1)_M$ global symmetry).  There are two quartic couplings that respect that symmetry, $(\Phi^i \Phi^\dagger_i)^2$ and $\Phi^i \Phi^i \Phi^\dagger_j \Phi^\dagger_j$,  and the $SO(2)_k$ theories with those two couplings have $SO$ duals.

The $USp(2)\cong SU(2)$ theory with one scalar ($N_f=1$) has a single gauge-invariant quartic coupling $(\Phi^a \Phi^\dagger_a)^2$ which preserves a global $SO(3)$ symmetry.  The same theory with $N_f=2$ has several gauge-invariant quartic couplings.  One of them is $SO(5)$ invariant, $(\Phi^{ai} \Phi^\dagger_{ai})^2$.  However, this theory does not participate in the $USp$ dualities.  There are two distinct gauge-invariant quartic couplings that preserve an $SO(3) \times O(2) \subset SO(5)$ global symmetry, the previous one and $\Phi^{ai} \Phi^\dagger_{aj} \Phi^{bj} \Phi^\dagger_{bi}$.  These two couplings are assumed to be present in the theories with $SU/U$ duals.

In the following, we analyze in detail these low-rank cases.

\subsec{$U(1)_k$ with $1$ $\Phi$}

We exploit that $U(1)_k$ with 1 $\Phi$ $\cong$ $SO(2)_k$ with 1 $\phi$. The $SO$ duality requires $N_f = 1$ if $k=2$ and $N_f\leq 2$ if $k>2$. The $SU/U$ duality requires $N_f = 1$. Therefore consider $N_f=1$ and $k\geq 2$. There is only one quartic term in the $U(1)_k \cong SO(2)_k$ scalar theory and the following fixed points are all the same:
\eqn\Specialbos{\eqalign{
SU(k)_{-{1\over 2}}  \ {\rm with \ 1}\ \Psi \qquad\longleftrightarrow&\qquad
U(1)_k {\rm\ with\ 1}\ \Phi \;\cong\; SO(2)_k {\rm\ with\ 1 }\ \phi \cr
U(k+1)_{-{1\over 2},{1\over 2}+k} \ {\rm with \ 1}\ \Psi \qquad\longleftrightarrow& \qquad\qquad\qquad\qquad\qquad\qquad\quad {\Big\updownarrow} \cr
U(k-1)_{-{1\over 2},{1\over 2}-k} \ {\rm with \ 1}\ \Psi \qquad\longleftrightarrow& \qquad\qquad\qquad\qquad\quad\;\;\; SO(k)_{-{3\over 2}} \ {\rm with\  1} \ \psi
}}
In the generic case the fixed point has $U(1) \rtimes \Z_2^{\bf C} \cong O(2)$ symmetry, which is a quantum symmetry in the fermionic $SO(k)_{-\frac32}$ theory.
In the special case $k=2$ the symmetry becomes $SO(3)$, which is visible in the $SU(2)_{-\frac12}$ fermionic theory while it is a quantum symmetry in all other descriptions. This case is precisely the one in \figDualities, indeed the scalar theory is the third example in \specialcases.

\subsec{$U(1)_{-N+{N_f\over 2}}$ with $N_f$ $\Psi$}

We exploit that $U(1)_{-N + \frac{N_f}2}$ with $N_f$ $\Psi$ $\cong$ $SO(2)_{-N + \frac{N_f}2}$ with $N_f$ $\psi$.
The $SO$ duality requires $N_f\leq N-1$. Then the following fixed points coincide:
\eqn\Specialferm{\eqalign{
SU(N)_1  \ {\rm with \ } N_f\ \Phi \;\longleftrightarrow\;\;\;&\;
U(1)_{\!\!\!\!\!{\vrule height 1.2 em width 0 pt} -N+{N_f\over 2}} {\rm\ with\  } N_f\ \Psi  \cong\; SO(2)_{\!\!\!\!\!{\vrule height 1.2 em width 0 pt} -N+{N_f\over 2}} {\rm\ with\  } N_f\ \psi \cr
U(N+1)_{1,-N} \ {\rm with \ }N_f\ \Phi \;\longleftrightarrow& \qquad\qquad\qquad\qquad\qquad\qquad\;\; {\Big\updownarrow} \cr
U(N-1)_{1,N} \ {\rm with \ }N_f\ \Phi \;\longleftrightarrow& \qquad\qquad\qquad\qquad\qquad\quad\; SO(N)_{2} \ {\rm with\  }N_f \ \phi
}}
In the generic case there is a $\big( U(N_f)/\Z_N \big)\rtimes \Z_2^{\bf C}$ symmetry, which is a quantum symmetry in the $SO(N)_2$ bosonic description.
In the special case $N=2$ and $N_f=1$, the fixed point coincides with \Specialbos\ with $k=2$ (this case is the one in \figDualities\ and in the third line of \specialcases). The symmetry becomes $SO(3)$, which is visible in the $SU(2)_1$ bosonic theory while it is a quantum symmetry in the other descriptions.

\subsec{$SU(2)_k$ with $1$ $\Phi$}

We exploit $SU(2)_k \cong USp(2)_k$. In the case $N_f=1$ both the $SU/U$ and $USp$ dualities are valid. The scalar theory has only one quartic gauge invariant, thus the two dualities share the same fixed point:
\eqn\sutwoonebos{\eqalign{
SU(2)_{k} \ {\rm with \ }1\ \Phi \qquad&\longleftrightarrow\qquad
USp(2k)_{-{1\over 2}}\ {\rm with \ }1\ \Psi \cr
{\Big\updownarrow} \qquad\qquad\quad &\qquad\cr
U(k)_{-{3\over 2}}\ {\rm with \ }1\ \Psi \quad\;\;\;&
}}
The two theories in the first row have manifest $SO(3)$ symmetry. The theory in the second row only has $U(1)_M \rtimes \Z_2^{\bf C}$ classically visible, thus it has enhanced quantum $SO(3)$ symmetry. In the special case $k=1$, the fixed point coincides with \Specialbos\ with $k=2$ (as in \figDualities\ and third line of \specialcases).

\subsec{$SU(2)_k$ with $2$ $\Phi$}

We could write the theory as $USp(2)_k$ with 2 $\Phi$, which has $N=1$ and $N_f=2$, however the $USp$ duality requires $N_f\leq N$ and so it is not valid. The $SU/U$ duality, instead, is valid.
In such a duality the scalar theory has two quartic terms, singlets under $\big( U(2)/ \Z_2 \big) \rtimes \Z_2^{\bf C} \cong SO(3) \times O(2)$. One term can be written as $\CO_{\bf 1}^2$, where $\cO_{\bf1}$ is the quadratic gauge-invariant $SO(5)$-invariant operator (the subscript indicates the $SO(5)$ representation). The other term is one of the components of $\cO_{\bf14}$ (a symmetric traceless rank-2 tensor of $SO(5)$) with the choice of coupling $\lambda_{\bf 14} \propto {\rm diag}(-3,-3,2,2,2)$, which breaks $SO(5)$ to $SO(3)\times O(2)$. (As we discuss later in Section 5, different signs of $\lambda_{\bf 14}$ could lead to two distinct fixed points. Here we choose the sign that produces the fixed point involved in the $SU/U$ duality.)	
Tuning $\lambda_{\bf 14}=0$ in the scalar theory gives a different fixed point with $SO(5)$ symmetry.
The flows are summarized as follows:
\eqn\flowsutwonftwo{\eqalign{
SU(2)_k \ {\rm with\ 2 }\ \Phi
\quad& \buildrel{\lambda_{\bf1}, \lambda_{\bf14} \strut}\over\longrightarrow \quad
{\rm CFT\ with }\ SO(3)\times O(2) \quad\longleftarrow\quad U(k)_{-1}\ {\rm with\ } 2\ \Psi\cr
\lambda_{\bf 1}{\Big\downarrow} \qquad\qquad& \cr
{\rm CFT\ with\ } SO(5) \;\;&
}}
This example, discussed at length in Section 5, does not develop quantum symmetries.

\subsec{$SU(2)_{-N+{N_f\over 2}}$ with $N_f$ $\Psi$}

Both $SU/U$ and $USp$ dualities require $N_f\leq N$. The two dualities have common fermionic theory and thus the fixed points are the same:
\eqn\sutwonffermion{\eqalign{
SU(2)_{-N+{N_f\over 2}} \ {\rm with \ }N_f\ \Psi
\quad&\longleftrightarrow\quad
USp(2N)_{1}\ {\rm with \ }N_f\ \Phi\cr
{\Big\updownarrow} \qquad\qquad\qquad&\qquad\cr
U(N)_{2}\ {\rm with \ }N_f\ \Phi \qquad&
}}
The fixed point has $USp(2N_f)/\Z_2$ symmetry, which is a quantum symmetry in the bosonic $U(N)_2$ theory.
When $N=N_f=1$, the fixed point coincides with \Specialbos\ with $k=2$ (as in \figDualities\ and the third line of \specialcases).


\subsec{Examples with Quantum $\mbf{SO}$(3) Symmetry and 't~Hooft anomaly matching}

Consider the examples with enhanced $SO(3)$ symmetry, specifically the family of CFTs in \sutwoonebos\ parametrized by $k$ and the family in \sutwonffermion\ with $N_f=1$ parametrized by $N$. We have already checked in Section \SecAnomaly\ that the 't~Hooft anomaly for the manifest symmetry matches across the various dualities. In the case of quantum symmetries, some description does not have the full symmetry $G^{\rm IR}$ manifest in the UV, and therefore we can only couple the UV theory to a background for the subgroup $G^{\rm UV} \subset G^{\rm IR}$. Still, we can check that the CS counterterms for $G^{\rm UV}$, when forced to be embeddable in $G^{\rm IR}$, present the same obstruction as the 't~Hooft anomaly for $G^{\rm IR}$. This in general provides a check of the quantum enhancement.
In the examples with enhanced $SO(3)$ symmetry, the 't~Hooft anomaly vanishes in all descriptions.

Combining the dualities in \Specialbos\ and \Specialferm\ we obtain six dual descriptions for the CFT with $SO(3)$ global symmetry that appeared in the third line of \specialcases\ and in \figDualities:
\eqn\SpecialwebDual{\eqalign{
U(1)_2 \ {\rm with}\ 1\ \Phi \quad&\longleftrightarrow\quad\;
SU(2)_1 \ {\rm with}\ 1\ \Phi \;\;\quad\longleftrightarrow\quad
U(3)_{1,-2} \ {\rm with}\ 1\ \Phi \cr
{\big\updownarrow} \qquad\qquad&\qquad\qquad\qquad\quad {\big\updownarrow} \qquad\qquad\qquad\qquad\qquad\qquad {\big\updownarrow} \cr
U(1)_{-\frac32} \ {\rm with}\ 1\ \Psi \quad&\longleftrightarrow\quad
SU(2)_{-\frac12} \ {\rm with}\ 1\ \Psi \quad\longleftrightarrow\quad
U(3)_{-\frac12, \frac52} \ {\rm with}\ 1\ \Psi
}}
The first two columns are special cases of the discussion above (and had already been considered in \AharonyJVV). The two theories in the last column can be coupled to a $U(1)$ background for the maximal torus of $SO(3)$ with Lagrangians
\eqn\countertermssutwo{\eqalign{
U(3)_{1,-2} \ {\rm with}\ 1\ \Phi \quad&\rightarrow\quad \CL = |D_a \Phi|^2 - |\Phi|^4 + \frac1{4\pi} \Tr \big( ada - {\textstyle \frac{2i}3} a^3 \big) - \frac3{4\pi} (\Tr a)d(\Tr a) \cr
&\qquad\qquad\quad + \frac1{2\pi} (\Tr a)dB + \frac1{4\pi} \big( {\textstyle \frac{K_s-3}2} \big) BdB \cr
U(3)_{-\frac12, \frac52} \ {\rm with}\ 1\ \Psi \quad&\rightarrow\quad \CL = i \bar\Psi \slashchar{D}_a \Psi + \frac3{4\pi} (\Tr a)d(\Tr a) + \frac{(\Tr a) dB}{2\pi} + \frac1{4\pi} \big( {\textstyle \frac{K_s+3}2} \big) BdB
}}
where the parameter $K_s$ is identified with the level of the $SU(2)_{K_s}/\Z_2 \cong SO(3)_{K_s/2}$ background in the upper middle description in \SpecialwebDual. The needed CS counterterms have been computed in \HsinBLU. In all six cases, the CS counterterms are well-defined for \hbox{$K_s + 1 \in 2\Z$}, providing a check of the dualities.


\newsec{Example with Quantum $\mbf{O}$(4) Symmetry: QED with Two Fermions
\seclab\SecQED}

In this section we consider three-dimensional QED, \ie\ $U(1)_0$, with two fermions of unit charge. As first observed in \XuLXA, this model enjoys self-duality.  The analysis of \HsinBLU\ paid more attention to global aspects of the gauge and global symmetries and to the Chern-Simons counterterms.  Here we continue that analysis and discuss in detail the global symmetry and its anomalies.  In particular, we will show that the IR behavior of this model has a global $O(4)$ symmetry and time-reversal invariance ${\bf T}$, but these symmetries have 't~Hooft anomalies.  As in previous sections, various subgroups or multiple covers of this symmetry are anomaly free and can be preserved in a purely $(2+1)d$ model.  We also add bulk terms to restore the full global symmetry.

\subsec{QED$_3$ with two fermions}

We consider a pair of dual UV theories flowing to the same IR fixed point.  As in \HsinBLU, we start with a purely $(2+1)d$ setting and study%
\foot{${\rm CS_{grav}}$ is a gravitational Chern-Simons term defined as $\int_{M = \partial X} {\rm CS_{grav}} = \frac1{192\pi} \int_X \Tr R \wedge R$. In this section we also use that the partition function of $U(N)_1$ is reproduced by the classical Lagrangian $-2N{\rm CS_{grav}}$. See \refs{\SeibergRSG,\SeibergGMD} for details.}
\eqn\AGaugedxa{\eqalign{
\! i\bar\Psi_1 &\slashchar{D}_{a+X}\Psi^1 + i\bar\Psi_2 \slashchar{D}_{a-X} \Psi^2 + {1\over 4\pi}  a d a
+ {1\over 2\pi}  a dY - {1\over 4\pi}YdY + 2{\rm CS_{grav}} \cr
&\quad\longleftrightarrow\quad
i\bar\chi_1 \slashchar{D}_{\tilde a+Y}\chi^1 + i\bar\chi_2\slashchar{D}_{\tilde a-Y}\chi^2 + {1\over 4\pi} \tilde ad\tilde a
+ {1\over 2\pi} \tilde a dX - {1\over 4\pi}XdX + 2{\rm CS_{grav}}
}}
where $a, \tilde a$ are dynamical $U(1)$ gauge fields (more precisely they are spin$_c$ connections \SeibergRSG) while $\Psi_{1,2}$, $\chi_{1,2}$ are complex fermions.

We would like to identify the global symmetry of the model.  The UV theory in the left side of \AGaugedxa\ has a global $SU(2)^X\times O(2)^Y$ symmetry.  The explicit background field $X$ couples to $U(1)^X\subset SU(2)^X$ and $Y$ couples to $U(1)^Y \subset O(2)^Y$.  The $\Z_2\subset O(2)^Y$ transformation $\bC^Y$ acts as charge conjugation: $\bC^Y (Y)=-Y$, $\bC^Y (a)= -a$, $\bC^Y (X)= X$, $\bC^Y (\Psi_i)= \epsilon_{ij} \bar \Psi^j$, hence it commutes with $SU(2)^X$.  Similarly, the UV theory in the right side has a $SU(2)^Y\times O(2)^X$ symmetry which includes a $\bC^X$ transformation.
We will soon see that they do not act faithfully.

Before we identify the global symmetry of the IR theory, we should find the precise global symmetry of the UV theories \AGaugedxa.  First we study how local operators transform under $SU(2)^X\times U(1)^Y$ in the left side of the duality \AGaugedxa.
A gauge-invariant polynomial made out of $\Psi^i$, $\bar\Psi_i$ and derivatives has even $U(1)^X$ charge corresponding to $SU(2)^X$ isospin $j^X \in \Z$, and it is neutral under $U(1)^Y$.  A monopole of $a$ has $U(1)^Y$ charge $Q^Y=1$ and $U(1)_a$ charge $1$. In order to make it gauge invariant, we must multiply it by a fermion, thus making the operator have $j^X= \half $.%
\foot{It is easy to see that the basic monopole operators can have spin zero.  More generally, our theory satisfies the spin/charge relation with $a$ the only spin$_c$ connection.  Therefore, all gauge-invariant local operators must have integer spin.}
More generally, it is easy to see that all gauge invariant operators have $2j^X+Q^Y\in 2\Z$.

As in the previous sections, this means that the dynamical $U(1)_a$ and the classical $SU(2)^X\times O(2)^Y$ should be taken to be $\big( U(1)_a \times SU(2)^X\times O(2)^Y \big)/\Z_2$ and the global symmetry that acts faithfully is $\big( SU(2)^X\times O(2)^Y \big)/\Z_2 $.

A similar argument can be used in the right hand side of \AGaugedxa\ showing that the global symmetry there is $\big( SU(2)^Y\times O(2)^X\big)/\Z_2 $.  The duality \AGaugedxa\ means that the IR theory should have the union of these two symmetries $SO(4) \cong \big( SU(2)^X\times SU(2)^Y \big)/\Z_2$.%
\foot{The additional conserved currents in the IR are provided by monopole operators of magnetic charge $\pm 2$ dressed by two fermion fields with flavor indices contracted, schematically $\frak M_{(-2)} \Psi\Psi$ and their conjugates. The fermions are contracted symmetrically with respect to Lorentz indices to give spin one, then by Fermi statistics flavor indices are antisymmetric giving a flavor singlet.}
Also, the duality means that the theory is invariant under a transformation $\Z_2^\CC$ that exchanges $X \leftrightarrow Y$, thus the global symmetry is really $O(4)\cong SO(4) \rtimes \Z_2^\CC$.  The Lagrangians in \AGaugedxa\ use only $U(1)^X\times U(1)^Y$ gauge fields and in terms of these the global symmetry is $\big( O(2)^X\times O(2)^Y \big)/\Z_2 \rtimes \Z_2^\CC$.

In addition, in the absence of background fields ({\it i.e.}\ as long as we consider correlators at separate points) the theory is clearly time-reversal invariant:%
\foot{When applying time reversal, one should be careful about the $\eta$-invariant from the regularization of the fermion path-integral \SeibergGMD.}
\eqn\tinvarff{\eqalign{
\bT&\bigg[ i\bar\Psi_1 \slashchar{D}_{a+X}\Psi^1 + i\bar\Psi_2 \slashchar{D}_{a-X} \Psi^2 + {1\over 4\pi}  a d a
+ {1\over 2\pi}  a dY - {1\over 4\pi}YdY + 2{\rm CS_{grav}} \bigg] \cr
&= i\bar\Psi_1 \slashchar{D}_{a+X}\Psi^1 + i\bar\Psi_2 \slashchar{D}_{a-X} \Psi^2
+{1\over 4\pi}  a d a + {1\over 2\pi}  a dY+ {2\over 4\pi} XdX+ {1\over 4\pi}YdY + 2{\rm CS_{grav}} \;,
}}
where $\bT(a)=a$, $\bT(X)=X$, $\bT(Y)=-Y$.  Of course, we can combine this transformation with $\bC^Y$ and/or with an element of $SU(2)^X$.
With a background, the theory is time-reversal invariant up to the anomalous shift $\frac{2}{4\pi} (XdX + YdY)$.  This anomaly should not be surprising.  The $U(1)^X$ symmetry is embedded into $SU(2)^X$ and in terms of that, the functional integral over $\Psi$ leads to an $\eta$-invariant (that can be described imprecisely as $SU(2)^X_{-\half}$) which has a time-reversal anomaly.  Note that in the other side of the duality this transformation must act as $\bT(\tilde a) = -\tilde a$.

Next, we would like to examine whether the $\Z_2$ quotient of $U(1)^X\times U(1)^Y$ is anomalous or not.  Since we should take the quotient $\big( U(1)_a\times U(1)^X \times U(1)^Y \big)/\Z_2$ (and similarly with $U(1)_{\tilde a}$), this means that the fluxes of $a,\tilde a,X,Y$ are no longer properly quantized, but $a\pm X$, $\tilde a \pm Y$ are properly quantized spin$_c$ connections and $X\pm Y$ are properly quantized $U(1)$ gauge fields.  A simple way to implement it is to change variables $a\to a-X$, $\tilde a\to \tilde a -Y$ in \AGaugedxa\ such that $a,\tilde a$ become ordinary spin$_c$ connections:
\eqn\AGaugedxc{\eqalign{
- \frac{2 YdY}{4\pi}+ i\bar\Psi_1 \slashchar{D}_{a}\Psi^1 +& i\bar\Psi_2 \slashchar{D}_{a-2X} \Psi^2 + \frac{ada}{4\pi} + \frac{ad(Y-X)}{2\pi}
+ \frac{(X-Y)d(X-Y)}{4\pi}  + 2{\rm CS_{grav}} \cr
&\quad  {\big\updownarrow} \cr
- \frac{2XdX}{4\pi}+ i\bar\chi_1\slashchar{D}_{\tilde a}\chi^1 +& i\bar\chi_2\slashchar{D}_{\tilde a-2Y}\chi^2 + \frac{\tilde ad\tilde a}{4\pi} + \frac{\tilde a d(X-Y)}{2\pi}
+ \frac{(X-Y)d(X-Y)}{4\pi}  + 2{\rm CS_{grav}} \;.
}}
Except for the first term in each side, namely $- \frac{2}{4\pi} YdY$ and $- \frac{2}{4\pi} XdX$, all the terms are properly normalized Chern-Simons terms under the quotient gauge group.

The existence of these terms means that the two dual UV theories \AGaugedxa\ have an 't~Hooft anomaly preventing us from taking the $\Z_2$ quotient.

We can change this conclusion by adding appropriate counterterms, {\it e.g.}\ $\frac2{4\pi} XdX$, to the two sides of the duality \AGaugedxa\ or equivalently \AGaugedxc.  Denoting the Lagrangians in these equations by $\CL_0(X,Y) \longleftrightarrow \CL_0(Y,X)$, we set
\eqn\shiftduality{
\CL_1(X,Y) = \CL_0(X,Y) +{2\over 4\pi }XdX \;.
}
This removes the first term in the right side of \AGaugedxc\ and makes also the left side consistent with the quotient.  Then, we can place the UV theory in $\big( U(1)^X\times U(1)^Y\big)/\Z_2$ backgrounds.  In the left side of the duality this term represents adding $SU(2)^X_1$ while in the right side this interpretation is meaningful only in the IR theory. After this shift, the IR theory can be placed in nontrivial $SO(4)$ backgrounds.  However, now the IR duality symmetry, which exchanges $X\leftrightarrow Y$, is anomalous:
\eqn\anomalousZt{
\CL_1(X,Y) \qquad\longleftrightarrow\qquad \CL_1(Y,X) +{2\over 4\pi }(XdX-YdY) \;,
}
{\it i.e.}\ under the $\Z_2^\CC$ transformation the IR theory is shifted by $SU(2)^X_{-1} \times SU(2)^Y_1$.

To summarize, the global symmetry that acts faithfully is $O(4)$, but we cannot couple the system to background $O(4)$ gauge fields.  Starting with \AGaugedxa\ we can couple it to $Pin^\pm(4)$ background fields,%
\foot{Since in the IR there are no operators transforming in spinor representations of $Spin(4)$, we can extend $O(4)$ to both $Pin^\pm(4)$.}
or starting with \shiftduality\ we can couple it to $SO(4)$ background fields.


\subsec{Mass deformations}

We can check the duality \AGaugedxa\ by deforming both sides with fermion bilinear operators in either the singlet or vector representation of the $SU(2)$ flavor symmetry factors.

The deformation by the $SO(4)$-singlet mass term $m\bar\Psi_i\Psi^i$ was discussed in \HsinBLU. The theory flows to the Lagrangians
\eqn\massdts{\eqalign{
- {2\over 4\pi}YdY \qquad {\it i.e.}\quad U(1)^Y_{-2}&\quad\subset\quad  SU(2)^Y_{-1}\qquad {\rm for}\ m>0 \;, \cr
- {2\over 4\pi}XdX \qquad {\it i.e.}\quad U(1)^X_{-2}&\quad\subset\quad  SU(2)^X_{-1}\qquad {\rm for}\ m<0 \;.
}}
This makes it clear that $m\bar\Psi_i\Psi^i$ is odd under $\Z_2\subset O(4)$.
The result of the deformation is consistent with the magnetic symmetry $U(1)^Y$, $U(1)^X$ on two sides of the duality being enhanced to $SU(2)^Y$, $SU(2)^X$ respectively.

The $SU(2)^X$ triplet mass term%
\foot{We thank D.~Gaiotto for a useful discussion about this deformation.}
$m(\bar \Psi_1\Psi^1-\bar\Psi_2\Psi^2)$ is in the $({\bf3},{\bf3})$ representation of $SU(2)^X\times SU(2)^Y$.
In fact, the duality \AGaugedxa\ can be derived by combining two fermion/fermion dualities involving a single fermion ({\it e.g.} see Section 6.3 of \HsinBLU), and from there one finds that the $SU(2)^X$ triplet mass term maps to the $SU(2)^Y$ triplet mass term $m(\bar \chi_1\chi^1-\bar\chi_2\chi^2)$. The representation $({\bf3},{\bf3})$ is completed by monopole operators $\frak M_{(-2)}\Psi\Psi$ and their conjugates, where now Lorentz indices are antisymmetric while flavor indices are symmetric. The triplet mass term explicitly breaks the symmetry to $U(1)^X\times U(1)^Y$.
Deforming the CFT \AGaugedxa\ by this mass term leads to the low energy Lagrangians
\eqn\lowenert{\eqalign{
 &{1\over 2\pi}a\,d(Y+X) -{1\over4\pi} (XdX +YdY) \qquad\qquad {\rm for}\ m>0 \;, \cr
 &{1\over 2\pi}a\,d(Y-X) -{1\over4\pi} (XdX +YdY) \qquad\qquad {\rm for}\ m<0  \;.
 }}
We see that the theory is not gapped: the photon $a$ is massless and its dual is the Goldstone boson of a spontaneously broken global symmetry.  From \lowenert\ we see that the unbroken symmetry is a diagonal mixture of $U(1)^X\times U(1)^Y$. Under both deformations \massdts\ and \lowenert\ we find consistency of the duality.

We could entertain the possibility that the symmetry of the CFT be $SO(5) \supset O(4)$. That would imply that at the fixed point the $O(4)$ invariant operator $\bar\Psi_i\Psi^i \bar\Psi_j \Psi^j$ sit in the same representation {\bf 14} of $SO(5)$ as $\bar\Psi_i (\sigma_3)^i_j \Psi^j$, and share the same dimension. As we just discussed, we can assume that the operator $\bar\Psi_i (\sigma_3)^i_j \Psi^j$, which is relevant in the UV, is relevant in the IR as well: this leads to a coherent picture. This would imply that also the 4-Fermi interaction is relevant in the IR, and since it is irrelevant in the UV, it would be a dangerously-irrelevant operator. Then, in order to reach the putative CFT with $SO(5)$ symmetry, one would need to tune the irrelevant operator $\bar\Psi_i\Psi^i \bar\Psi_j \Psi^j$ in the UV. The theory we have been discussing in this section---QED with two fermions---does not have such a tuning, and therefore it would not reach the $SO(5)$ fixed point even if the latter existed.

\subsec{Coupling to a $(3+1)d$ bulk}

We have seen that the IR behavior of the UV theories \AGaugedxa\ has an $O(4)$ global symmetry and time-reversal ${\bf T}$.  But these symmetries suffer from an 't~Hooft anomaly.  We cannot couple them to background gauge fields for these symmetries.  We saw that depending on the choice of counterterms we can have either $Pin^\pm(4)$ or $SO(4)$ background fields, but we cannot have $O(4)$ background fields and in either case we do not have time-reversal symmetry.

However, we can couple our $(2+1)d$ system to a $(3+1)d$ bulk and try to add background gauge fields in the bulk such that the full global symmetry is realized.

Let us start with the $O(4)\cong \big( SU(2)^X\times SU(2)^Y \big)/\Z_2 \rtimes \Z_2^\CC$ symmetry.  The bulk couplings of these gauge fields are characterized by two $\theta$-parameters, $\theta_X$ and $\theta_Y$.  Because of the $\Z_2$ quotient, they are subject to the periodicity
\eqn\thetaspinfour{
(\theta_X,\theta_Y)\sim (\theta_X+2\pi,\theta_Y+2\pi)\sim  (\theta_X+4\pi,\theta_Y)
}
and the semidirect product restricts to $(\theta_X,\theta_Y) \sim (\theta_Y,\theta_X)$.

Consider a bulk term $S_{\rm B}$ with $(\theta_X=-2\pi, \theta_Y=0)$.  For a closed four-manifold with $X$ and $Y$ being $Pin^\pm(4)$ gauge fields, this bulk term is trivial.  When $X$ and $Y$ are $O(4)$ gauge fields the partition function $e^{iS_{\rm B}}$ is $\pm 1$ and depends only on $w_2$ of the gauge fields.  (More precisely, the sign is determined by the Pontryagin square $\cP(w_2)/2$.)  This means that even for $O(4)$ gauge fields the partition function is independent of most of the details of $X$ and $Y$ in the bulk.

The $\Z_2^\CC$ transformation, which exchanges $X$ and $Y$, shifts the bulk term $S_{\rm B}$ by the term $(\theta_X=2\pi, \theta_Y=-2\pi)$.  On a closed four-manifold this shift has no effect on the answers.  But in the presence of a boundary it shifts the boundary Lagrangian  by the Chern-Simons terms of $\big( SU(2)_1\times SU(2)_{-1} \big)/\Z_2$.  In other words, in the presence of a boundary the bulk term $S_{\rm B}$ has an anomaly under $\Z_2^\CC$.

Starting with the boundary theory \AGaugedxa\ we add the boundary term in \shiftduality\ and the bulk term $S_{\rm B}$.  Naively, this did not change anything.  The bulk term might be thought of as an $SU(2)^X_{-1}$ boundary Chern-Simons term and therefore it seems like it removes the term added in \shiftduality.  However, because of the quotients this conclusion is too fast.  Instead, the bulk term is meaningful for $SO(4)$ fields and has an anomaly under $\Z_2^\CC$.  The boundary term we added in \shiftduality\ made the boundary theory meaningful for $SO(4)$ fields and created an anomaly under $\Z_2^\CC$.  Together, we have a theory with a bulk and a boundary with the full $O(4)$ symmetry.

Now that we have achieved an $O(4)$ symmetry we can try to add additional terms to restore time-reversal symmetry.  We would like to add a bulk $O(4)$ term that even with a boundary does not have a $\Z_2^\CC$ anomaly, but such that it compensates the anomaly in time reversal \tinvarff. Clearly, we need to add a bulk term $S_{\rm B}'$ with $(\theta_X=\pi,\theta_Y=\pi)$.  Without a boundary this term is ${\bf T}$ and $\Z_2^\CC$ invariant.  With a boundary it does not have an anomaly under $\Z_2^\CC$ but it has a ${\bf T}$ anomaly which exactly cancels that of the boundary theory \tinvarff.  Note that the time-reversal anomaly \tinvarff\ was not modified by adding the boundary term in \shiftduality\ and the bulk term $S_{\rm B}$ with $(\theta_X=-2\pi,\theta_Y=0)$.  These two terms almost completely cancel each other.  To summarize, the theory with the added boundary term in \shiftduality\ and a bulk term $S_{\rm B} + S_{\rm B}'$ with $(\theta_X=-\pi,\theta_Y=\pi)$ has the full symmetry of the problem.

We should make a final important comment.  As we said above, the bulk term $S_{\rm B}$ with $(\theta_X=-2\pi,\theta_Y=0)$ leads to dependence only on some topological information of the bulk fields.  Instead, the bulk term $S_{\rm B}'$ with $(\theta_X=\pi,\theta_Y=\pi)$ depends on more details of the bulk fields.


\newsec{Example with Global $\mbf{SO}$(5) Symmetry}

In this section we would like to study in some detail the theory
\eqn\usptwok{
USp(2)_k \cong SU(2)_k {\rm \ with\ two\ scalars} \;,
}
and the relation with its $SU/U$ dual $U(k)_{-1}$ with two fermions.
Since $N_f=2$, there are various quartic terms we can include in the potential, and depending on the choice we reach different IR fixed points. We will use mass deformations to check the duality, and exploit the `t~Hooft anomaly matching for general $SU/U$ dualities discussed in Section \SecAnomaly.

\subsec{A family of CFTs with $SO(5)$ global symmetry}

Let us first consider $USp(2)_k$ with two $\Phi$. As a $USp$ theory, it has maximal global symmetry $SO(5) \cong USp(4)/\Z_2$. We can classify relevant deformations accordingly. We describe the scalars through complex fields $\varphi_{ai}$ with $a=1,2$, $i=1,\dots,4$ subject to the reality condition $\varphi_{ai}^{\phantom{*}} \epsilon^{ab} \Omega^{ij} = \varphi_{bj}^*$ (where $\Omega$ is the $USp(4)$ symplectic invariant tensor). The quadratic gauge invariants are collected into the antisymmetric matrix $M_{ij} = \varphi_{ai} \varphi_{bj} \epsilon^{ab}$, which decomposes under $SO(5)$ as
\eqn\decomposquadric{
\CO_{\bf 1} = - \Tr \Omega M \;,\qquad\qquad \CO_{\bf 5} = M - \frac14 \Omega \, \CO_{\bf 1} \;.
}
Here the subscript is the $SO(5)$ representation and we suppress the indices.
Given the decomposition $({\bf1} + {\bf5}) \otimes_S ({\bf1} + {\bf5}) = {\bf1} \oplus {\bf5} \oplus {\bf1} \oplus {\bf14}$, in principle there are four quartic gauge invariants: $\cO_{\bf1}^2$, $\cO_{\bf1} \cO_{\bf5}$, $\Tr \Omega \cO_{\bf5} \Omega \cO_{\bf5} \equiv \cO_{\bf5}^2$, and $\cO_{\bf14}$ constructed as
\eqn\Ofourteen{
(\cO_{\bf14})_{ijkl} = (\cO_{\bf5})_{ij} (\cO_{\bf5})_{kl} + \frac1{20} \big( \Omega_{ij} \Omega_{kl} - 2 \Omega_{ik} \Omega_{jl} + 2 \Omega_{il} \Omega_{jk} \big) \cO_{\bf5}^2 \;.
}
However, since the gauge group has rank one, it turns out that $\cO_{\bf1}^2 = 4 \cO_{\bf5}^2$ and so there is only one quartic singlet.

In $USp(2)_k$ with two $\Phi$ we insist on $SO(5)$ global symmetry: we turn on $\cO_{\bf1}$ and $\cO_{\bf1}^2$ with a fine-tuning on $\cO_{\bf1}$ and we assume that it flows to a nontrivial fixed point $\CT_0^{(k)}$. Such a fixed point has $SO(5)$ global symmetry. We can couple the theory to $SO(5)$ background gauge fields as
\eqn\sofivebackground{
\frac{USp(2)_k \times USp(4)_L}{\Z_2} \ \ {\rm with}\ \Phi {\rm\ in}\ ({\bf 2}, {\bf 4}) \;,
}
with some CS counterterm with level $L$. The conditions to have a well-defined $(2+1)d$ action are $L\in \Z$ and $k - 2L \in 2\Z$. As we discussed in Section 2.3, for $k$ even the equations can be solved, but for $k$ odd they cannot and there is an `t~Hooft anomaly. The anomaly is captured by the bulk term
\eqn\sofiveanomaly{
S_{\rm anom} = k \, \pi \int_{\CM_4} \frac{\CP(w_2)}2 \;.
}
For a closed manifold $\CM_4$ this is trivial for $k$ even, in the sense that $e^{iS_{\rm anom}}=1$, but it is $\pm 1$ for $k$ odd.  For $\CM_4$ with a boundary this term is anomalous for odd $k$ and corrects the anomaly in the boundary theory.

\ifig\figRGflow{RG diagram and fixed points of $SU(2)_k$ with two scalars. The blue and brown lines are RG flows that preserve the global symmetry of the UV fixed point they emanate from.  The green lines are deformations that break the $SO(5)$ global symmetry.  The solid green and brown lines separate between phases with different IR behavior ($\S^1$, $\S^2$, and gapped with a TQFT), with an $\S^4$ theory (with a Wess-Zumino term with coefficient $k$) along the brown solid line.  There is no phase transition along the dotted brown line, as the IR physics is gapped there.}
{\epsfxsize4.5in\epsfbox{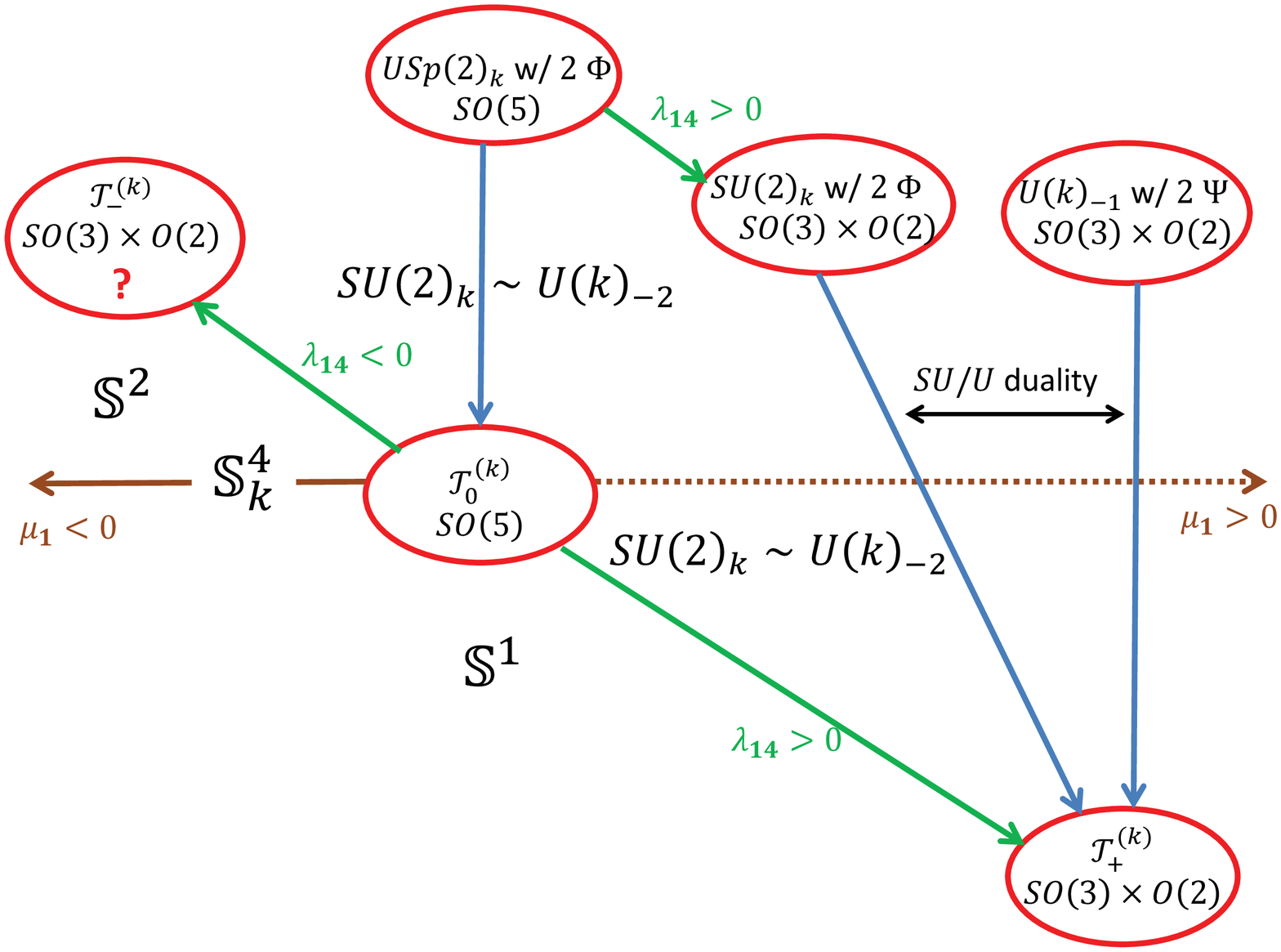} }

We can consider relevant deformations of $\CT_0^{(k)}$ that preserve $SO(5)$, see \figRGflow, by turning on $\mu_{\bf1}\cO_{\bf1}$, which is an equal mass for all scalars. With positive mass-squared $\mu_{\bf1}>0$ we flow to the TQFT $SU(2)_k$. With negative mass-squared $\mu_{\bf1}<0$  some of the scalars condense Higgsing $SU(2)_{\rm dyn}$ completely. To figure out the breaking pattern, notice that we start with 8 real fields, one is an overall scale, 3 are eaten by the Higgs mechanism, so at most there can be 4 massless fields. However the largest subgroup of $SO(5)$ is $SO(4)$, so there must be at least 4 Goldstone bosons. Thus the only possible breaking pattern is $SO(5) \to SO(4)$, leaving an $\S^4 \cong SO(5)/SO(4)$ sigma model in the IR. More directly, the potential $V = \mu_{\bf1} \cO_{\bf1} + \frac12 \cO_{\bf1}^2$ has minima at $\cO_{\bf1} = -\mu_{\bf1}$.
By the algebraic relation ${\cal O}_{\bf 5}^2={1\over 4} {\cal O}_{\bf 1}^2$ we find that also ${\cal O}_{\bf 5}$ condenses, and it spontaneously breaks the symmetry to $SO(4)$, producing the Goldstone modes living on $\S^4$.
Alternatively, the equation ${\cal O}_{\bf 1} = -\mu_{\bf1}$ describes an $\S^7$, which is an $SU(2)$ Hopf fibration over $\S^4$, therefore gauging $SU(2)$ leaves the $\S^4$ NLSM.

As was shown in \RabinoviciMJ\ (we review it in Appendix A), the $\S^4$ NLSM has a Wess-Zumino interaction term $k S_{\rm WZ}$ that originates from the level $k$ Chern-Simons term in the UV. The Wess-Zumino term can be written as
\eqn\WZterm{
k\, S_{\rm WZ} = 2\pi k \int_{\wt\CM_4} \sigma^*(\omega_4)
}
where the integral is over a four-manifold with boundary, $\sigma$ are the NLSM fields, and $\omega_4$ is the volume form of $\S^4$ normalized to total volume $1$. We can couple the theory to $SO(5)$ background gauge fields (gauging in general dimension was discussed in \HullMS). From our derivation in Appendix A it is clear that for odd $k$, the WZ action depends on how the $SO(5)$ background fields are extended to the bulk, and the dependence is captured by the very same term \sofiveanomaly.%
\foot{We thank Todadri Senthil for pointing out to us the relevance of $\CP(w_2)$ in this context and for mentioning \WangTXT.} This is 't~Hooft anomaly matching along the RG flow.

Let us stress that the far IR limit of the $\S^4$ NLSM is given by 4 free real scalar fields. At higher energies there are irrelevant interactions that turn it into the $\S^4$ NLSM with WZ term. As we show in Appendix A, at the same scales there are also other irrelevant (higher-derivative) interactions that break time reversal $\bf T$ (for $k>0$). Therefore such an $\S^4$ NLSM has only $SO(5)$ global symmetry, as the UV theory.

\subsec{Two families of CFTs with $SO(3) \times O(2)$ global symmetry}

Let us now consider $SU(2)_k$ with two $\Phi$ ({\it i.e.}\ the same gauge group and matter content as before), but imposing only $SU(2) \times U(1)$ symmetry on the quartic terms, as it is the case in the $SU/U$ dualities. Then there is another quartic deformation we can add in the UV:
\eqn\extraquartic{
\wt\cO_{({\bf14})} = (\cO_{\bf14})_{ijkl} \, \eta^{jk} \eta^{il} \;,
}
where $\eta$ is a $U(2)$-invariant tensor. This operator is contained in $\cO_{\bf14}$ from the decomposition ${\bf14} \to {\bf1}_0 \oplus {\bf5}_0 \oplus {\bf3}_1 \oplus {\bf3}_{-1} \oplus {\bf1}_2 \oplus {\bf1}_{-2}$ under $USp(4)\to SU(2) \times U(1)$. It turns out that the preserved symmetry acting faithfully is $SO(3) \times O(2)$.%
\foot{The reduced symmetry is $\big( O(3) \times O(2) \big)/\Z_2$ embedded into $SO(5)$, which includes charge conjugation $\Z_2^{\bf C}$. Equivalently, it is $\big( U(2)/\Z_2 \big) \rtimes \Z_2^{\bf C}$ embedded into $USp(4)/\Z_2$ as follows: a $2\times 2$ unitary matrix $T$ is mapped to $\left( {T \atop 0} {0 \atop T^*} \right)$, the quotient is by $-\unit$, and charge conjugation is mapped to $\Omega = \left( {0 \atop \unit} {-\unit \atop 0} \right) \in USp(4)$. The $U(2)$-invariant tensor is then $\eta = \left( {0 \atop \unit} {\unit \atop 0} \right)$, odd under $\Z_2^{\bf C}$.}
If we want to avoid the appearance of directions in field space where the potential is unbounded from below, the absolute value of the coefficient $\lambda_{\bf14}$ of $\wt\cO_{({\bf14})}$ should not be too large compared to that of $\cO_{\bf1}^2$ (which is positive). Since at $\lambda_{\bf14} =0$ there is a phase transition with enhanced $SO(5)$ symmetry, we expect two different RG flows for $\lambda_{\bf14} \gtrless 0$ that can lead to two families $\CT_\pm^{(k)}$ of fixed points with $SO(3)\times O(2)$ global symmetry.

According to \anomaly, the theory $SU(2)_k$ with two $\Phi$ for odd $k$ has an anomaly when coupled to $U(2)/\Z_2 \cong SO(3)\times SO(2)$ backgrounds. This anomaly directly follows from \sofiveanomaly, if we restrict $SO(5)$ backgrounds to $SO(3)\times SO(2)$.

We can learn about the properties of the fixed points $\CT_\pm^{(k)}$, if they exist, by looking at the RG diagram in \figRGflow. In particular, we can predict what $\CT_\pm^{(k)}$ flow to, if we deform them by $\cO_{\bf1}$, which is the only other relevant deformation invariant under the symmetries (notice that $\cO_{\bf5}$ does not contain $SO(3)\times O(2)$ singlets). This should be the same as first deforming $\CT_0^{(k)}$ by $\cO_{\bf1}$, as in Section 5.1, and then by $\wt\cO_{({\bf14})}$. The TQFT $SU(2)_k$ is not affected by $\wt\cO_{({\bf14})}$, because $SO(5)$ only acts on massive particles and thus $\wt\cO_{({\bf14})}$ is decoupled. In the $\S^4$ NLSM we use coordinates $\rho_{1,\dots,5}$ with $\sum_{I=1}^5 \rho_I^2 = 1$.  Then we have $\wt\cO_{({\bf14})} = - 3 (\rho_1^2 + \rho_2^2) + 2( \rho_3^2 + \rho_4^2 + \rho_5^2$). If we deform the potential by $\wt\cO_{({\bf14})}$ with positive coefficient, we flow to an $\S^1$ NLSM, while a negative coefficient leads to an $\S^2$ NLSM. We conclude that, for $\mu_{\bf1}<0$, the $\mu_{\bf 1}\cO_{\bf1}$ deformation of $\CT_+^{(k)}$ gives an $\S^1$ NLSM, while the $\mu_{\bf 1}\cO_{\bf1}$ deformation of $\CT_-^{(k)}$ gives an $\S^2$ NLSM. Notice that when the NLSM maps are restricted to an equatorial $\S^1$ or $\S^2$, the WZ term \WZterm\ vanishes.

We can provide two different descriptions of $\CT_+^{(k)}$ through the $SU/U$ duality
\eqn\dualityone{
SU(2)_k \ {\rm with}\ 2\ \Phi \qquad\longleftrightarrow\qquad U(k)_{-1} \ {\rm with}\ 2\ \Psi \;,
}
where the theory on the left has $\cO_{\bf1}^2$ and $\wt\cO_{({\bf14})}$ quartic couplings both with positive coefficient.\foot{As in \HsinBLU, $U(1)\subset U(k)$ is a spin$_c$ connection and we must add a transparent line to the theories in order for the duality to be valid.  This transparent line does not affect the critical behavior.} This gives evidence that the fixed points $\CT_+^{(k)}$ exist for all $k>0$. One can check that the $SO(3) \times O(2)$ invariant mass deformation of $U(k)_{-1}$ with two $\Psi$ gives $U(k)_{-2}$ (dual to $SU(2)_k$) for negative fermion mass, and $U(k)_0$ (whose low energy limit is the $\S^1$ NLSM) for positive mass. Moreover, as discussed in Section \SecAnomaly, the fermionic theory correctly reproduces the 't~Hooft anomaly.

What about $\CT_-^{(k)}$? For $k=1$ a natural candidate for a dual description is
\eqn\candidate{
\CT_-^{(1)} \;: \qquad U(1)_2 \ {\rm with}\ 2 \ \Phi \;.
}
This theory has $\big( U(2)/\Z_2 \big) \rtimes \Z_2^{\bf C} \cong SO(3) \times O(2)$ global symmetry acting faithfully, and it reproduces the anomaly of $SU(2)_1$ with 2 $\Phi$. Besides, the theory has  unique $SO(3) \times O(2)$ invariant quadratic and quartic terms. Upon invariant positive mass-squared deformation it flows to $U(1)_2$, which is dual to $SU(2)_1$. For negative mass-squared, the minima of the potential form an $\S^3$ with the $U(1)$ Hopf fiber gauged, and thus the theory flows to an $\S^2$ NLSM. Since the CS level in \candidate\ is even, there is no topological Hopf term \WilczekCY\ in the $\S^2$ NLSM, reproducing the result from the deformation of $\S^4$. This gives evidence that the fixed point $\CT_-^{(1)}$ could exist.

Interestingly, the description \candidate\ of $\CT_-^{(1)}$ and the fermionic description in \dualityone\ of $\CT_+^{(1)}$ almost fit in the dualities \SUUduality-\SOSPduality\ but fail to be dual because their parameters are outside the allowed region. For instance the two theories
\eqn\failedduality{
SO(2)_2 \ {\rm with}\ 2 \ \phi \qquad {\buildrel {\rm NOT} \over \longleftrightarrow} \qquad SO(2)_{-1} \ {\rm with}\ 2 \ \psi
}
fail to give a dual pair (as advocated in \AharonyJVV\ by a different argument). In this example $\CT_\pm^{(1)}$ appear in the same RG diagram, but are indeed distinct.

\subsec{A family of RG flows with $O(4)$ global symmetry}

We can consider a different deformation of the $SO(5)$ invariant theories $\CT_0^{(k)}$, obtained by using a quartic operator in $\cO_{\bf14}$ that preserves an $O(4)$ . We will call this operator $\wh\cO_{({\bf14})}$. It can be written in terms of a $Spin(4)$ invariant tensor $\eta'$ as%
\foot{To define $\eta'$ it is convenient to use a different basis than before, namely $\Omega = \left( {\omega \atop 0} {0 \atop \omega} \right)$ with $\omega = \left( {0 \atop 1} {-1 \atop 0} \right)$. We embed $Spin(4) \rtimes \Z_2^{\bf C}$ into $USp(4)$ as $\left( {T_1 \atop 0} {0 \atop T_2} \right)$ where $T_{1,2} \in SU(2)$, while the $\Z_2^{\bf C}$ charge conjugation is ${\bf C} = \left( {0 \atop \unit} {\unit \atop 0} \right)$. Quotient by $-\unit$ gives an embedding of $O(4)$ into $USp(4)/\Z_2$. Then $\eta' = \left( {\omega \atop 0} {0 \atop -\omega} \right)$ is invariant under a $Spin(4)$ and is odd under $\Z_2^{\bf C}$.}
\eqn\newextraquartic{
\wh\cO_{({\bf14})} = (\cO_{\bf14})_{ijkl} \, \eta^{\prime jk} \eta^{\prime il} \;.
}
Deforming $\CT_0^{(k)}$ by $\wh\cO_{({\bf14})}$ breaks $SO(5) \to O(4)$.
Thus, we study the theory $USp(2)_k$ with 2 $\Phi$ and quartic deformations $\cO_{\bf1}^2$ and $\wh\cO_{({\bf14})}$. More easily, this is $SU(2)_k$ with two scalars and a potential $V = \alpha \big( |\Phi_1|^4 + |\Phi_2|^4 \big) + 2\beta |\Phi_1|^2 |\Phi_2|^2$ with $\alpha\neq \beta$. In order to have a theory with a potential bounded from below, $\cO_{\bf1}^2$ should have positive coefficient while the coefficient $\wh\lambda_{\bf14}$ of $\wh\cO_{({\bf14})}$ should not be too large in absolute value. As before, we expect two different RG flows for $\wh\lambda_{\bf14} \gtrless 0$, separated by $\CT_0^{(k)}$ with enhanced $SO(5)$ symmetry.

The only other $O(4)$ invariant relevant deformation is $\cO_{\bf1}$, which is an equal mass for all scalars ($\CO_{\bf5}$ does not contain $O(4)$ invariants), and we can study the combined effect of $\cO_{\bf1}$ and $\wh\cO_{({\bf14})}$ on $\CT_0^{(k)}$---in a way similar to what we did in \figRGflow. With positive mass-squared we flow to the TQFT $SU(2)_k$, which is not affected by $\wh\cO_{({\bf14})}$ because the latter is decoupled. With negative mass-squared we flow to deformations of the $\S^4$ NLSM. In the NLSM coordinates, $\wh\cO_{({\bf14})} = - (\rho_1^2 + \rho_2^2 + \rho_3^2 + \rho_4^2) + 4 \rho_5^2$. Therefore, $\wh\lambda_{\bf14}>0$ leads to an $\S^3$ NLSM, while $\wh\lambda_{\bf14}<0$ leads to two gapped vacua with spontaneous breaking of $\Z_2^{\bf C}$. The WZ term $k S_{\rm WZ}$ in the $\S^4$ NLSM descends to a $\theta$-term $\pi k Q$ in the $\S^3$ NLSM, where $Q \in \Z$ is the wrapping number in $\pi_3(\S^3)=\Z$ (in other words $\theta = k\pi$).%
\foot{Restricting the NLSM maps $\sigma$ to an equatorial $\S^3$ in $\S^4$, the WZ term gives $0$ on a map that does not wrap $\S^3$, and $\pi$ on a map that wraps $\S^3$ once. By linearity, $S_{\rm WZ} = \pi Q(\sigma)$.}

In the presence of the deformation $\wh\cO_{({\bf14})}$, with either sign of its coupling $\wh\lambda_{\bf14}$, a tuning on $\cO_{\bf1}$ may or may not lead to a fixed point. At the moment we do not have candidate dual descriptions for those fixed points, and we leave the question open.

\topinsert
\centerline{\vbox{\halign{
\hfil#\hfil&\qquad\qquad#\hfil&\qquad\qquad#\hfil \cr
\noalign{\hrule\smallskip}
\omit\bf Global symmetry &  \kern-2em Quadratic $\Phi^2$ & \kern2em Quartic $\Phi^4$ \cr
\noalign{\smallskip\hrule\smallskip}
$SO(5)$ & $\cO_{\bf1}$ & $\cO_{\bf1}^2$ \cr
$SO(3)\times O(2)$ & $\cO_{\bf1}$ & $\cO_{\bf1}^2$, $\wt\cO_{({\bf14})}$ \cr
$O(4)$ & $\cO_{\bf1}$ & $\cO_{\bf1}^2$, $\wh\cO_{({\bf14})}$ \cr
$SO(4)$ & $\cO_{\bf1}$, $\cO_{({\bf5})}$ & $\cO_{\bf1}^2$, $\wh\cO_{({\bf14})}$, $\cO_{\bf1}\cO_{({\bf5})}$ \cr
$U(1)^2 \rtimes \Z_2^{\bf C}$  & $\cO_{\bf1}$, $\cO_{({\bf5})}$ & $\cO_{\bf1}^2$, $\wt\cO_{({\bf14})}$, $\wh\cO_{({\bf14})}$, $\cO_{\bf1}\cO_{({\bf5})}$ \cr
\noalign{\smallskip\hrule}
}}}
\noindent\centerline{{\bf Table 1:} Relevant deformations of $\CT_0^{(k)}$ depending on the preserved global symmetry.}
\smallskip
\endinsert

There are more general deformations of $\CT_0^{(k)}$ we can consider, depending on the amount of symmetry we want to preserve. A few examples, some of which we have already discussed, are in Table 1. By $\cO_{({\bf5})}$ we mean the specific component of $\cO_{\bf5}$ that is a singlet under the preserved symmetry group under consideration.

For instance, if we want to preserve only $SO(4) \subset SO(5)$, in terms of the two $SU(2)$ doublets $\Phi_1$ and $\Phi_2$ we can turn on the following relevant deformations: there are two mass terms $|\Phi_1|^2$ and $|\Phi_2|^2$, and three quartic terms $|\Phi_1|^4$, $|\Phi_2|^4$, $|\Phi_1|^2 |\Phi_2|^2$. This is a different basis than the one in Table 1. With many operators at our disposal, the precise breaking pattern depends on the ratios between the various terms.

\subsec{Relation with a Gross-Neveu-Yukawa-like theory}

We can compare the $USp(2)_k$ theory with two scalars with a different model, discussed in \SenthilJK, which also exhibits $SO(5)$ global symmetry and a phase described by the $\S^4$ NLSM with WZ term.

Consider a Gross-Neveu-Yukawa-like theory (GNY) with 5 real scalars, $4k$ complex fermions and schematic Lagrangian \SenthilJK
\eqn\LagGNY{
\CL = (\partial \phi)^2 + \bar\Psi \slashchar{\partial} \Psi - \phi^4 + \phi^a \bar\Psi \Gamma_a \Psi \;.
}
The scalars transform in the vector representation {\bf5} of $Spin(5)$, the fermions in $k$ copies of the spinor representation {\bf4}, and $\Gamma_a$ are gamma matrices of $Spin(5)$. The Lagrangian \LagGNY\ enjoys a $\big( USp(4) \times USp(2k) \big) /\Z_2$ global symmetry, and the quartic interaction is the only one preserving that symmetry. In addition, the theory also preserves a time-reversal $\Z_2^{\bf T}$ symmetry under which $\phi$ is odd.

With a tuning, the Lagrangian \LagGNY\ is expected to flow to a fixed point with the full global symmetry. The tuning is on the scalar mass deformation (while the fermion mass is odd under ${\bf T}$ and is thus set to zero by imposing that symmetry). We could also think of the fixed point as the IR limit of the $O(5)$ Wilson-Fisher fixed point with $4k$ complex decoupled fermions perturbed by the relevant operator $\phi^a \bar\Psi \Gamma_a \Psi$.

As discussed in \SenthilJK, if we deform \LagGNY\ by a negative scalar mass-squared, the scalars condense breaking spontaneously $SO(5) \to SO(4)$ and leading to an $\S^4$ NLSM. In addition, because of the Yukawa interaction the fermions become massive. Integrating them out produces a WZ interaction $k S_{\rm WZ}$ \AbanovQZ. Deformation by a positive mass-squared leads to $4k$ complex massless free fermions.%
\foot{At intermediate energies one finds a Gross-Neveu-like model of $4k$ complex massless fermions with quartic interactions, which however are irrelevant and disappear in the IR.}

The GNY fixed point \LagGNY\ and the fixed point $\CT_0^{(k)}$ of the $USp(2)_k$ theory with two scalars discussed above, despite sharing the $\S^4$ NLSM phase with a WZ term, are clearly different.  Even their global symmetries are different.  The GNY fixed point has a $\big( USp(4)\times USp(2k) \big)/\Z_2$ symmetry and $\bT$-reversal symmetry, while the fixed point $\CT_0^{(k)}$ has only an $SO(5)$ global symmetry.  In fact, even the $\S^4$ NLSM phases are slightly different. The one obtained from \LagGNY\ has the time-reversal symmetry ${\bf T}$, preserved by $S_{\rm WZ}$, while the one from $SU(2)_k$ with two scalars has higher-derivative corrections that break ${\bf T}$, because time-reversal symmetry is not present in the UV.  In addition, the phase obtained by positive mass-squared is different in these two models.


\bigskip
\noindent{\bf Acknowledgments}

We would like to thank Dan Freed, Anton Kapustin, Zohar Komargodski, Todadri Senthil, Stephen Shenker, Juven Wang, and Edward Witten for useful discussions, and especially Ofer Aharony for collaboration at the early stage of this work. FB was supported in part by the MIUR-SIR grant RBSI1471GJ ``Quantum Field Theories at Strong Coupling: Exact Computations and Applications'', by the INFN, and by the IBM Einstein Fellowship at the Institute for Advanced Study.
The work of PH was supported by Physics Department of Princeton University.
NS was supported in part by DOE grant DE-SC0009988.  NS thanks the Hanna Visiting Professor Program and the Stanford Institute for Theoretical Physics for support and hospitality during the completion of this work.


\appendix{A}{Derivation of the Wess-Zumino term in the 3D {$\S^{\bf4}$} NLSM}

Here we show that when $SU(2)_k$ with two scalars flows to the $\S^4$ NLSM by mass deformation, the Chern-Simons term induces a Wess-Zumino interaction at level $k$ at low energies \RabinoviciMJ.

Insisting on $SO(5)$ global symmetry and turning on a negative mass-squared, the minima of the potential lie along $\sum_{a,i} |\Phi_{ai}|^2 = \lambda$ ($a=1,2$, $i=1,2$) which is $\S^7$ (here $\lambda$ is a mass scale). The $SU(2)$ action corresponds to the Hopf fibration $SU(2) \to \S^7 \to \S^4$, thus what we gauge is the $SU(2)$ fiber. Recall that $SU(2)$ bundles over $\S^4$ are completely classified by $\pi_3\big( SU(2) \big) = \Z$ which is the second Chern class, and $\S^7$ has minimal class:
\eqn\HopfChernclass{
\frac1{8\pi^2} \int_{\S^4} \Tr G \wedge G = 1 \;,
}
where $G = dC - i C^2$ is the curvature of the $SU(2)$ bundle and $C(\Phi)$ is a function of $\Phi$.

The 3D gauge theory has a CS term
\eqn\CSUV{
S_{\rm CS} = \frac k{4\pi} \int_{\S^3} \Tr \Big( ada - \frac{2i}3 a^3 \Big) = \frac k{4\pi} \int_{\wt\CM_4} \Tr F \wedge F \;, \qquad\qquad \S^3 = \partial \wt\CM_4 \;,
}
where $\S^3$ is the topology of spacetime and the second definition is the proper one. At low energies the scalars are constrained to $\sum |\Phi|^2 = \lambda$ and we integrate out the gauge field. Starting with the schematic Lagrangian
\eqn\initialL{
\CL = \big| D_\mu \Phi \big|^2 + \frac{k}{4\pi} \CL_{\rm CS}(a) \;,
}
the equation of motion for $a$ is
\eqn\eomforA{
0 = a_\mu \Phi\Phi^\dagger + \Phi\Phi^\dagger a_\mu - i \Phi \partial_\mu \Phi^\dagger + i \partial_\mu\Phi \, \Phi^\dagger + \frac{k}{2\pi} \epsilon_{\mu\nu\rho} F^{\nu\rho} \;.
}
This equation contains $a_\mu$ as well as its first derivative, and it is non-linear. If we drop the last term, the equation is simply $J_\mu \equiv - i \Phi {\buildrel \leftrightarrow \over D}_\mu \Phi^\dagger = 0$ setting to zero the $SU(2)$ gauge current. This means that $a_\mu$ is identified with the connection $C(\Phi)$ of the $SU(2)$ bundle over $\S^4$. To take into account the last term, we notice that the first four terms in \eomforA\ are of order $\lambda$ (because $|\Phi|^2 \sim \lambda$) while the last term is of order $\lambda^0$ and it contains a derivative. We can then solve the equation as a series expansion in $\lambda^{-1}$, and since $\lambda^{-1}$ is dimensionful, the series is actually a derivative expansion. Thus in the IR limit we have
\eqn\IRexpansion{
a_\mu dx^\mu = C(\Phi) + \dots
}
where the dots are higher-derivative corrections.

Having identified the field strength $F$ in \CSUV\ over the extended spacetime manifold $\wt\CM_4$ with the curvature $G(\Phi)$ of the Hopf fibration (up to higher-derivative corrections), we obtain
\eqn\reductionCStoWZ{
S_{\rm CS} \,\to\, k\,S_{\rm WZ} + \dots = 2\pi k \int_{\wt\CM_4} \omega_4(\Phi) + \dots \;,
}
where $\omega_4$ is the volume form on $\S^4$ normalized to have integral $1$. Notice that \eomforA, because of the last term, is not invariant under time reversal ${\bf T}$. Therefore the higher-derivative corrections to \IRexpansion\ do not transform homogeneously under ${\bf T}$, and they break ${\bf T}$ in \reductionCStoWZ.

Finally, consider coupling the UV theory to $SO(5)$ background fields, namely consider the theory $\big( SU(2)_k \times USp(4)_L \big)/\Z_2$ with a bifundamental scalar. As discussed in Section 2.3, for odd $k$ the action has a sign dependence on the extension of the $SO(5)$ background fields to $\wt\CM_4$. By \reductionCStoWZ, this implies that also the WZ term coupled covariantly to $SO(5)$ background fields \HullMS\ has the same anomalous dependence.


\appendix{B}{Comments on Self-Dual QED with Two Fermions}

Building on the interesting fermion/fermion duality of  \refs{\SonXQA\PotterCDN-\WangGQJ}, the authors of \XuLXA\ proposed the self-duality of a $U(1)$ theory with two fermions.  This was later generalized in \ChengPDN\ to the self-duality of a $U(1)$ gauge theory with two fermions, one with charge $1$ and one with charge $k$ odd.  As emphasized in \refs{\SeibergRSG,\SeibergGMD,\HsinBLU}, the coefficients in the Lagrangians in \refs{\SonXQA\PotterCDN-\WangGQJ} are improperly quantized.  This was fixed in \SeibergGMD\ by adding more fields and more terms to the Lagrangian.  Then, a proper derivation of the self-duality of the theory with $k=1$ was given in \HsinBLU.  That perspective was also consistent with the spin/charge relation and described the proper coupling of background gauge fields.  Here we will present a similar derivation of the self-duality of the theory with generic odd $k$.  This will lead us to a more detailed analysis of the global symmetries and 't~Hooft anomalies of the problem.

We start with the fermion/fermion duality of \SeibergGMD:
\eqn\ffKo{
i\bar\Psi\slashchar{D}_{A}\Psi \quad\longleftrightarrow\quad
i\bar\chi\slashchar{D}_{a}\chi + {1\over 2\pi}adu - {2\over 4\pi}udu + {1\over 2\pi}udA-{1\over 4\pi} AdA -2\CSg \;.
}
Next, we follow the steps in \HsinBLU. We take a product of the theory in \ffKo\ and of its time-reversed version in which we substitute $A \to kA - 2X$ ($X$ is a background $U(1)$ gauge field):
\eqn\AUngaugedx{\eqalign{
i\bar\Psi_1\slashchar{D}_{A}\Psi^1 +i\bar\Psi_2\slashchar{D}_{kA-2X}\Psi^2 \quad\longleftrightarrow\quad
& i\bar\chi_1\slashchar{D}_{a_1}\chi^1 +i\bar\chi_2\slashchar{D}_{a_2}\chi^2 + {1\over 2\pi} \big( a_1du_1-a_2du_2 \big) \cr
& + {2\over 4\pi} \big( u_2du_2 - u_1 du_1 \big) + {1\over 4\pi} a_2da_2 \cr
& + {1\over 2\pi}u_1dA + {1\over 2\pi}u_2d \big( 2X-kA \big) - {1\over 4\pi} AdA \;.
}}
Note that for odd $k$ this is consistent with the spin/charge relation.
We add the following counterterms, $\frac{1}{2\pi}Ad \big( Y-kX \big) + {1\over 4\pi } \big( XdX - YdY \big) +N \big( {1\over 4\pi} AdA +2\CSg \big)$, to the two sides of the duality. Here $N = (k^2+1)/2$ and $Y$ is a background $U(1)$ field. The specific counterterms and the value of $N$ were picked such that we can integrate out most of the fields on the right hand side. Then we can promote $A$ to a dynamical field (more precisely, a spin$_c$ connection) $a$.  On the left hand side we find
\eqn\AUngaugedxL{
i\bar\Psi_1\slashchar{D}_{a}\Psi^1 +i\bar\Psi_2\slashchar{D}_{ka-2X}\Psi^2+{N\over 4\pi} ada+\frac{1}{2\pi}ad \big( Y-kX \big) +{1\over 4\pi } \big( XdX-YdY \big)+2N\CSg \;.
}
We will call this Lagrangian $\CL_0(X,Y)$.
On the right hand side there are several gauge fields, but we can integrate most of them out.  We redefine $a=a'+2 u_1$ and $ u_2=u_2'+k u_1 +{k+1\over 2} a'$, then the Lagrangian is linear in $u_1$ and it can be integrated out to set $a_1=ka_2-2Y$. Finally we can integrate out $a'$ to find
\eqn\AUngaugedxR{
i\bar\chi_1\slashchar{D}_{k \tilde a - 2Y}\chi^1 +i\bar\chi_2\slashchar{D}_{\tilde a}\chi^2
+{N\over 4\pi}\tilde ad\tilde a+{1\over 2\pi}\tilde a d \big(X- k Y \big)
+{1\over 4\pi} \big(  YdY-XdX \big) +2N\CSg \;,
}
where we relabeled $a_2=\tilde a$. Note that all terms are properly quantized with $\tilde a$ being a spin$_c$ connection. We see that \AUngaugedxL\ and \AUngaugedxR\ are related by relabeling the dynamical fields and by exchanging $X\leftrightarrow Y$.  This establishes the self-duality of the model, namely $\CL_0(X,Y) \longleftrightarrow \CL_0(Y,X)$.%
\foot{In order to compare with \ChengPDN, for every fermion coupled with $\slashchar{D}_\CA$ we should add the terms $-{1\over 8\pi} \CA d \CA - \CSg $.  This turns \AUngaugedxL\ and \AUngaugedxR\ into
\eqn\AUngaugedxLW{\eqalign{
&i\bar\Psi_1\slashchar{D}_{a}\Psi^1 +i\bar\Psi_2\slashchar{D}_{ka-2X}\Psi^2+{1\over 2\pi}  Yda -{1\over 4\pi} ( XdX+YdY) \qquad \longleftrightarrow \qquad \cr
&\qquad i\bar\chi_1\slashchar{D}_{k \tilde a  - 2Y}\chi^1 +i\bar\chi_2\slashchar{D}_{\tilde a}\chi^2 +{1\over 2\pi}Xd\tilde a  -{1\over 4\pi} ( XdX+YdY)
}}
where we removed the gravitational Chern-Simons term.
Up to the last counterterm (which we cannot remove because of the spin/charge relation) this agrees with the equations in \ChengPDN.}

As a check, for $k=1$ we can substitute $a\to a+X$ in \AUngaugedxL, $\tilde a \to\tilde a +Y$ in \AUngaugedxR\ and subtract the counterterm ${1\over 2\pi} Xd Y$ from both sides, to find the same duality \AGaugedxa\ as in \HsinBLU.

Let us examine the global symmetry of the problem.  First, there is a $U(1)^X\times U(1)^Y$. Second, there is a charge-conjugation symmetry acting as $\bC(a) = -a$, $\bC(X) = -X$, $\bC(Y) = -Y$ (and $\bC(\tilde a) = -\tilde a$ in the dual). We will denote the combined group for these two symmetries as $S\big( O(2)^X \times O(2)^Y \big)$. Third, because of the duality there is the $\Z_2^\CC$ transformation that exchanges $X \leftrightarrow Y$. Fourth, there is a time-reversal transformation with $\bT(a) = a$, $\bT(X) = X$, $\bT(Y) = -Y$ (and $\bT(\tilde a) = -\tilde a$) that acts on the theory as
\eqn\Atimereversal{
\bT \Big[ \CL_0(X,Y) \Big] = \CL_0(X,Y) + \frac{2}{4\pi} (XdX + YdY) - 2 (k^2 - 1) {\rm CS_{grav}} \;,
}
{\it i.e.}\ it is a symmetry up to an anomalous shift of CS counterterms. Next we determine the symmetry that acts faithfully. Operators constructed out of polynomials in $\Psi^i$, $\bar \Psi_i$ and derivatives have even $U(1)^X$ charge and vanishing $U(1)^Y$ charge.  Similarly, operators made out of polynomials in $\chi^i$, $\bar \chi_i$ and derivatives have even $U(1)^Y$ charge and vanishing $U(1)^X$ charge.  We can also consider monopole operators of $a$ or $\tilde a$: they have odd $U(1)^X$ and odd $U(1)^Y$ charge. Hence the symmetry that acts faithfully on the space of operators is $S \big( O(2)^X\times O(2)^Y \big)/\Z_2$. Including $\Z_2^\CC$ and time reversal, we find the symmetry group $S\big( O(2)^X \times O(2)^Y \big)/\Z_2 \rtimes \Z_2^\CC \rtimes \Z_2^\bT$.

Let us consider background gauge fields for the symmetry group that does not act on spacetime. Because of the $\Z_2$ quotient, we allow background fields $X,Y$ with $\int {dX\over 2\pi} \mod 1 = \int {dY\over 2\pi} \mod 1 = \half$.  The restriction on the $U(1)^X\times U(1)^Y$ charges of local operators should make such backgrounds consistent.  However, one can check (\eg\ by defining ordinary $U(1)$ fields $ Z_\pm = X\pm Y$) that the two sides \AUngaugedxL-\AUngaugedxR\ of the duality are not well defined in the presence of such fluxes. This is an anomaly.

As in the other cases, in particular the one in Section 4, we have different options.
\item{1.} We can leave the $(2+1)d$ Lagrangian $\CL_0(X,Y)$ as it is, but then we can only couple it to $S\big( O(2)^X \times O(2)^Y \big) \rtimes \Z_2^\CC$ background fields with no fractional fluxes.
\item{2.} We add to the two sides \AUngaugedxL-\AUngaugedxR\ of the duality the Chern-Simons counterterms $-{1\over 4\pi} (XdX-YdY)$. These counterterms violate the spin/charge relation.  Now we can have $S \big( O(2)^X\times O(2)^Y \big)/\Z_2$ backgrounds, but $\Z_2^\CC$ is violated.
\item{3.} We can attach the theory to a $(3+1)d$ bulk, add suitable bulk terms and obtain a well-defined theory on general backgrounds, but whose partition function depends on the extension of the background fields to the bulk.

Let us explore the third option. The $\theta$-parameters of $S\big( O(2)^X \times O(2)^Y \big)/\Z_2 \rtimes \Z_2^\CC$ are subject to the periodicities
\eqn\Aperiodicities{
(\theta_X, \theta_Y) \sim (\theta_X + 8\pi, \theta_Y) \sim (\theta_X + 4\pi, \theta_Y + 4\pi) \;,
}
and the restrictions $(\theta_X, \theta_Y) \sim (-\theta_X, - \theta_Y) \sim (\theta_Y, \theta_X)$ up to periodicities. We add a boundary term $-\frac1{4\pi} (XdX - YdY)$, {\it i.e.}\ we consider the boundary theory
\eqn\ALone{
\CL_1(X,Y) = \CL_0(X,Y) - \frac1{4\pi} \big( XdX - YdY \big) \;,
}
and also add a bulk term $S_{\rm B}$ with $(\theta_X = 2\pi, \theta_Y = -2\pi)$. Now the boundary theory is well defined on $S\big( O(2)^X \times O(2)^Y \big)/\Z_2$ backgrounds. The $\Z_2^\CC$ transformation is anomalous, and maps $\CL_1(X,Y) \to \CL_1(X,Y) + \frac2{4\pi}(XdX - YdY)$ (making use of the duality), however this is precisely offset by an opposite anomalous transformation of $S_{\rm B}$.

In order to preserve time-reversal as well, we add another boundary term $S_{\rm B}'$ with $(\theta_X = 2\pi, \theta_Y = 2\pi)$ and also $- \frac{k^2-1}{192\pi} \int \Tr R \wedge R$. In the presence of a boundary, the variation of $S_{\rm B}'$ under $\bT$ precisely cancels the one of $\CL_0$ (while the variations of the added boundary term to get $\CL_1$ and of $S_{\rm B}$ cancel among themselves).


\appendix{C}{More 't Hooft anomalies}

We list here the 't~Hooft anomalies for other cases discussed in the main text.
Consider the theories $U(N)_k$ with $N_f$ scalars and $SU(k)_{-N + \frac{N_f}2}$ with $N_f$ fermions. The global symmetry is $U(N_f)/\Z_k$ and charge conjugation that we will neglect. Following the same steps as in Section 2.2, one finds that for generic choices of the CS counterterms and with the same conventions as in \suknPsiwithbackground\ and \unkPhiwithbackground, the anomaly is
\eqn\USUanomaly{
S_{\rm anom} = 2\pi  \int_{\CM_4} \bigg[ \frac{N}k \, \frac{\CP(w_2^{(k)})}2 - \frac{L}{N_f} \, \frac{\CP(w_2^{(N_f)})}2 + \frac{J}{D^2} \frac{\wt F^2}{8\pi^2} \bigg] \;.
}
Here $d = \gcd(k,N_f)$, $D = \lcm(k,N_f) = \frac{kN_f}d$, $F$ is the field strength of $U(1) \subset U(N_f)$ while $\wt F = DF$ is the well-defined and integer field strength of the $U(1)/\Z_D$ bundle, $w_2^{(N_f)}$ is the second Stiefel-Whitney class of the $PSU(N_f)$ bundle and $w_2^{(k)}$ is defined by the constraint
\eqn\constraintUSU{
\frac{\wt F}{2\pi} = \frac{N_f}d\, w_2^{(k)} + \frac kd \, w_2^{(N_f)} \quad\mod D \;.
}
With the choice $J \in D\Z$, using the square of the previous relation the anomaly simplifies to
\eqn\USUanomalyI{
S_{\rm anom} = 2\pi \int_{\CM_4} \bigg[ \frac{J + Nk}{k^2} \; \frac{\CP(w_2^{(k)})}2 + \frac{J - N_f L}{N_f^2} \; \frac{\CP(w_2^{(N_f)})}2 + \frac{J}{kN_f} w_2^{(k)} \cup w_2^{(N_f)} \bigg] \;.
}
The case $k=0$ is special and the formulae above do not directly apply.

So, consider the theory $U(N)_0$ with $N_f$ scalars. In this case the global symmetry is $PSU(N_f) \times U(1)_M$, as well as charge conjugation and time reversal that we neglect. The scalars are coupled to a $U(N_f)$ gauge field $B$ (where $U(1) \subset U(N_f)$ is dynamical) and a dynamical gauge field $b$, with $N_f \Tr db = N\Tr dB$. The coupling to the magnetic $U(1)$ background field $B_M$ is described by the ill-defined expression ${N\over 2\pi N_f} (\Tr B) d B_M$ which needs to be moved to the bulk. This highlights that the global symmetry suffers from an 't~Hooft anomaly. Including a CS counterterm at level $L$ for $SU(N_f)$ (which could be set to zero), the anomaly is characterized by the bulk term
\eqn\Uzeroanomaly{
S_{\rm anom} = 2\pi \int_{\CM_4} \bigg[ \frac{N}{N_f} \, w_2^{(N_f)} \cup \frac{dB_M}{2\pi} - \frac{L}{N_f} \, \frac{\cP(w_2^{(N_f)})}2 \bigg] \;,
}
where we have identified $\frac1{2\pi} \Tr dB = w_2^{(N_f)} \mod N_f$. This expression can be regarded as a singular limit of \USUanomaly.

Similarly, the theory $U(k)_{\frac{N_f}2}$ with $N_f$ fermions has global symmetry $U(N_f)/\Z_{N_f}$, besides charge conjugation that we neglect. The expression \SUUanomaly\ for the anomaly does not directly apply (since $N=0$). Following similar steps as before, we find that the anomaly is characterized by the bulk term
\eqn\Uzerofermionanomaly{
S_{\rm anom} = 2\pi \int_{\CM_4} \bigg[ \frac{k}{N_f} \, w_2^{(N_f)} \cup \frac{dB_M}{2\pi} - \frac{L}{N_f} \, \frac{\cP(w_2^{(N_f)})}2 \bigg] \;.
}

The other time-reversal invariant theory is $U(k)_0$ with $N_f$ fermions, which requires $N_f$ to be even. The UV symmetry is $U(N_f)/\Z_{N_f/2}$ together with charge conjugation and time reversal. Applying \SUUanomalyI\ with $N=N_f/2$, the anomaly is
\eqn\anomftinv{
S_{\rm anom} = 2\pi  \int_{\CM_4} \bigg[ - {2k\over  N_f} \, \frac{\CP(w_2^{(N_f/2)})}2 - \frac L{N_f} \, \frac{\CP(w_2^{(N_f)})}2 + \frac{J}{N_f^2} \frac{\wt F^2}{8\pi^2} \bigg]
}
where $\wt F$ satisfies \correlationbundles\ with $d=N_f/2$, $D=N_f$.
Besides, under time reversal there is an anomalous shift by $\big(SU(N_f)_{-2L-k}\times U(1)_{-2K_f}\big)/\Z_{N_f}$ where $K_f={4J-2N_fk\over N_f^2}$.
For the special case $k=1$, $N_f=2$ we can choose the counterterms $L=J=0$ such that there is no anomaly for the $U(2)=\big( SU(2)\times U(1)\big)/\Z_2$ symmetry, but there is a time-reversal anomaly that shifts the theory by $\big( SU(2)_{-1}\times U(1)_2\big)/\Z_2$. We elaborate more on the anomaly for $U(1)_0$ with two fermions in Section 4.


\listrefs
\end